\documentclass[aps,prd,showpacs,twocolumn,superscriptaddress,sort&compress,floatfix]{revtex4-1}

\usepackage{graphicx}
\usepackage{amsmath}

\newcommand{\Tr}{\mathrm{Tr}}

\usepackage{microtype}

\begin{document}

\title{Effects of a dressed quark-gluon vertex in vector heavy-light mesons\\
	and theory average of $B_c^*$ meson mass}


\author{M. G\'{o}mez-Rocha}
\email[]{gomezr@ectstar.eu}
\affiliation{ECT*, Villa Tambosi, 38123 Villazzano (Trento), Italy}

\author{T. Hilger}
\email[]{thomas.hilger@uni-graz.at}
\affiliation{Institute of Physics, University of Graz, NAWI Graz, A-8010 Graz, Austria}

\author{A. Krassnigg}
\email[]{andreas.krassnigg@uni-graz.at}
\affiliation{Institute of Physics, University of Graz, NAWI Graz, A-8010 Graz, Austria}

\date{\today}

\begin{abstract}
We extend earlier investigations of heavy-light pseudoscalar mesons to the vector case,
using a simple model in the context of the Dyson-Schwinger-Bethe-Salpeter
approach. We investigate the effects of a dressed-quark-gluon vertex in a systematic
fashion and illustrate and attempt to quantify corrections beyond the phenomenologically
very useful and successful rainbow-ladder truncation. In particular we investigate dressed
quark photon vertex in such a setup and make a prediction for the experimentally
as yet unknown mass of the $B_c^*$, which we obtain at $6.334$ GeV well in line with
predictions from other approaches. Furthermore, we combine a comprehensive 
set of results from the theory literature. The theory average for the mass of the 
$B_c^*$ meson is $6.336\pm0.002$ GeV.

\end{abstract}


\maketitle

\section{ Introduction }\label{sec:intro}

The Dyson-Schwinger-Bethe-Salpeter-equation (DSBSE) approach is a modern nonperturbative
framework based on continuum quantum field theory 
\cite{Roberts:2007jh,Fischer:2006ub,Alkofer:2000wg,Sanchis-Alepuz:2015tha} 
and is thus complementary to lattice-regularized 
QCD \cite{Dudek:2007wv,Lang:2011mn,Liu:2012ze,Thomas:2014dpa,Flynn:2015mha,Lang:2015hza} 
and other modern approaches to the strong-interaction sector of the standard model of elementary particle physics.

Modern DSBSE studies with phenomenological background mostly use a setup where a simple truncation
is combined with a sophisticated effective model interaction, see 
\cite{Maris:1999nt,Holl:2003dq,Holl:2004fr,Krassnigg:2004if,Holl:2005vu,Alkofer:2005ug,Eichmann:2008kk,Eichmann:2008ef,Eichmann:2008ae,%
Krassnigg:2009zh,Eichmann:2009qa,Alkofer:2009jk,Krassnigg:2010mh,Dorkin:2010ut,Blank:2010bz,Blank:2010pa,%
Mader:2011zf,Blank:2011ha,UweHilger:2012uua,Popovici:2014pha,Hilger:2014nma,Fischer:2014xha,Fischer:2014cfa,%
Hilger:2015hka,Hilger:2015ora,Hilger:2015zva,Raya:2015gva} and references therein.
Beyond the most popular rainbow-ladder (RL) truncation, systematic schemes exist to explore the infinite
system of Dyson-Schwinger equations (DSEs) in a symmetry-preserving fashion \cite{Bender:1996bb,Sanchis-Alepuz:2015tha}. 
In a concrete, numerical setup \cite{Bhagwat:2007rj,Krassnigg:2008gd,Blank:2010bp,Blank:2010sn,Blank:2011qk},
one faces increasing complexity  
\cite{Watson:2004jq,Watson:2004kd,Fischer:2005en,Fischer:2008wy,Fischer:2009jm,Williams:2009wx,%
Williams:2014iea,Sanchis-Alepuz:2014wea,Chang:2009zb,Heupel:2014ina,Sanchis-Alepuz:2015qra,Williams:2015cvx,%
Fu:2015tdu,Binosi:2016rxz,Qin:2016fbu} 
such that simple models are of an obvious advantage, e.\,g., 
\cite{Horvatic:2007qs,Horvatic:2007wu,Horvatic:2010md,Roberts:2011wy,GutierrezGuerrero:2010md,Bedolla:2015mpa,Segovia:2016iaf}
and references therein.

A particularly simple effective interaction \cite{Munczek:1983dx} is also employed in our present work,
which was used in the past to study certain classes of diagrams or particular effects of interest 
\cite{Bender:1996bb,Alkofer:2000wg,Bender:2002as,Krassnigg:2003dr,Bhagwat:2004hn,Holl:2004qn,%
Matevosyan:2006bk,Matevosyan:2007cx,Jinno:2015sea}. These can then easily serve as both a
testing ground for and a means to estimate missing effects in a setup using a more sophisticated 
effective interaction.

In this work we continue an investigation of a systematically dressed quark-gluon vertex (QGV) 
which consistently enters both the quark DSE and the meson Bethe-Salpeter-equation (BSE) via their
respective integral-equation kernels \cite{Bender:2002as,Bhagwat:2004hn,Gomez-Rocha:2014vsa,Gomez-Rocha:2015qga}.
Following up on \cite{Gomez-Rocha:2015qga}, our focus remains on heavy-light mesons, which probe
the underlying equations and their building blocks such as the QGV in different ways.
For example, dressing effects for the quark propagator have been questioned and tested for the case of b-quarks 
\cite{Nguyen:2009if,Souchlas:2010zz,Nguyen:2010yh}, since one can make use of simplifying 
assymptions about the heavy-quark propagator based on the large value of the quark mass
\cite{Ivanov:1997iu,Ivanov:1997yg,Ivanov:1998ms,Blaschke:2000zm,Bhagwat:2006xi,Ivanov:2007cw,ElBennich:2009vx,ElBennich:2010ha}.
Ultimately, one goal is to check heavy-quark symmetry predictions \cite{Neubert:1993mb} as,
e.\,g., in relativistic Hamiltonian dynamics \cite{Keister:1991sb,Krassnigg:2003gh,Krassnigg:2004sp,Polyzou:2015rra} 
as well as reduced versions of the BSE \cite{Blankenbecler:1965gx,Gross:1969rv}, 
where heavy quarks have been under renewed investigation recently 
\cite{GomezRocha:2012zd,Gomez-Rocha:2014aoa,Li:2015zda,Leitao:2015cxa}.
Another goal is to prepare, e.\,g., investigations of the spectral difference of parity partners
in analogy to recent progress with QCD sum rules 
\cite{Thomas:2007es,Hilger:2009kn,Hilger:2010zb,Hilger:2010cn,Hilger:2011cq,Hilger:2012db,Buchheim:2014rpa,%
Gubler:2014pta,Buchheim:2015yyc,Buchheim:2015xka,Gubler:2015yna}.

The article is organized as follows: In Sec.~\ref{sec:setup} we briefly sketch the setup used for the quark
DSE, the QGV, and the meson BSE. Results and discussion are presented in Sec.~\ref{sec:results}; 
conclusions follow in Sec.~\ref{sec:conclusions}. 
Technical details are collected in the appendices.

\section{Setup}\label{sec:setup}

Since this work is an extension of \cite{Bender:2002as,Bhagwat:2004hn,Gomez-Rocha:2015qga}, we only
very briefly sketch the relevant formulae, mostly in order to be able to understand and interpret the results
presented as well as to connect to the new details presented 
in the appendices. For a more complete presentation of our particular setup and approach, see 
\cite{Gomez-Rocha:2015qga}. More details on the case of equal-mass constituents can be found in 
\cite{Bhagwat:2004hn}, and the truncation scheme and basic assumptions are laid out in \cite{Bender:2002as}.
Our calculations are performed in Euclidean momentum space.

\subsection{Quark DSE}

Solution of a bound-state problem in the DSBSE formalism requires knowledge of the building blocks and
their interactions. In our case the meson BSE requires us to know the quark propagator for both
the heavy and the light quark under consideration, and the quark-gluon interaction as well as
the gluon propagator. We go \emph{in medias res} by assuming the simplification inherent in 
the effective interaction of \cite{Munczek:1983dx}, namely the Munczek-Nemirovsky (MN) gluon-momentum dependence
\begin{equation} \label{eq:mnmodel} 
D_{\mu\nu}(k) \sim \mathcal{G}^2 \, \delta^4(k)\,,
\end{equation} 
where $D_{\mu\nu}$ is the renormalized dressed gluon propagator and $\mathcal{G}$ an effective coupling
constant, which sets the scale of the model. This transforms all integral equations into algebraic equations. In addition,
since this model is UV finite, all renormalization constants are $=1$.

In particular, the quark DSE reads
\begin{equation}\label{eq:quarkdse}
 S^{-1}(p)= i \gamma\cdot p+m_q + \gamma_\mu S(p)\Gamma_\mu^\mathcal{C}(p)\;,
\end{equation}
where the renormalized dressed quark propagator $S$ has the form
\begin{eqnarray} 
\label{eq:quarkprop} 
S(p)^{-1}&=&i\gamma\cdot p A(p^2)+B(p^2)\\
&=&A(p^2)\left(i\gamma\cdot p + M(p^2) \right)\;
\end{eqnarray} 
with the dressing functions $A$ and $B$ or, alternatively, $A$ and $M$; $m_q$ is the current-quark
mass, and flavor is inherent to the solution depending on $m_q$.

The renormalized dressed QGV is written as $\Gamma^a_\nu$ with the color index ${}^a$, which we write explicitly as
$\Gamma_\mu^a(p)=\frac{\lambda^a}{2} \Gamma_\mu(p)$.
Furthermore, we have set $\mathcal{G}=1$ in Eq.~(\ref{eq:quarkdse}) and the following, thereby obtaining all 
dimensioned quantities in appropriate units of $\mathcal{G}$.
The model parameter $\mathcal{C}$ introduced in Eq.~(\ref{eq:quarkdse}) and its meaning are best
illustrated via the DSE for the QGV, following \cite{Bhagwat:2004hn} obtained as 
the effective equation 
\begin{equation}\label{eq:qgvdse}
\Gamma_\mu^\mathcal{C}(p)=\gamma_\mu-\mathcal{C}\,\gamma_\rho\, S(p)\,\Gamma_\mu^\mathcal{C}(p)\,S(p)\,\gamma_\rho\;,
\end{equation}
where the dependence on $\mathcal{C}$ stems from the effective combination of the abelian and non-abelian correction terms
in the QGV DSE, and the value of $\mathcal{C}$ is chosen in accordance with, e.\,g., lattice QCD or phenomenology.

Concrete possible values are: $\mathcal{C}=-1/8$, corresponding to abelian-only dressing \cite{Bender:2002as};
$\mathcal{C}=0$ corresponding to RL truncation; $\mathcal{C}=0.51$, used in \cite{Bhagwat:2004hn}
as a result from fitting to lattice quark propagators \cite{Bowman:2002kn,Bowman:2002bm,Bhagwat:2003vw}. 
Herein, we fix $\mathcal{C}=0.51$ throughout for easy comparison and direct connection to the earlier studies
of \cite{Bhagwat:2004hn,Gomez-Rocha:2015qga}.

To define our truncation scheme \cite{Bender:2002as}, we iterate eq.~(\ref{eq:qgvdse}) such that the bare QGV serves 
as a starting value $\Gamma_{\mu,0}^\mathcal{C}(p)=\gamma_\mu$ and the recursion relation is
\begin{equation}\label{eq:qgvrecursion}
\Gamma_{\mu,i}^\mathcal{C}(p)=-\mathcal{C}\,\gamma_\rho\, S(p)\,\Gamma_{\mu,i-1}^\mathcal{C}(p)\,S(p)\,\gamma_\rho\;.
\end{equation}
At a given order $n$ in this scheme one has for the QGV
\begin{equation}\label{eq:summedvertex}
\Gamma_\mu^\mathcal{C}(p)=\sum_{i=0}^n \Gamma_{\mu,i}^\mathcal{C}(p).
\end{equation}
and the fully dressed result for the QGV is obtained by $n\rightarrow\infty$. Note that the flavor
content of Eqs.~(\ref{eq:qgvdse}) and (\ref{eq:qgvrecursion}) is implicitly carried by the factors of $S$.

\subsection{Meson BSE}\label{sec:bse}

The meson BSE in the current setup is simplified in a similar fashion to the quark DSE, namely via
the effective interaction's property (\ref{eq:mnmodel}). The solution of the BSE, the Bethe-Salpeter amplitude
(BSA) is often combined with the quark propagators in the integration kernel to the so-called Bethe-Salpeter 
wave function $\chi$ and we have
\begin{equation} \label{eq:chi} 
\chi(P): =S(q_+)\, \Gamma(P) \,S(q_-)\;. 
\end{equation} 
The meson flavor is determined by the quark flavors of the two factors of $S$, and the total meson momentum
is the only remaining variable, since the quark and antiquark momenta are reduced to $q_+=\eta P$ and $q_-=-(1-\eta)P$. 

The momentum partitioning parameter $\eta\in [0,1]$ is in principle arbitrary in any covariant computation 
as a result of the freedom in the definition of the quark-antiquark relative momentum such that observables are 
independent of $\eta$. However, our particular model interaction is oversimplifying in the sense that not
all possible covariant structures of the BSA are retained. As a result, there is a dependence on $\eta$, which is
a model artifact and must be properly analyzed in any study using this particular interaction. 
Such an analysis was already performed in Ref.~\cite{Munczek:1983dx} and also in our previous work
on pseudoscalar mesons in \cite{Gomez-Rocha:2015qga}; for our present study, this analysis is presented in
App.~\ref{sec:etadependence}. In the presence of such a detailed analysis, this model 
artifact does not destroy the model's capacity to elucidate our investigation's goals. Furthermore, it is 
easily quantified and thus well under control.

For the unequal-mass case in our setup, the BSE reads, see \cite{Gomez-Rocha:2015qga} and App.~\ref{sec:ProofAposteriori},
\begin{eqnarray}\nonumber 
\Gamma^M(P) &=& -\frac{1}{2}\left[ \gamma_\mu \chi^M(P)\,\Gamma_\mu^\mathcal{C}(q_-)  \right.\\ \nonumber
&+&  \gamma_\mu  S(q_+)  \, \Lambda^M_\mu(P) +  \Gamma_\mu^\mathcal{C}(q_+)\chi^M(P)\gamma_\mu\\
&+& \left. \Lambda^M_\mu(P)  S(q_-)  \gamma_\mu  \right]\;. \label{eq:mngenbsekernel}
\end{eqnarray} 
The superscript label ${}^M$ denotes the type of meson under study, since the structure of the 
correction term $\Lambda^M_\mu$ depends on the structure of the corresponding BSA. Herein we consider
the vector meson case, for which all details are given appropriately in the appendices.

The quark momenta $q_\pm$ in this equation denote the flavor content and, in particular, the mass
ordering among the quarks in that the heavier quark is associated with the subscript ${}_+$.

While the first term on the r.h.s.\ of Eq.~(\ref{eq:mngenbsekernel}) is straight-forward to construct from a
given QGV, the construction of the second term is based on a recursion relation analogous to the
one for the QGV. Correction terms are summed up to a particular order $n$ to get $\Lambda^M$ as
\begin{equation}\label{eq:lambda}
\Lambda^M_\nu(P) = \sum_{i=0}^n \Lambda_{\nu,i}^M(P)\,,
\end{equation} 
and the full result is then obtained by $n\rightarrow\infty$.

The recursion relation reads \cite{Bender:2002as,Bhagwat:2004hn}:
\begin{eqnarray} \nonumber 
\frac{1}{\mathcal{C}}\Lambda_{\nu,n}^M(P) &= & -\gamma_\rho \chi^M(P)\Gamma_{\nu,n-1}^\mathcal{C}(q_-)S(q_-) \gamma_\rho\\ \nonumber
&-& \gamma_\rho S(q_+) \Gamma_{\nu,n-1}^\mathcal{C}(q_+) \chi^M(P) \gamma_\rho \\
&-& \gamma_\rho S(q_+) \Lambda_{\nu,n-1}^M(P) S(q_-) \gamma_\rho\;, \label{eq:recbiglambda}  
\end{eqnarray} 
where quark flavors and properties in the factors of $S$ and $\Gamma_\nu$ are given via the subscripts ${}_\pm$
in their argument, as described above.

Evaluating the recursion relations to a desired order, one uses the initial condition \cite{Bender:2002as}
\begin{equation}
\Lambda_{\nu,0}^M(P) = 0 \;.
\end{equation} 
In the pseudoscalar case for equal-mass quarks and $\eta=1/2$ this implies \cite{Bender:2002as}
\begin{equation}
\Lambda_{\nu,0}^\mathrm{P}(P) = 0 \quad\Rightarrow\quad \Lambda_{\nu}^\mathrm{P}(P) \equiv 0\;,
\end{equation} 
which was used as a testing case for our general setup in \cite{Gomez-Rocha:2015qga}.
In the vector case, however, no symmetry exists to enable such a cancellation and thus an appropriate
testing case is the equal-mass result presented in \cite{Bender:2002as}. Further details on the construction
of $\Lambda_{\nu}^\mathrm{V}(P)$ are technical and thus collected in App.~\ref{sec:vekerneldetails}.

\section{Results and Discussion}\label{sec:results}

We investigate the effect of QGV dressing on vector-meson ground-state masses
in the scheme described above as a representative way to apply systematic corrections to
the often and well used RL truncation.

As mentioned above, our simplified model leads to an artificial dependence on the 
momentum-partitioning parameter $\eta$, which one must study, but nontheless not put
in the center of attention. We present the dependence on $\eta$ in detail in App.~\ref{sec:etadependence}
and produce corresponding error bars in our comparison to experimental data below in Fig.~\ref{fig:expcomp};
however, other than this we focus on one particular representative value for $\eta$ and compare 
our results for the various dressing stages in the scheme in physically meaningful ways.

The study of mesons with unequal-mass constituents was started in Ref.~\cite{Gomez-Rocha:2015qga}
for the pseudoscalar case. While we presented also some detailed analysis of the 
quark propagator dressing functions there, we will not repeat those here. Instead, our focus is
the vector-meson case in general and two interesting items in particular: First, we study
the dressed quark-photon vertex by solving the inhomogeneous BSE for the first time in the
scheme under consideration here. Second, we predict the mass of the $B_c^*$ meson via
a pseudoscalar-vector-splitting analysis.
Overall, our results allow not only qualitative, but also quantitative statements.

Our model parameters are fixed to the values used earlier in \cite{Bhagwat:2004hn} and 
\cite{Gomez-Rocha:2015qga}: 
$\mathcal{C}=0.51$, $\mathcal{G}=0.69$, and the current-quark masses are 
$m_u=0.01$ GeV, $m_s=0.166$ GeV, $m_c=1.33$ GeV, $m_b=4.62$ GeV. For the light isovector case we assume
isospin symmetry and the equality of the current-quark masses of the $u$ and $d$ quarks.

Note that this set of parameters was originally found to fit quarkonium vector-meson masses throughout the entire 
quark-mass range. As a result, our numbers presented below in Fig.~\ref{fig:expcomp} are not aimed at
nor to be understood in the sense of a pure theory-experiment comparison. While in some cases
agreement is excellent and the use of splittings is a perfectly fine example of a valid technique under
our circumstances, we would like to stress the emphasis on the size of dressing effects as they 
are produced here.

\subsection{Meson BSE}

We present results for vector-meson ground-state masses. In the figures in this section, we plot meson masses as 
functions of the order $n$ in our truncation scheme. In addition, we discuss the differences of the various masses
from the fully dressed result at every $n$ below. Before we discuss the results and figures in detail, we remark
that it is possible that the homogeneous BSE doesn't have a solution for a particular setup, configuration or
set of parameters. In such a case the corresponding data point's place in the figure is left empty.

\begin{figure*}[t]
  \includegraphics[width=0.9\textwidth, clip=true]{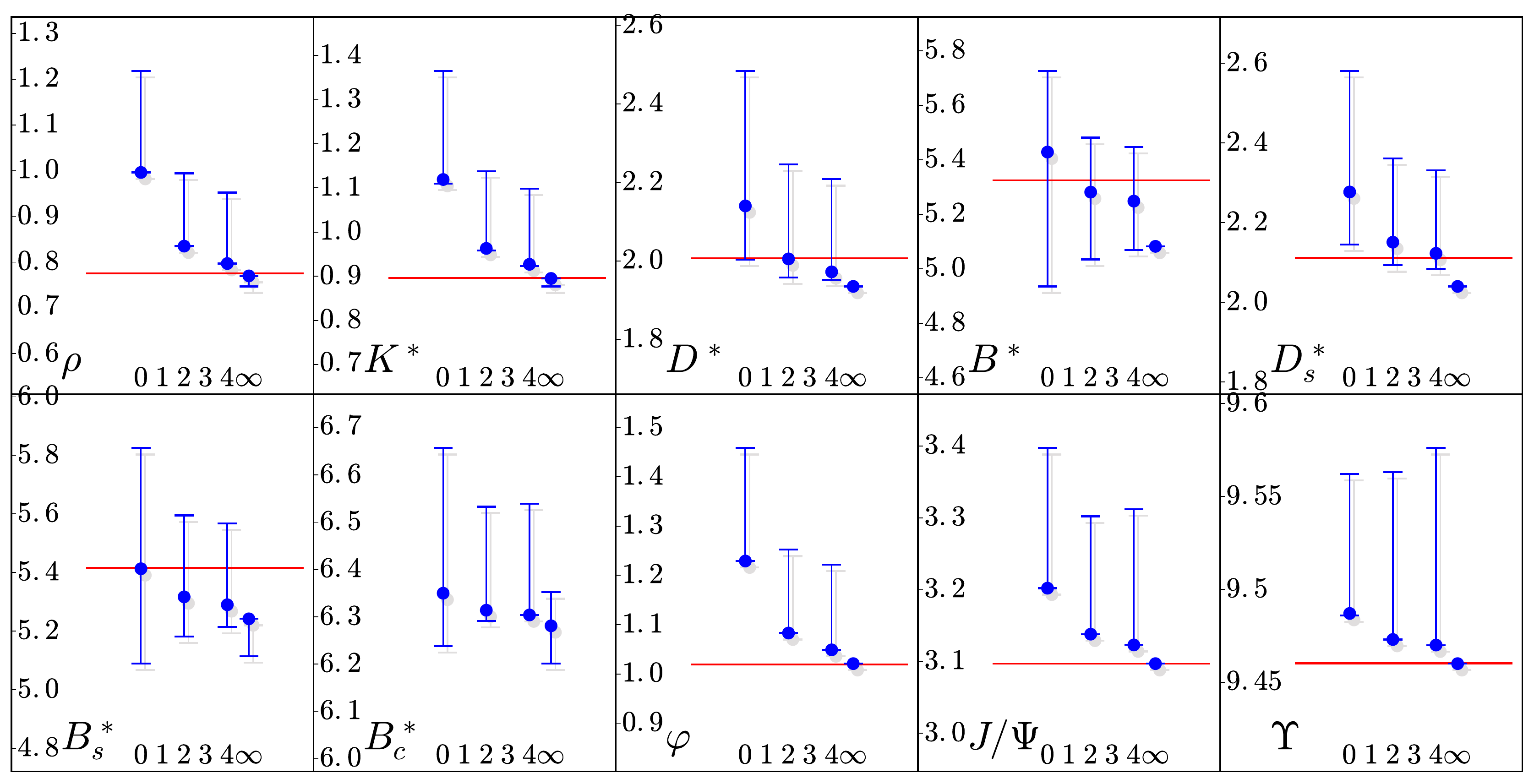}
\caption{\label{fig:expcomp}
Bound-state masses for $\rho$ meson, $\varphi$, $J/\Psi$, $\Upsilon$, and all different flavored vector ground states 
as a function of $n$, given in GeV. 
The dependence on $\eta$ is illustrated via the error bars. Calculated results are given by blue dots; experimental
data are represented by horizontal lines. \cite{Olive:2014rpp}}
\end{figure*}

Let us look at the convergence of the results with $n$ and the comparison with experimental data first. These results
are presented in Fig.~\ref{fig:expcomp} in several boxes, one for each quark-flavor combination.
The filled circles in the plots are our results for each $n$, where available, for a fixed value of $\eta$ in each case.
In particular: $\eta=0.5$ for the $\rho$, $\varphi$, $J/\Psi$, and $\Upsilon$, $0.6$ for the $K^*$, $0.75$ for the $B_c^*$, 
$0.8$ for the $D^*$ and $D_s^*$, $0.9$ for the $B_s^*$, and $0.95$ for the $B^*$.

The actual $\eta$ dependence for each case is encoded in the form of a systematic error in our results in Fig.~\ref{fig:expcomp}: 
the error bars are plotted from the lowest to the largest value of the mass result for any given $n$. Thus, they are
asymmetric and the value of $\eta$ chosen for the data point, as defined below, can be also either
the smallest or the largest value available at this $n$. 

It should be noted here that we chose each $\eta$ via the requirement to find a solution of the BSE for all $n$, 
if possible. While this doesn't seem to work for odd values of $n$, we are able to find $\eta$ values such that 
a solution can be obtained for $n=\infty$ in addition to the even values of $n$. As it turns out (see also the 
figures in App.~\ref{sec:etadependence}) this corresponds to a value of $\eta$ where the dressing effects for the
meson under consideration are close to minimal with respect to their range as functions of $\eta$. The asymmetric
values given above also make sense in correlation to the asymmetry of the quark-antiquark-mass content in each meson.
The various aspects of $\eta$ and their influence on the quark-propagator dressing functions have been discussed
in detail in our previous investigation for the pseudoscalar meson-case in Ref.~\cite{Gomez-Rocha:2015qga}.
One may, at this point, speculate that an actual minimization of the dressing effects over the $\eta$-parameters space
would lead to values very similar to the ones quoted here.

The largest error bars resulting from the $\eta$ range appear to be of the order of 20\%, which is in rough agreement with 
our previous work \cite{Gomez-Rocha:2015qga} as well as the analysis in the original \cite{Munczek:1983dx}, 
where the authors quote changes smaller than 15\%. However, our detailed analysis presented in the plots in 
Figs.~\ref{fig:etaeven} and \ref{fig:etauneven} in App.~\ref{sec:etadependence} clearly indicate that 
the extreme values $\eta=0$ and $1$ produce
those masses with the largest deviations from the data point chosen for experimental comparison.
In contrast to the observation regarding close-to-minimal dressing effects for our chosen values of $\eta$,
one could state here that at the boundaries of the $\eta$ interval $[0,1]$ dressing effects appear to be
maximal instead. This effect can also be traced back to the extended domain probed by extreme $\eta$ values
in the quark-propagator dressing functions that are involved via the quark momenta squared $q_\pm^2$, which
are directly proportional to $\eta^2$ and $(1-\eta)^2$, respectively. It is on these extended domains that
dressing effects are larger than close to or in the spacelike domain \cite{Gomez-Rocha:2015qga}.

In this sense it is certainly correct to state that the error bars in 
Fig.~\ref{fig:expcomp} should represent the entire range of $\eta$ observed in our calculations; on the
other hand it also means that in practice the extreme $\eta$ values have to be taken with a grain of salt
in the sense that they may not be representative to an amount that actually justifies the size of these
error bars and we in general regard them as overestimates of more suitably defined systematic errors.
In addition, we remark that the figures in App.~\ref{sec:etadependence} also show cases where very few or
even only one of the $\eta$ values on our standard grid produce a solution of the corresponding BSE.
These cases are easily recognized by their small error bars, 
which we chose not to rescale or blow up artificially. Note that it is possible that solutions exist for
values of $\eta$ that are not part of our standard grid.

In terms of the comparison to experiment and the convergence behavior we find a clear pattern of 
higher, even $n$ lowering the meson mass with the fully dressed result again being lower than the result 
for our largest finite $n$ presented here, namely $n=4$.
For odd $n$ in general no solutions were found. We note at this point why we do not find solutions in the 
odd-$n$ cases: our solutions of the homogeneous
BSE are obtained by finding zeros of the appropriate determinant. It turns
out that for the odd-$n$ cases, the determinant becomes complex at and below
some particular negative value for $P^2$. If a zero is found above this value (which is
the case for some of the pseudoscalar cases studied in \cite{Gomez-Rocha:2015qga}), we have a solution. 
For the present investigation of vector-mesons, which are heavier than their 
pseudoscalar counterparts, it appears that no zero of the determinant exists on the 
domain where it is still real. 

Experimental values for the quarkonia were fitted via the quark masses, which is evident from the
corresponding subplots in Fig.~\ref{fig:expcomp}. For the $K^*$, agreement of the fully dressed result and
the experimental mass value is excellent; in the other cases, experimental values are underestimated by
our results. An experimental value for the mass of the $B_c^*$ meson is still missing, and we predict
a value below via the use of the pseudoscalar-vector mass splitting.

\begin{figure*}[t]
  \includegraphics[width=0.9\textwidth, clip=true]{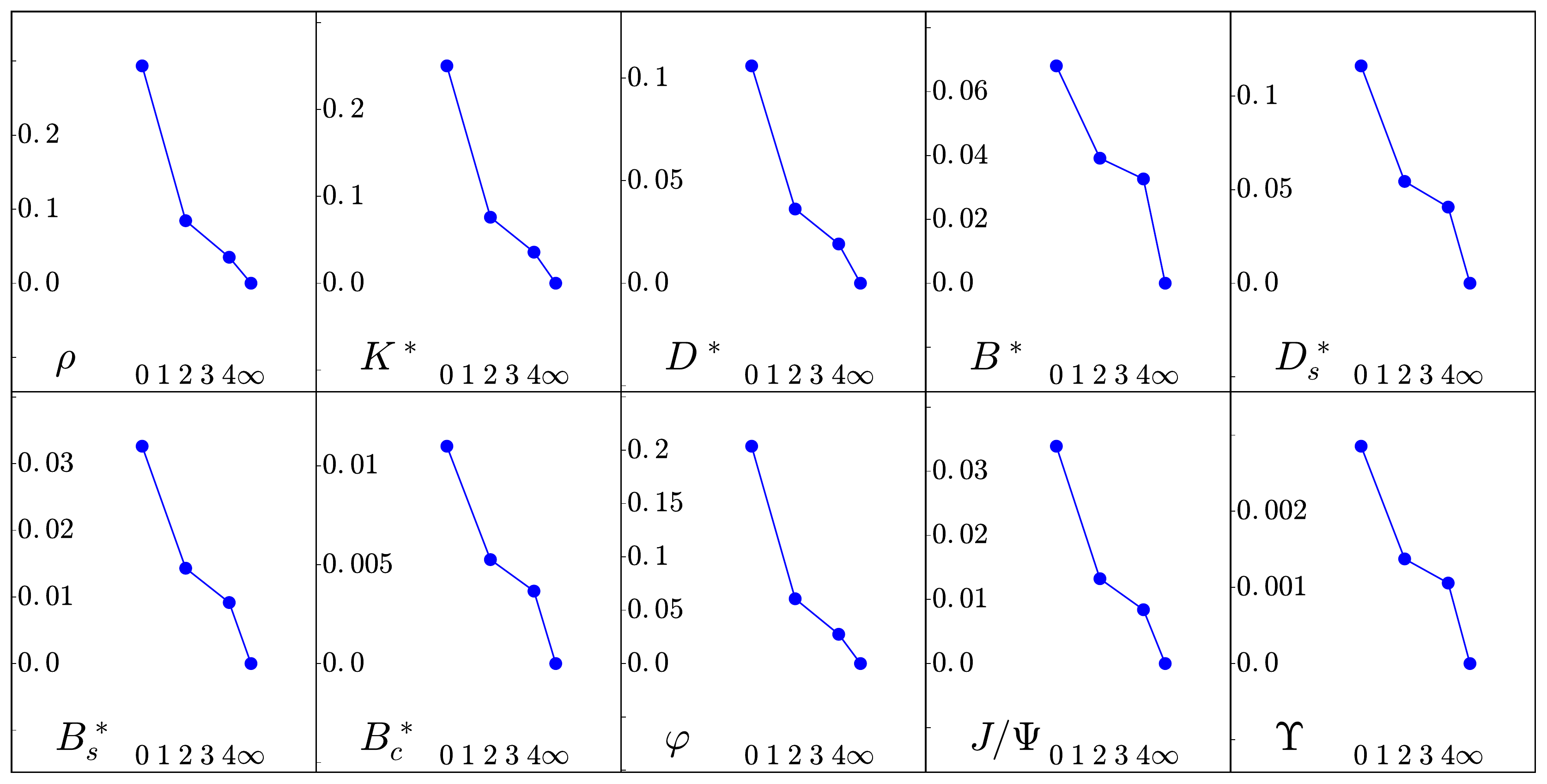}
\caption{\label{fig:reldiff}
Relative mass differences $\Delta m_H^{\mathrm{rel}\,n}$ to fully dressed result for $\rho$ meson, $\varphi$, $J/\Psi$, $\Upsilon$, 
and all different flavored vector ground states as a function of $n$ analogous to Fig.~\ref{fig:expcomp}.
Note that by definition Eq.~(\ref{eq:reldiffmeson}) one obtains $\Delta m_H^{\mathrm{rel}\,\infty}=0$ in each case. }
\end{figure*}

Next we have a look at the relative differences of meson masses at each value of $n$ compared to $n=\infty$,
defined via
\begin{equation}\label{eq:reldiffmeson}
\Delta m_H^{\mathrm{rel}\,n}:=\frac{\Delta m_H^n}{m_H^{n\rightarrow\infty}}
:=\frac{m_H^{n}-m_H^{n\rightarrow\infty}}{m_H^{n\rightarrow\infty}} \;,
\end{equation} 
which is dimensionless. Note that instead of comparing the difference to the fully dressed result here, one may also 
divide by the RL result; however, such a construction is uniquely related to Eq.~(\ref{eq:reldiffmeson}) and, since the 
differences are small, this choice does not affect our discussion.

\begin{table}[b]
\caption{Absolute and relative mass differences for the vector mesons
with all possible flavor combinations, together with the corresponding pseudoscalar values adapted from 
Ref.~\cite{Gomez-Rocha:2015qga}. $\Delta m_H^0$ is given in GeV, the other quantities are dimensionless 
(see text).  \label{tab:diffscollected}}
\begin{ruledtabular}
  \begin{tabular}{ l  c  c  c  c }
 $H$ & $\Delta m_H^0$ & $\Delta m_H^{\mathrm{rel}\,0}$ & $\Delta m_H^0 (\mathrm{P})$ \cite{Gomez-Rocha:2015qga} & 
 $\Delta m_H^{\mathrm{rel}\,0} (\mathrm{P})$ \cite{Gomez-Rocha:2015qga}  \\ \hline
 $\rho$ &  0.226 & 0.294  & 0.011 & 0.078 \\ 
 $\varphi$ &  0.208 & 0.204  & \ldots & \ldots \\ 
 $J/\Psi$ &  0.105 & 0.034  & 0.048 & 0.016 \\ 
 $\Upsilon$ &  0.027 & 0.003  & 0.016 & 0.002 \\ 
 $K^*$ &  0.224 & 0.250  & 0.031 & 0.072 \\ 
 $D^*$ &  0.205 & 0.106  & 0.059 & 0.034 \\ 
 $B^*$ &  0.346 & 0.068  & 0.124 & 0.025 \\ 
 $D_s^*$ &  0.237 & 0.116  & 0.074 & 0.039 \\ 
 $B_s^*$ &  0.171 & 0.033  & 0.099 & 0.019 \\ 
 $B_c^*$ &  0.122 & 0.020  & 0.150 & 0.024 \\ 
  \end{tabular}
\end{ruledtabular}
\end{table}

The results for $\Delta m_H^{\mathrm{rel}\,n}$ are plotted Fig.~\ref{fig:reldiff}. 
In addition, the values for $\Delta m_H^{\mathrm{rel}\,0}$ are tabulated in the second data column in 
Tab.~\ref{tab:diffscollected} together with the absolute differences in mass $\Delta m_H^0$ for a given meson $H$, which 
is obtained between fully dressed and RL result, given in the first data column of 
Tab.~\ref{tab:diffscollected}.  Note that all values in this table are also $\eta$ dependent, and we calculate
the ones presented here at the $\eta$ values given above for each meson case.

The results follow the expected pattern that, where heavier quarks are involved, the dressing effects 
tend to be smaller. While such a statement is certainly true regarding the relative differences, the vector
case is not as clear in this regard as the pseudoscalar one, if one considers the absolute differences.

More precisely, we find that in comparable cases like the bottom-flavored mesons, the absolute differences
are the smaller the heavier the other quark flavor is. The largest absolute difference of almost $350$ MeV 
from RL truncation to the fully-dressed case is found, expectedly, for the most unbalanced system, the $B^*$ 
meson case, whose value is more than twice as large
as for the corresponding pseudoscalar, the $B$. 

The smallest $\Delta m_H$, on the other hand, unsurprisingly as well nonetheless, is found for the 
bottomonium case of the $\Upsilon$, where we find only $27$ MeV; still this is almost twice as large as
in the pseudoscalar counterpart, the $\eta_b$. These ratios are of interest, in particular, since the
values of the hyperfine splitting in heavy quarkonia was an issue of recent debate.

\begin{figure*}[t]
  \includegraphics[width=0.9\columnwidth]{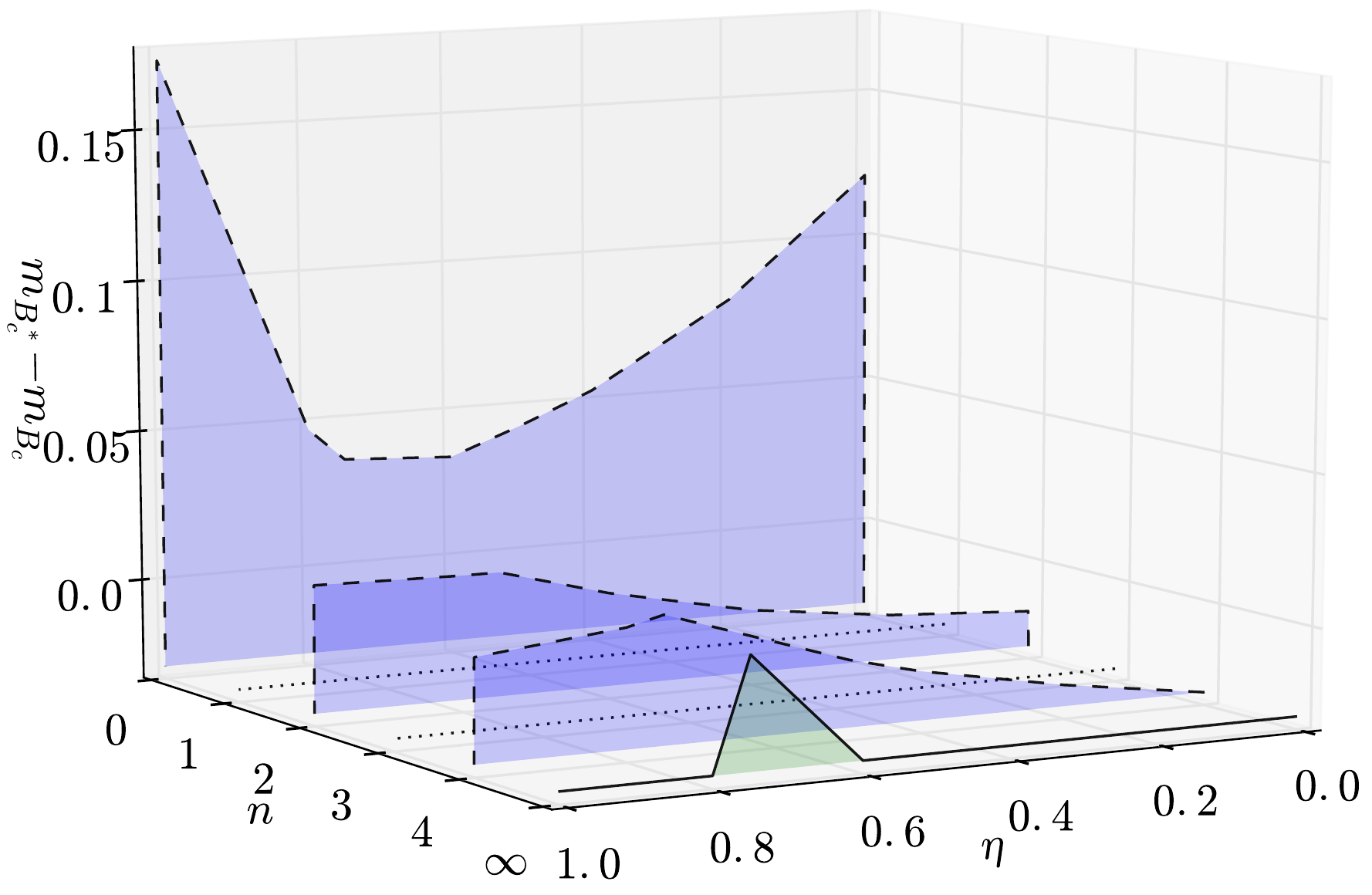}
  \includegraphics[width=0.9\columnwidth]{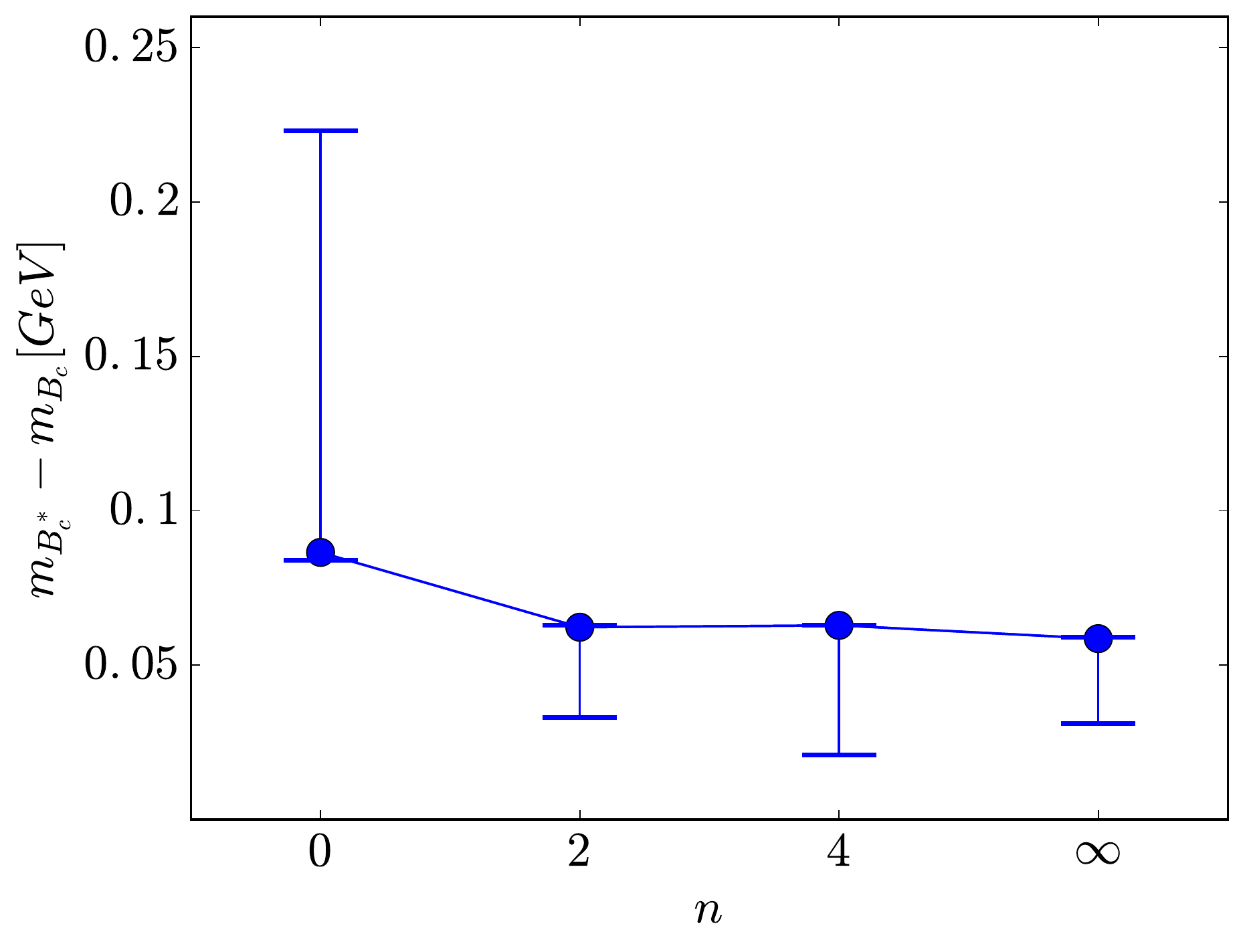}
\caption{\label{fig:bcstar}
Left panel: $B_c^*-B_c$ mass splitting as a function of $n$  and $\eta$ in our scheme.
Right panel: $B_c^*-B_c$ mass splitting as a function of $n$ in our scheme for fixed $\eta=0.75$ (blue circles); 
error bars indicate variations with respect to $\eta$.}
\end{figure*}

Overall, we find that relative dressing effects are of the order of 30\% for the $\rho$ and $K^*$, and go down to
a few percent for the $B_c^*$ or even below one percent for the $\Upsilon$. The sizes of relative
dressing effects increase with a decrease of either the meson mass or the sum of the quark masses
in the meson, which is a natural outcome and interpretation. 

Regarding absolute dressing effects we find that these are significantly more pronounced in the 
vector-meson case than the pseudoscalar one. We also see
that a two-loop vertex dressing ($n=2$) already covers half or more of the dressing effect of the full
vertex as compared to the RL result, with the remaining difference---except for the light-meson cases---being
below 5\%.

On another general note, absolute as well as relative dressing effects appear to be of the same order
of magnitude for mesons from the categories with equal- or unequal-mass constituents. 

In terms of interpretation of RL studies in general we can state that effects are sizeable and worth studying,
but at the same time they are systematic and do not a priori destroy the validity and predictive power
of a sophisticated and well-controlled RL investigation, which can be useful by utilizing, e.\,g., mass splittings,
trends or cases protected by the symmetries of the theory in a careful and comprehensive manner.

\subsection{Mass of the $B_c^*$}

Next, we make a prediction for the mass of the $B_c^*$ meson. It's value is as yet unknown
experimentally and has been predicted in the literature, e.\,g., in the quark model (QM)
\cite{Eichten:1980mw,Stanley:1980zm,Buchmuller:1980su,Godfrey:1985xj,Gershtein:1987jj,Kaidalov:1987gk,Kwong:1990am,%
Baker:1991ty,Chen:1992fq,Itoh:1992sd,%
Eichten:1994gt,Bagan:1994dy,Zeng:1994vj,Roncaglia:1994ex,Kiselev:1994rc,Gupta:1995ps,Fulcher:1998ka,Ebert:2002pp,Ikhdair:2003tt,Ikhdair:2003ry,%
Godfrey:2004ya,Ikhdair:2004hg,Ebert:2011jc}, 
light-front quark model (LFQM) \cite{Frederico:2002vs,Choi:2009ai,Choi:2015ywa}, 
reductions of the BSE (BSR) \cite{AbdElHady:1998kc,Baldicchi:2000cf,Ikhdair:2004tj},
with the nonrelativistic renormalization group (NRG) \cite{Penin:2004xi}, 
QCD sum rules (QCDSR) \cite{Aliev:1992vp,Gershtein:1994jw,Wang:2012kw}, 
an RL study in the DSBSE approach (MT-RL) \cite{Fischer:2014cfa},
and lattice QCD (LAT) \cite{Davies:1996gi,Chiu:2007bc,Gregory:2009hq,Dowdall:2012ab,Burch:2015pka}. 
In Tab.~\ref{tab:bcstarmass} we compare these results from the literature and add our
own, ignoring error bars in each case.

\begin{table*}[t]
\caption{Comparison of predictions of the $B_c^*$ meson mass.   \label{tab:bcstarmass}}
\begin{ruledtabular}
  \begin{tabular}{ c c c c c c c }
   MN Full & MN RL & MT-RL\cite{Fischer:2014cfa} & LAT \cite{Davies:1996gi} & LAT \cite{Chiu:2007bc} & LAT \cite{Gregory:2009hq} & 
   			LAT \cite{Dowdall:2012ab} \\
   6.334  & 6.362 & 6.419 & 6.320 & 6.315 & 6.330 & 6.332 \\\hline
   LAT \cite{Burch:2015pka} & NRG \cite{Penin:2004xi} & BSR \cite{AbdElHady:1998kc} & BSR \cite{Baldicchi:2000cf} & BSR \cite{Ikhdair:2004tj} & 
   			QCDSR \cite{Aliev:1992vp} & QCDSR \cite{Gershtein:1994jw}  \\
   6.329  & 6.323 & 6.406 & 6.345 & 6.316 & 6.300 & 6.317 \\\hline
   QCDSR \cite{Wang:2012kw} & LFQM \cite{Frederico:2002vs} & LFQM \cite{Choi:2009ai} & LFQM \cite{Choi:2015ywa} & QM \cite{Eichten:1980mw} & 
			QM \cite{Stanley:1980zm} & QM \cite{Buchmuller:1980su} \\
   6.334  & 6.346 & 6.310 & 6.330 & 6.339 & 6.346 & 6.340 \\\hline
   QM \cite{Godfrey:1985xj} & QM \cite{Gershtein:1987jj} & QM \cite{Kaidalov:1987gk} & QM \cite{Kwong:1990am}  & 
   			QM \cite{Baker:1991ty} & QM \cite{Chen:1992fq} & QM \cite{Itoh:1992sd} \\
   6.340  & 6.329 & 6.370 & 6.321 & 6.372 & 6.344 & 6.328 \\\hline
   QM \cite{Eichten:1994gt} & QM \cite{Bagan:1994dy} & QM \cite{Zeng:1994vj} & QM \cite{Roncaglia:1994ex} & QM \cite{Kiselev:1994rc} & 
   			QM \cite{Gupta:1995ps} & QM \cite{Fulcher:1998ka} \\
   6.337  & 6.330 & 6.340 & 6.320 & 6.317 & 6.308 & 6.341 \\\hline
   QM \cite{Ebert:2002pp} & QM \cite{Ikhdair:2003tt} & QM \cite{Ikhdair:2003ry} & QM \cite{Godfrey:2004ya} & QM \cite{Ikhdair:2004hg} & 
   			QM \cite{Ebert:2011jc} & \\
   6.332  & 6.324 & 6.325 & 6.338 & 6.329 & 6.333 &  
  \end{tabular}
\end{ruledtabular}
\end{table*}

\begin{figure*}[t]
  \includegraphics[width=0.99\textwidth]{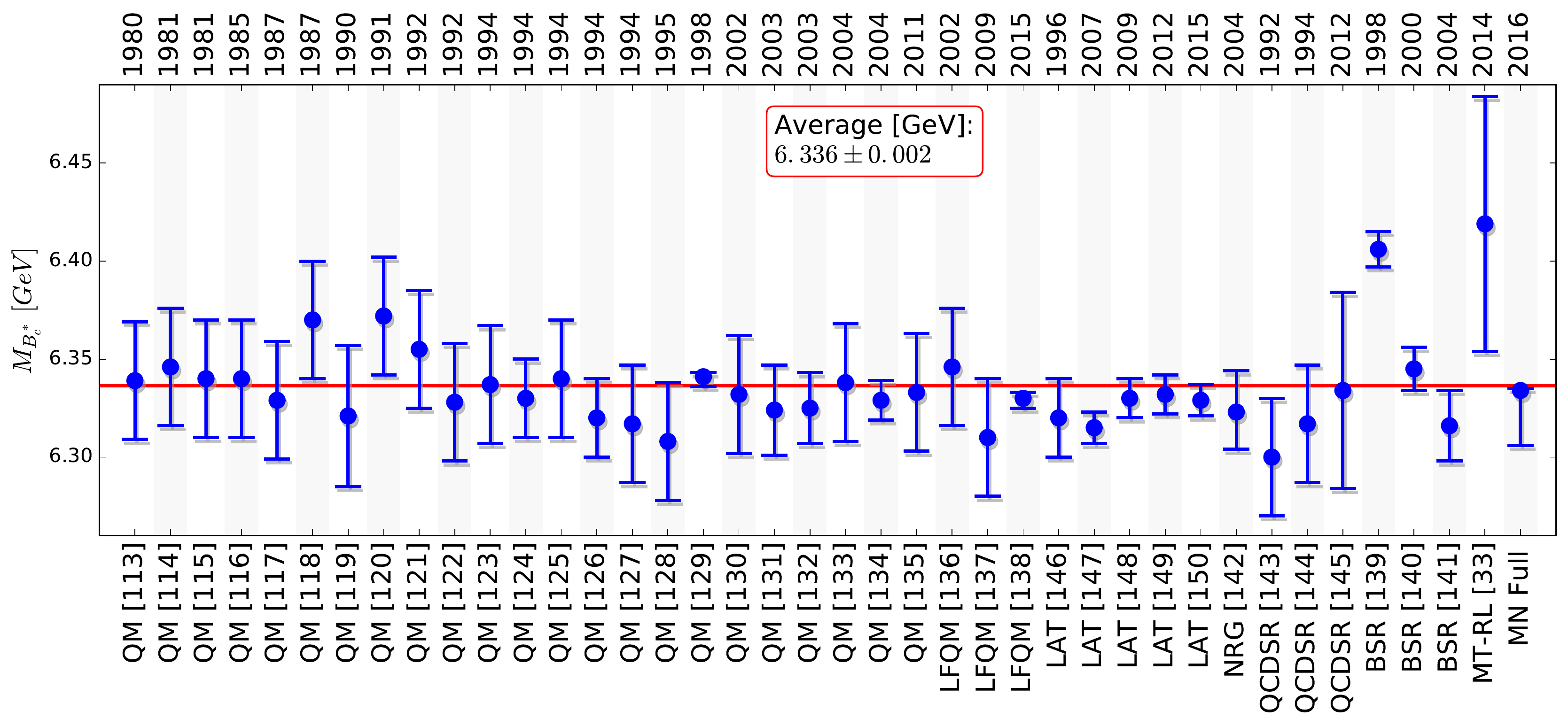}
\caption{\label{fig:bcstaraverage}
$B_c^*$ mass values including error bars (blue data points) taken from the references as listed on the
lower axis in the figure, corresponding to Tab.~\ref{tab:bcstarmass}. Years of appearance are 
given on the upper axis for each data point. The horizontal red line is the average; its error
is about the size of the thickness of the line. Value and error of the average are provided in the insert.}
\end{figure*}

We present two of our own values in this table, namely one for our RL case in column two and the result for the fully dressed
setup in column one. The RL result is included to allow better comparison with regard to the 
RL study and result in \cite{Fischer:2014cfa}, which uses a more sophisticated effective interaction, the
MT-model \cite{Maris:1999nt}. 

Our full result agrees very nicely with the predictions from the various approaches and studies.
We obtain the number via calculating the hyperfine mass splitting between the $B_c^*$ and $B_c$ mesons and
adding it to the experimentally measured mass of the $B_c$, which is $6.275\pm 0.001$ GeV \cite{Olive:2014rpp}. 
We note here that for the DSBSE study of \cite{Fischer:2014cfa} we have adjusted their published result in a similar 
manner, i.\,e., computed their value of the splitting and added it to the experimental pseudoscalar mass.

Investigating the $\eta$ dependence of this splitting for each $n$ shows again a situation where a very small range
of $\eta$ values is available at $n=\infty$ and all values at our chosen $\eta$ are largest or smallest available.
This is illustrated in the left and right panels of Fig.~\ref{fig:bcstar}, where again the dressing 
effects appear close to minimized by our choice of $\eta$.
Plotting our splitting as a function of $n$ in our scheme in the right panel of Fig.~\ref{fig:bcstar} we observe that
RL truncation overestimates it at $0.087^{+0.136}_{-0.003}$ GeV and corrections reduce its value to the full result of 
$0.059^{+0.001}_{-0.028}$ GeV. 
The error bars, also plotted in the figure, again represent the results' dependence on $\eta$ with the interpretation as a
systematic error as explained above. Incorporating the error bar into our result for the $B_c^*$ mass, we arrive at
$6.334^{+0.001}_{-0.028}$ GeV.

To obtain a better picture of the overall comparison of the various results among each other, we have collected them in 
Fig.~\ref{fig:bcstaraverage}. The references together with their characteristic as noted in Tab.~\ref{tab:bcstarmass}
are given below each data point, while the year of the study is shown above.

For the data points plotted we used the central value of each calculation together with an error bar as follows:
For those results, where an error bar is explicitly given in the reference, we include it as provided; where no error bar is
provided, we choose a default size for an appropriate error bar via an argument from Ref.~\cite{Kwong:1990am},
where the authors list an error of $\pm 0.036$ GeV for their quark-model result in order to ``get an idea of systematic 
errors inherent in quark models''. Concretely, we set the default error to $\pm 0.03$ GeV, which provides a reasonable
picture as well as a solid basis for the next step: to arrive at an interesting estimate of the overall theory prediction for the mass
of the $B_c^*$, we perform a standard weighted average of all values and errors, whose result is inserted in Fig.~\ref{fig:bcstaraverage}
and also plotted underneath the data as a horizontal red line. For the two cases of asymmetric errors we
treated the average of the upper and lower error as a symmetric error instead to simplify the procedure.
The averaged result is $6.336\pm0.002$ GeV, which may serve as
a more suitable number to compare to than the individual theoretical results.

\subsection{Quark-Photon Vertex}
\begin{figure*}[t]
  \includegraphics[width=\columnwidth]{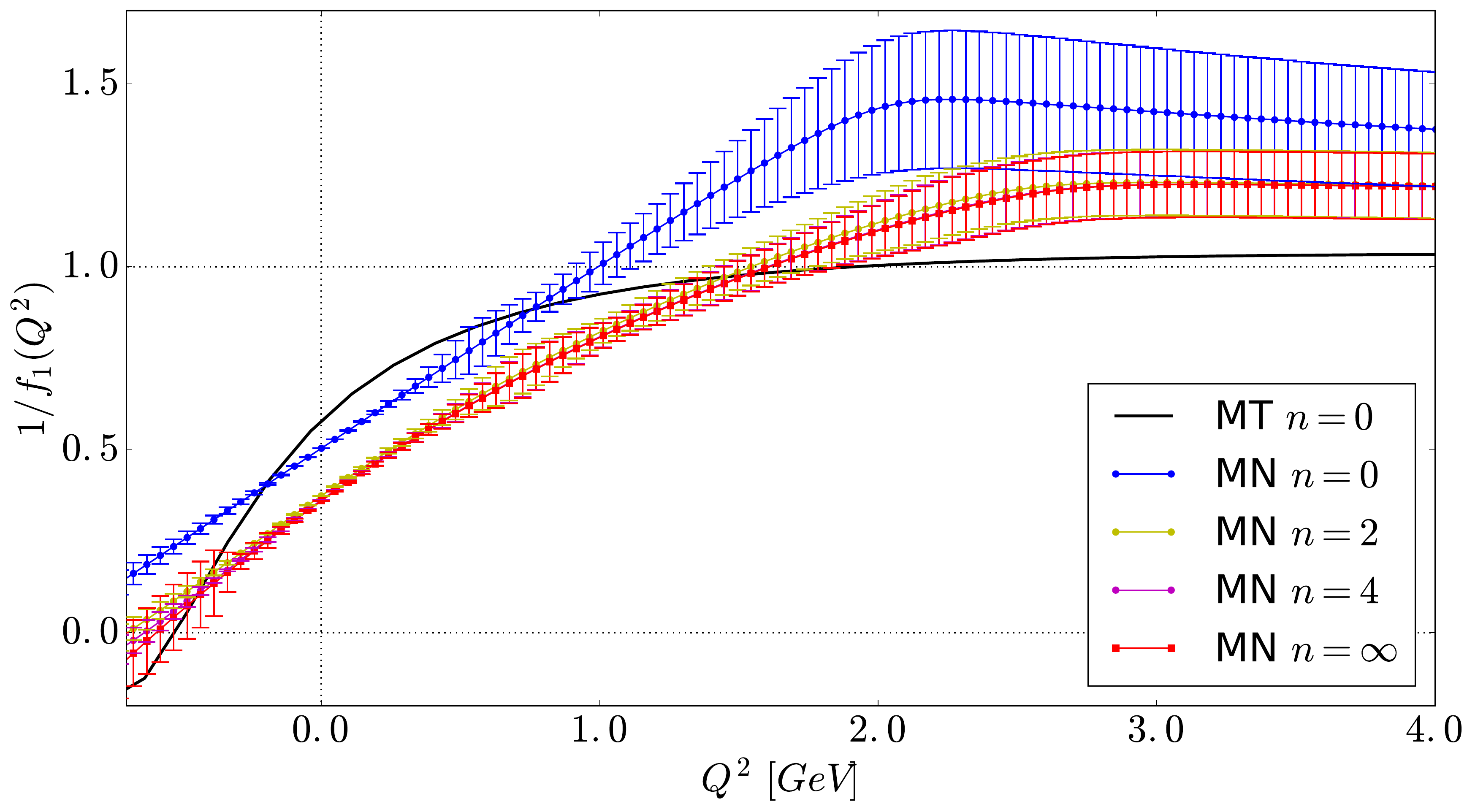}
  \includegraphics[width=\columnwidth]{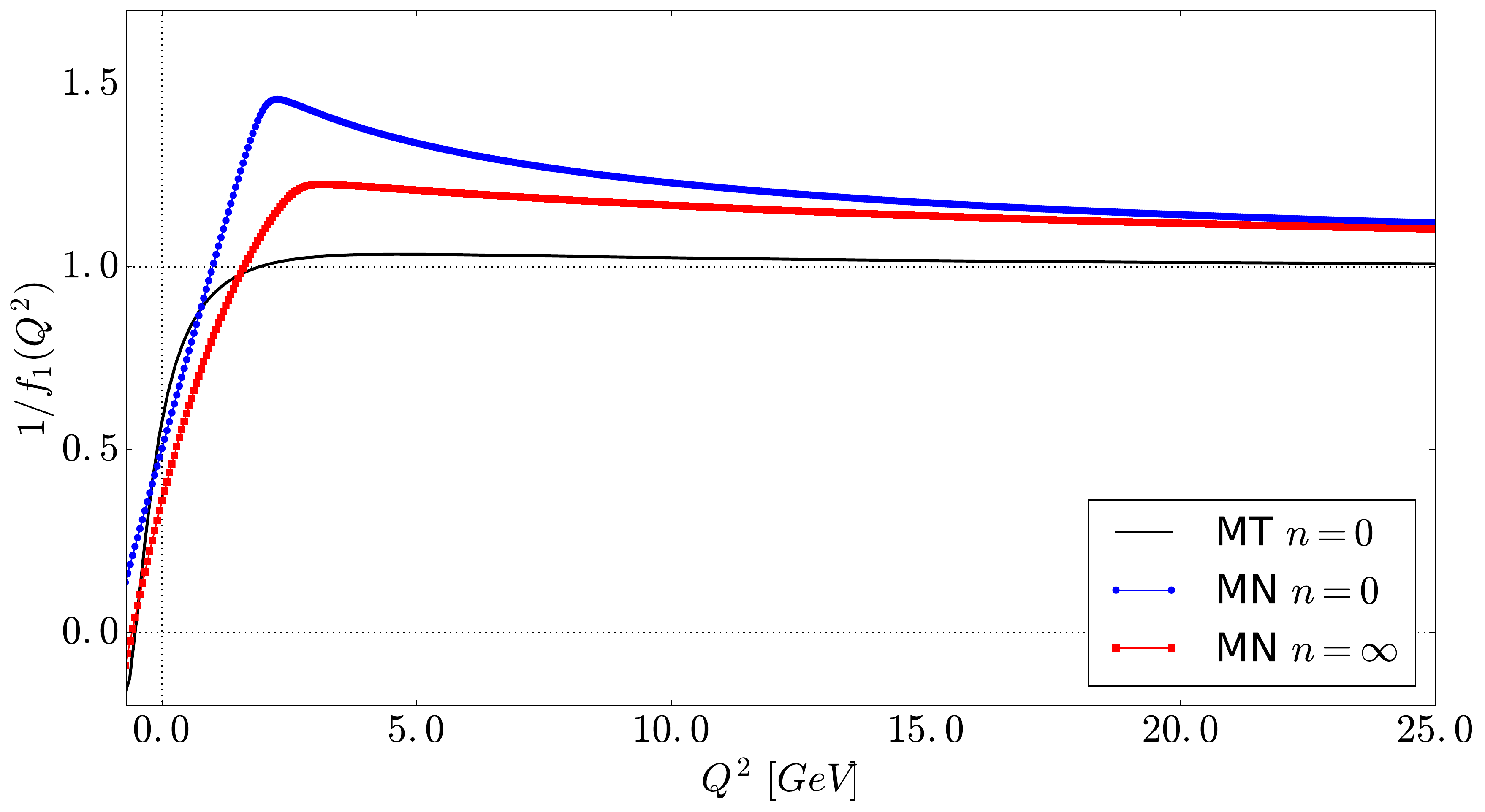}
\caption{\label{fig:qpv}
Inverse of transversal component $1/f_1$ of the quark-photon vertex for different $n$ in the MN model
studied herein together with a sophisticated 
(Maris-Tandy, MT) model \cite{Maris:1999nt} result in RL truncation for comparison, see text. \emph{Left panel}: Detailed view in region around
$Q^2=0$; \emph{Right panel}: large-scale view without error bars on the MN curves to emphasize the asymptotic behavior.}
\end{figure*}
A case of immediate interest in the investigation of the quantum numbers $J^{PC}=1^{--}$ is the dressed quark-photon
vertex \cite{Maris:1999bh,Maris:2002mz,Maris:2005tt,Eichmann:2014qva}. 
It can be obtained consistently from the inhomogeneous version of the vector BSE, which is a straight-forward
computation once the BSE kernel has been defined and computed \cite{Bhagwat:2007rj,Blank:2010bp,Blank:2010sn}.

The vertex has the general structure 
\begin{eqnarray}
\Gamma_\mu(Q;k)=\Gamma_\mu^L(Q;k) + \sum_{i=1}^8 T^i_\mu(Q;k) f_i(Q;k)\;,
\end{eqnarray}
where the arguments are the relative quark-antiquark momentum $k$ and the (photon) total momentum $Q$,
the eight covariants $T_\mu^i$ are transverse with respect to $Q$, and the longitudinal part $\Gamma_\mu^L(Q;k)$
is fixed via the vector Ward-Takahashi identity and can be written in the Ball-Chiu construction 
\cite{Ball:1980ay} as the longitudinal projection with respect to $Q$ of
\begin{eqnarray}\nonumber
\Gamma_\mu^\mathrm{BC}(Q;k)&=&i\;\gamma_\mu \Sigma_A(Q;k)\\ &+& 2k_\mu[i\;k\cdot\gamma\;\Delta_A(Q;k)+\Delta_B(Q;k)]\;.
\end{eqnarray}
In particular,
\begin{eqnarray}
\Sigma_A(Q;k)&=&\frac{A(k_+^2)+A(k_-^2)}{2}\;,\\
\Delta_A(Q;k)&=&\frac{A(k_+^2)-A(k_-^2)}{k_+^2-k_-^2}\;,\\
\Delta_B(Q;k)&=&\frac{B(k_+^2)-B(k_-^2)}{k_+^2-k_-^2}\;,
\end{eqnarray}
with the (anti)quark momenta $k_\pm$ defined analogously as in the homogeneous BSE above as
$k_+=k+\eta Q$ and $k_-=k-(1-\eta)Q$, which entails
\begin{eqnarray}
k_+^2 &=& k^2+2\eta\;k\cdot Q+\eta^2Q^2\;,\\
k_-^2 &=& k^2-2(1-\eta)\;k\cdot Q+(1-\eta)^2Q^2\;,\\
k_+^2-k_-^2 &=& (2\eta-1)(2\;k\cdot Q+Q^2)\;.
\end{eqnarray}

In our case, after the simplification via Eq.~(\ref{eq:mnmodel}), we remain with
\begin{eqnarray}
k_+^2 &=& \eta^2Q^2\;,\\
k_-^2 &=& (1-\eta)^2Q^2\;,\\
k_+^2-k_-^2 &=& (2\eta-1)Q^2\;.
\end{eqnarray}
and thus
\begin{eqnarray}
\Sigma_A(Q)&=&\frac{A(\eta^2Q^2)+A((1-\eta)^2Q^2)}{2}\;,\\
\Delta_A(Q)&=&\frac{A(\eta^2Q^2)-A((1-\eta)^2Q^2)}{(2\eta-1)Q^2}\;,\\
\Delta_B(Q)&=&\frac{B(\eta^2Q^2)-B((1-\eta)^2Q^2)}{(2\eta-1)Q^2}\;.
\end{eqnarray}
In the case of the quark-photon vertex, the quark and antiquark in the BSE have equal flavor and mass due
to the nature of the electromagnetic interaction. For the standard setting in such a case, $\eta=0.5$, we obtain
\begin{eqnarray}
k_+^2 = k_-^2 &=& Q^2/4\;,\\
\Sigma_A(Q)&=&A(Q^2/4)\;,\\
\Delta_A(Q)&=&A'(Q^2/4)\;,\\
\Delta_B(Q)&=&B'(Q^2/4)\;,
\end{eqnarray}
so that under normal circumstances with finite values of $A'(Q^2/4)$ and $B'(Q^2/4)$, the Ball-Chiu vertex
reduces to
\begin{eqnarray}
\Gamma_\mu^\mathrm{BC}(Q)&=&i\;\gamma_\mu \;A(Q^2/4)\;.
\end{eqnarray}
As we discuss herein, explicitly in App.~\ref{sec:vekerneldetails}, for the transverse vector covariants, only  2 are left 
nonzero by the model's simplifications and one can easily solve the inhomogeneous BSE to obtain the corresponding solutions.

In Fig.~\ref{fig:qpv} we plot the nonzero amplitudes as functions of the total momentum squared to illustrate
the size of dressing effects for the dressed quark-photon vertex in our (MN) scheme. The quark mass is chosen
to be the light-quark mass. The most prominent sets to look at
are the case $n=0$, which corresponds to the rainbow-ladder-truncation result and is depicted by the blue disks, 
and $n=\infty$, which represents the result from the fully dressed QGV and is depicted by the red boxes. In addition,
to highlight the rapid convergence of this function in our scheme, we also plotted the cases $n=2$ and $n=4$, which
are almost on top of the $n=\infty$ result; however,
in order not to overcrowd the figure, odd values of $n$ are left out here. 

In short, the difference between the RL
truncated result and any of the dressed versions is sizeable, while all dressed solutions among themselves are hard to distinguish,
and differences are minor. In absence of the dependence on a relative momentum squared, the behavior seen here may well be interpreted
as the prototype of variation of the $P^2$-dependence of elements of the quark-photon vertex beyond RL truncation in the sense
that already the first order in our scheme provides a result close to the fully dressed case.

In each case, we have, as required by our own statements, studied the model-artificial $\eta$ dependence of the MN results
and depicted the variation via error bars on each of the curves. As it turns out, such a dependence is stronger for larger
values of $Q^2$ and negligible around $Q^2=0$. The central curve is always given by the natural choice of $\eta=0.5$.

To complete the picture, we also plot the corresponding component for a dressed quark-photon vertex obtained with 
a sophisticated model interaction (the Maris-Tandy/MT model \cite{Maris:1999nt}) in RL truncation in analogy to the 
study in \cite{Maris:1999bh}. More precisely, we plot the inverse of the zeroth Chebyshev moment at zero relative momentum squared
as a function of $Q^2$, which corresponds to our MN-RL curve and serves as a baseline to impose putative dressing effects 
as they are studied here. This kind of comparison is supported as a result of the calculational restrictions due to the 
truncation scheme's adherence to the symmetry requirements of the theory represented by the relevant WTIs, which
are respected in both the MT and MN cases, as discussed above.

In the figure we have also plotted three dotted lines for ease of orientation, namely: a horizontal line at $1/f_1=0$, which
clearly shows the position of zeros in each curve, i.\,e., the $\rho$-pole positions---note that such a pole contribution is
present in every single case; a vertical line at $Q^2=0$, which marks the transition from the timelike to the spacelike region
of photon momentum; finally, a horizontal line at $1/f_1=1$, which indicates the limit of the asymptotic behavior of all curves
for $Q^2\rightarrow\infty$, i.\,e., the perturbative limit in which all curves agree by construction.

To better illustrate both the details of the various curves as well as their asymptotic behavior, we provide two panels in
Fig.~\ref{fig:qpv}: the left panel shows a detailed view of the region around $Q^2=0$, includes error bars as well as multiple
curves from the MN truncation scheme. The right panel on the other hand shows only three curves without error bars and very
nicely documents them approaching the perturbative limit.

\section{Conclusions}\label{sec:conclusions}

We have extended earlier DSBSE studies in a systematic truncation scheme using a simple
effective-interaction model together with a dressed QGV to the unequal-mass case of vector
mesons. After a general analysis of dressing effects in both the quarkonia and the various
flavored mesons, we focused on two items of special interest, namely the mass of the 
$B_c^*$ meson and the dressed quark-photon vertex.

The general pattern of dressing effects confirms expectations where dressing effects
beyond RL truncation are the stronger, the lighter the involved quark content is. We
found, rather importantly, that such effects are more pronounced in the vector-meson case
than in the pseudoscalar case studied earlier. This entails that mass (such as hyperfine)
splittings are modified significantly by corrections in a systematic truncation scheme such
as the one presented here.

Using such a calculated splitting between the $B_c^*$ and the $B_c$ mesons, we predict the
mass of the former and put our result in the context of other predictions available in the
literature. Our number, $6.334^{+0.001}_{-0.028}$ GeV compares well with the rest of the literature,
and our comparison to the RL result sheds some light on possible changes of corresponding
results at higher order in a systematic scheme such as the one presented here. 

In addition we have provide an average of a comprehensive set of results from the theory
literature. The averaged result for the mass of the $B_c^*$ meson is $6.336\pm0.002$ GeV.

To obtain results for the dressed quark-photon vertex, we present solutions for the
inhomogeneous vector-vertex BSE for the first time in the context of a truncation scheme.
Our simple-model convergence picture is contrasted to an RL calculation with a more
sophisticated model interaction and analogies are discussed in detail.

Our results support both studies of corrections to RL truncation as well as sophisticated and
well-controlled RL studies as such, since they can be performed with a much more comprehensive
scope in mind. Furthermore, we have once again demonstrated the strength of the use of
mass splittings as tools with predictive power in our approach.

\begin{acknowledgments}
We acknowledge valuable interactions with W.~Lucha, S.~Prelovsek, and Z.~G.~Wang.
This work was supported by the Austrian Science Fund (FWF) under project no.\ P25121-N27.
\end{acknowledgments}

\appendix

\section{$\eta$-dependence of meson masses}\label{sec:etadependence}

\begin{figure*}[t]
  \includegraphics[width=0.9\columnwidth]{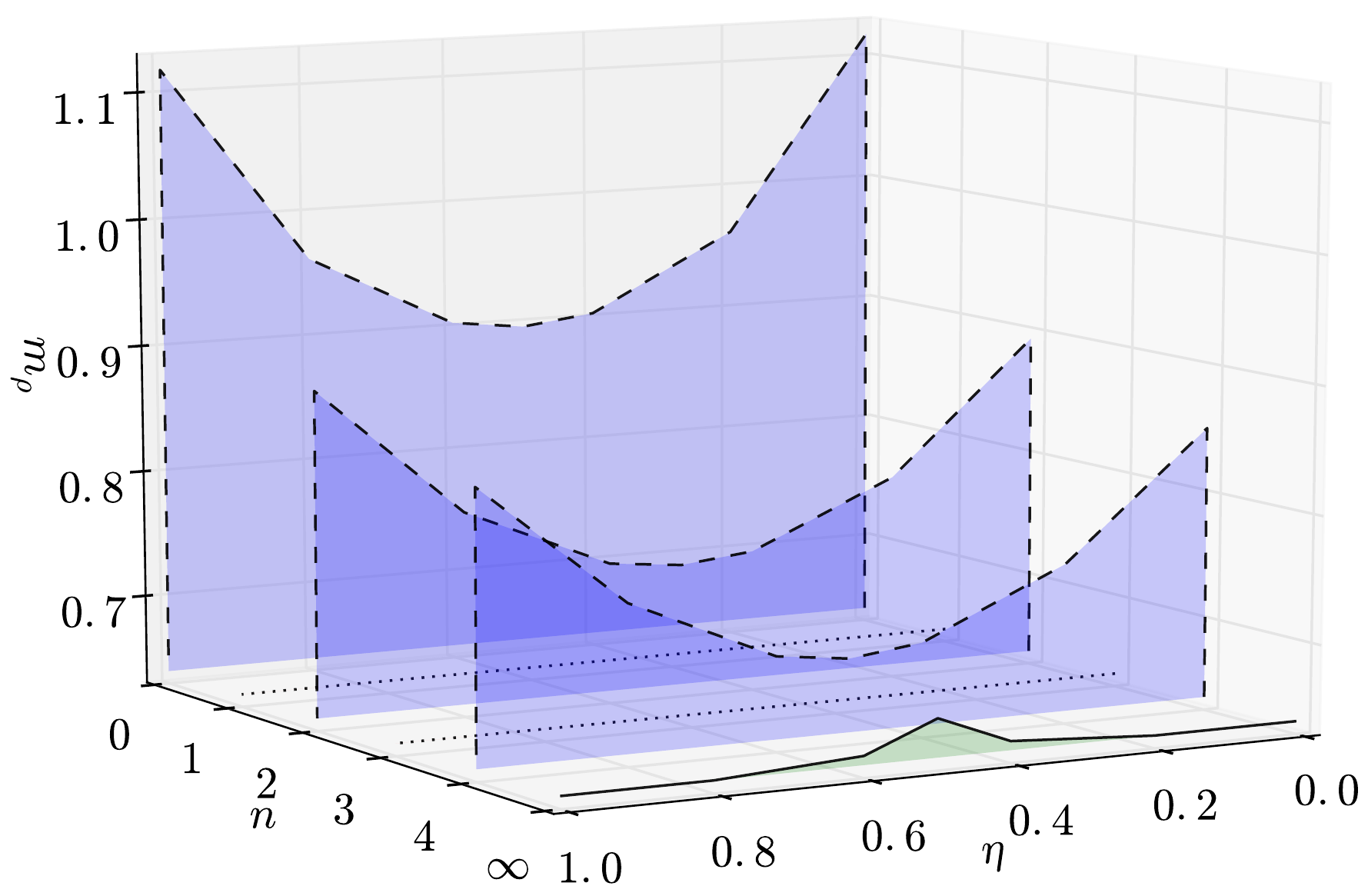}
  \includegraphics[width=0.9\columnwidth]{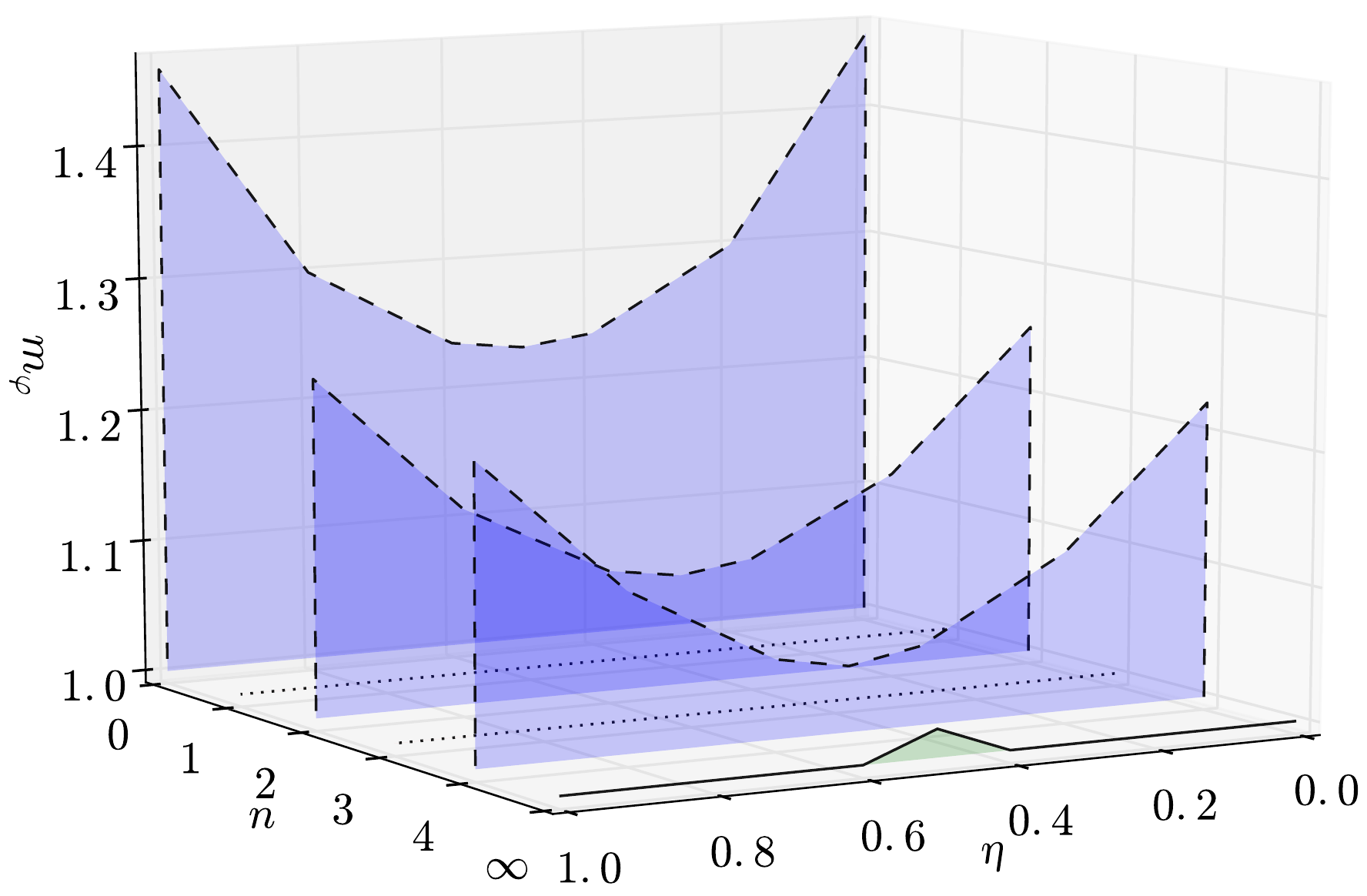}
  \includegraphics[width=0.9\columnwidth]{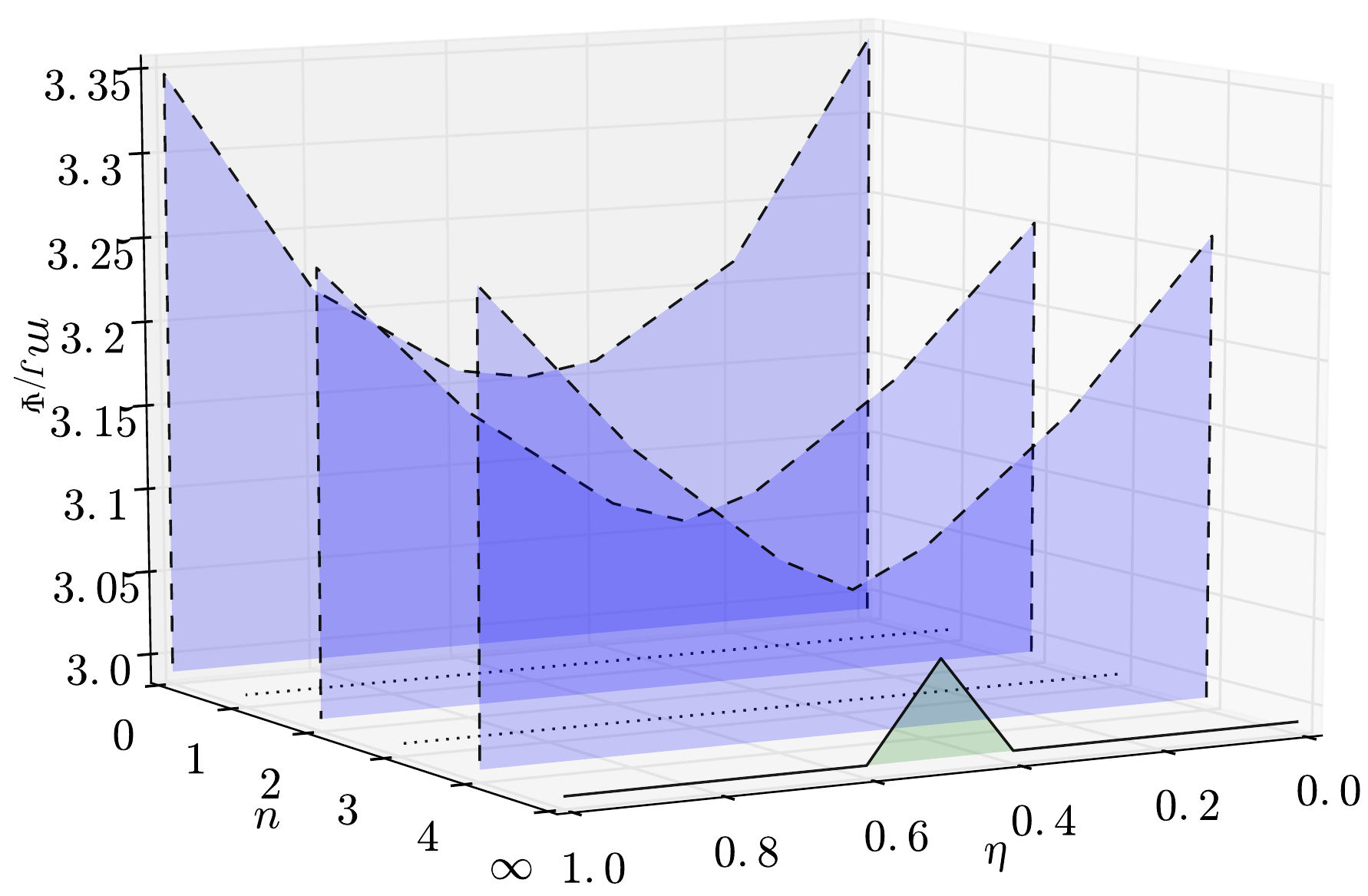}
  \includegraphics[width=0.9\columnwidth]{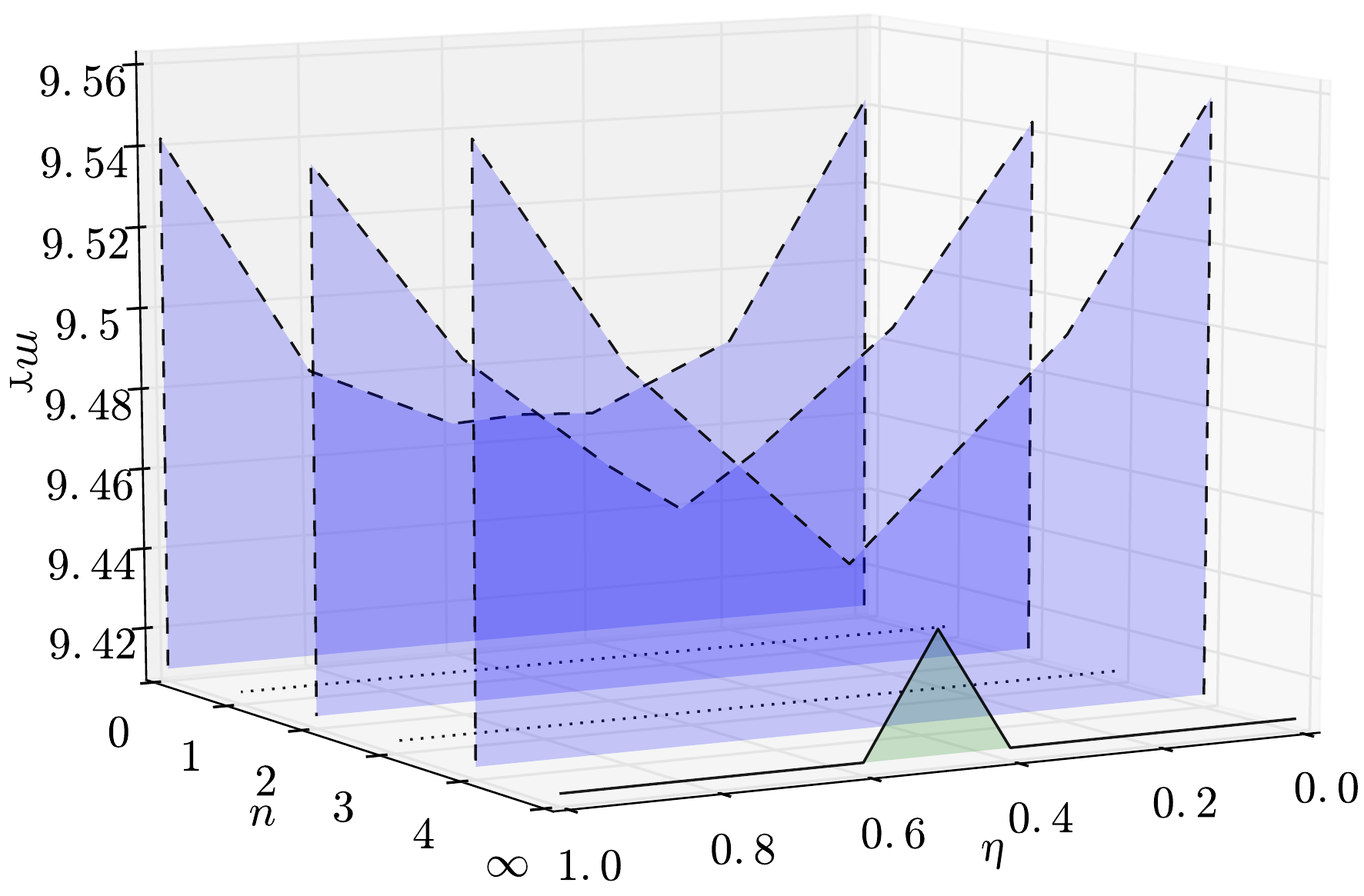}
\caption{\label{fig:etaeven}
Meson bound-state masses as functions of $n$ and $\eta$, given in GeV. Even $n$ are depicted by dashed
lines, odd ones by dotted lines, and the fully summed result by a solid line. If no solution is found,
no surface is plotted at the corresponding $n$. Left upper panel: $\rho$; right upper panel: $\varphi$;
eft lower panel: $J/\Psi$; right lower panel: $\Upsilon$.}
\end{figure*}

\begin{figure*}[t]
  \includegraphics[width=0.9\columnwidth]{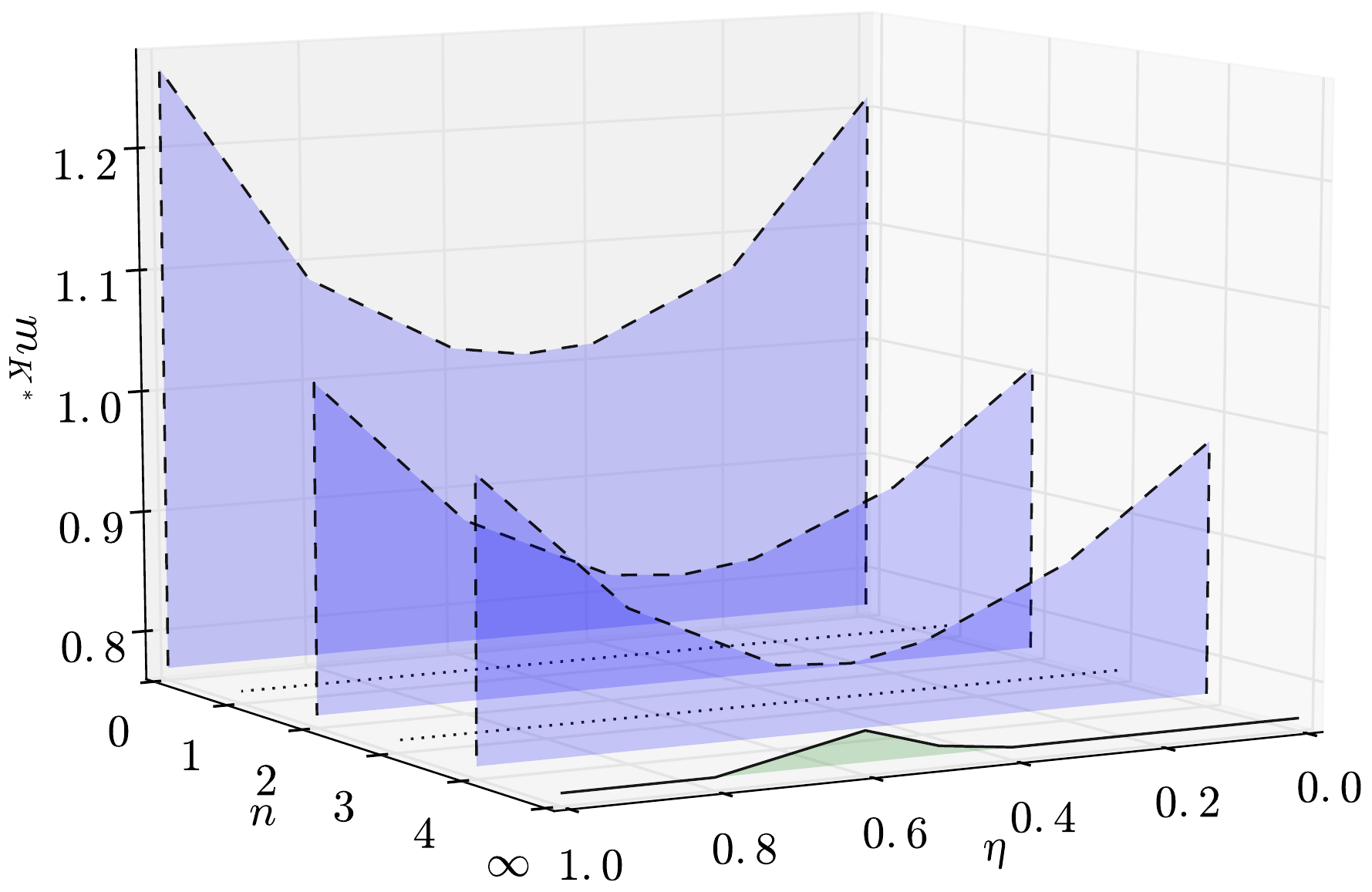}
  \includegraphics[width=0.9\columnwidth]{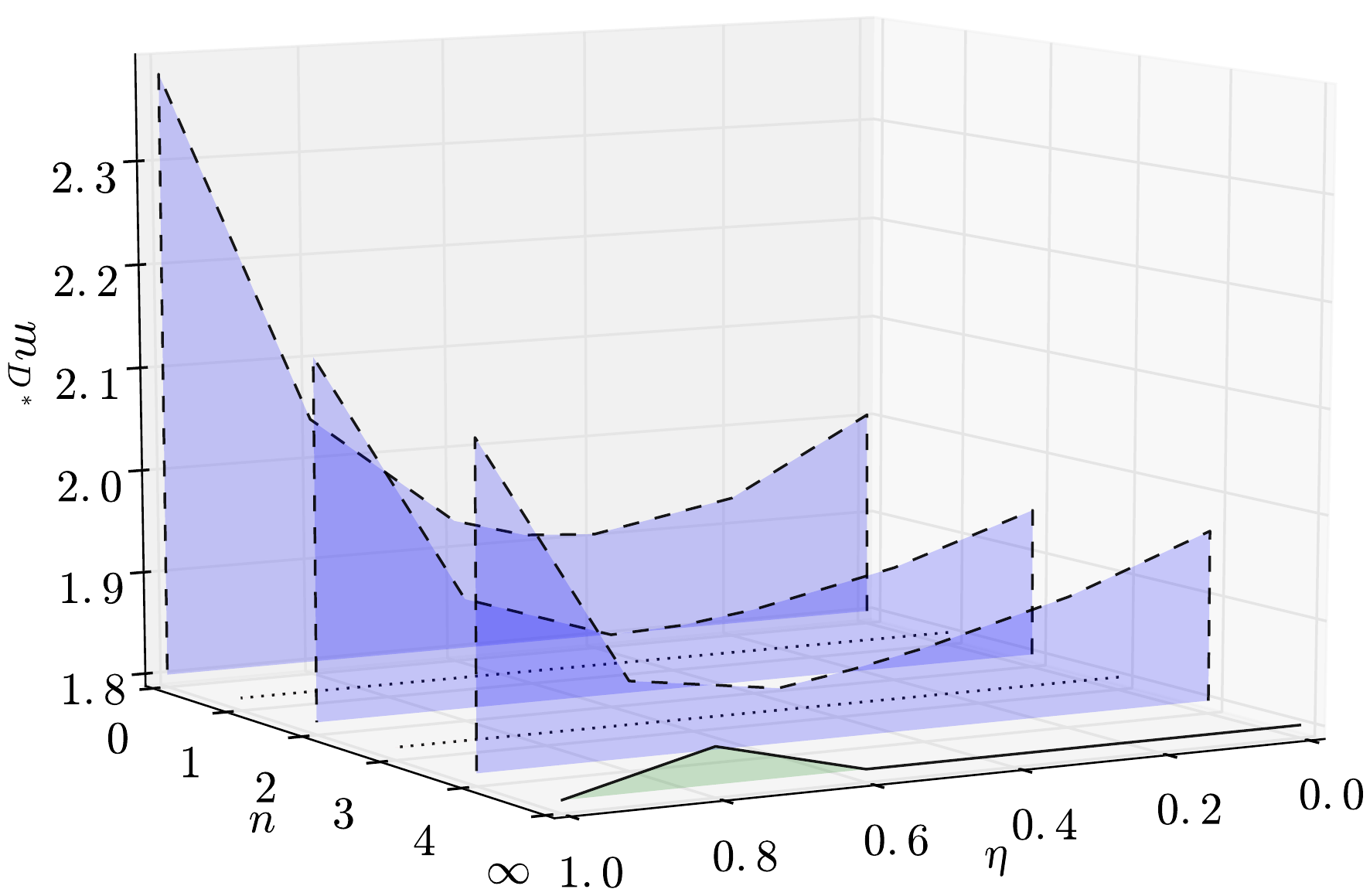}
  \includegraphics[width=0.9\columnwidth]{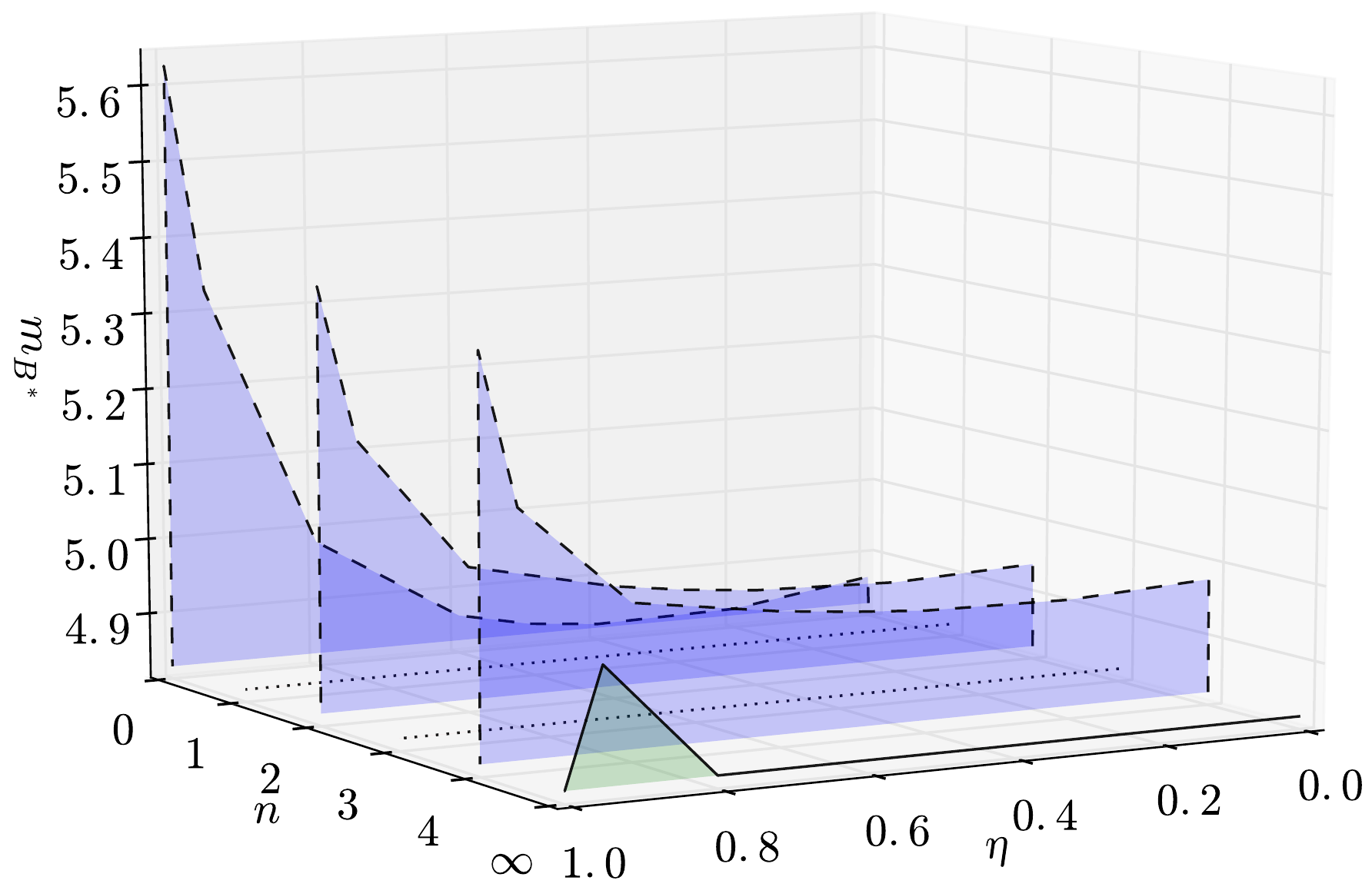}
  \includegraphics[width=0.9\columnwidth]{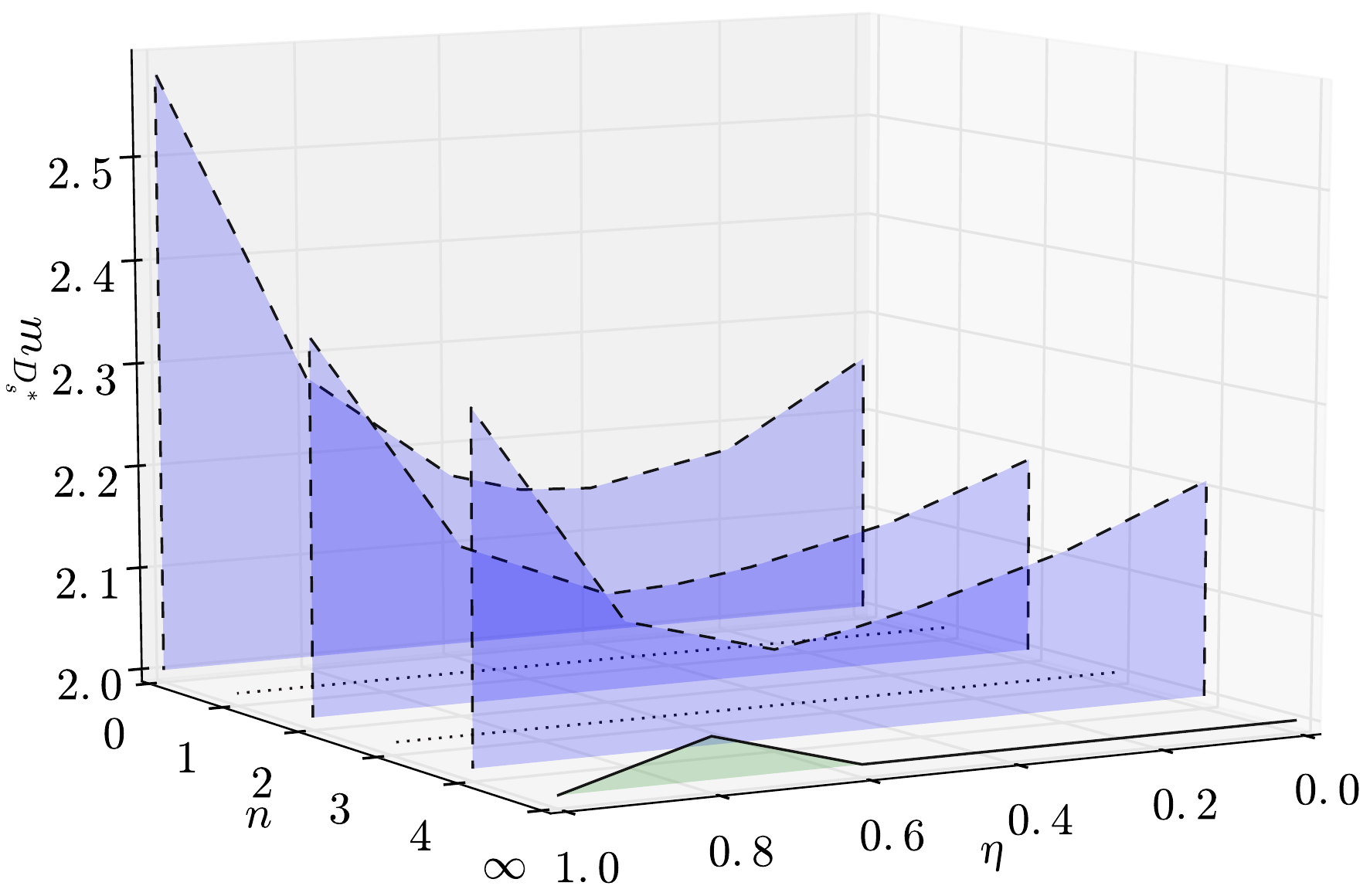}
  \includegraphics[width=0.9\columnwidth]{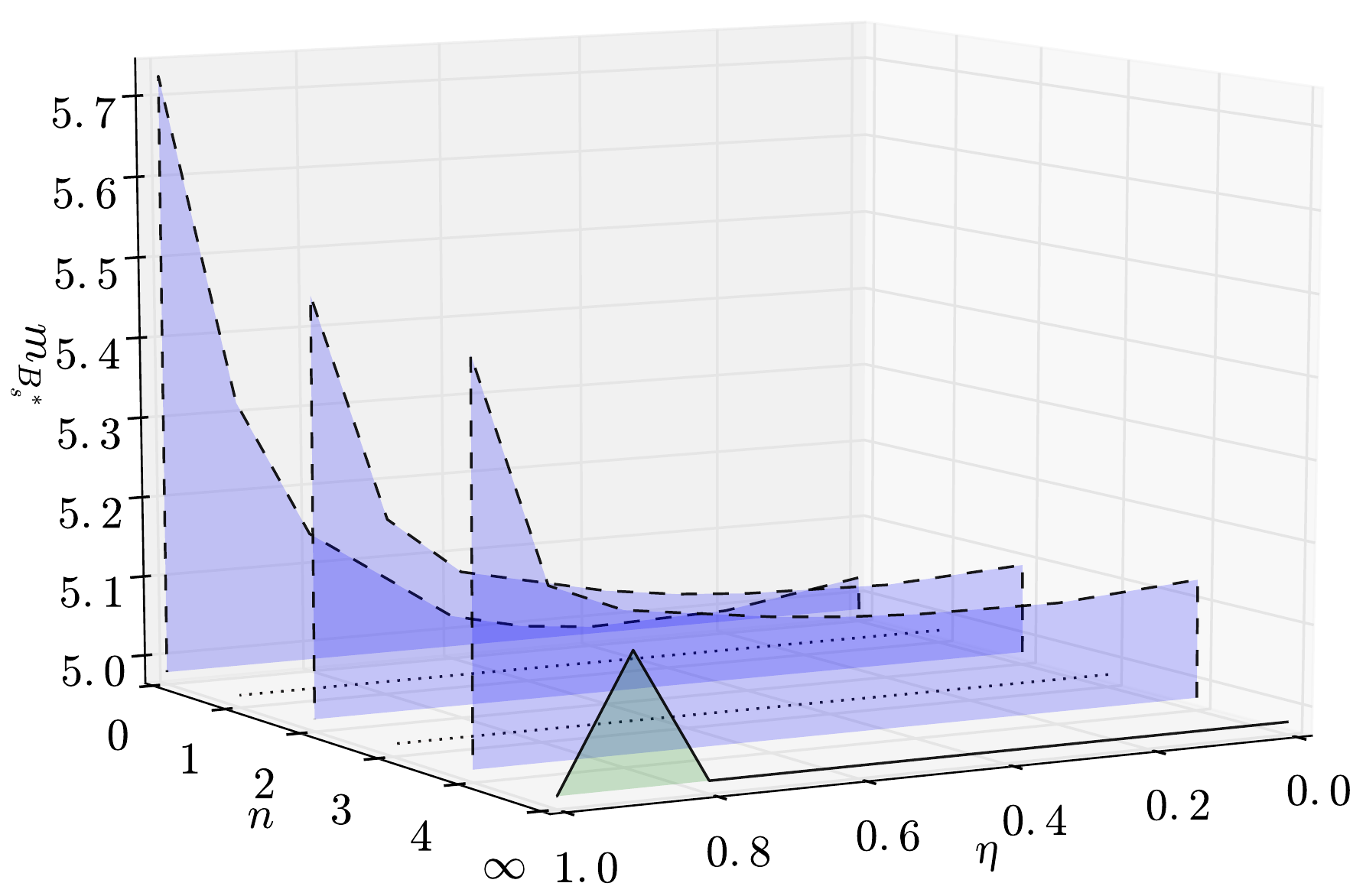}
  \includegraphics[width=0.9\columnwidth]{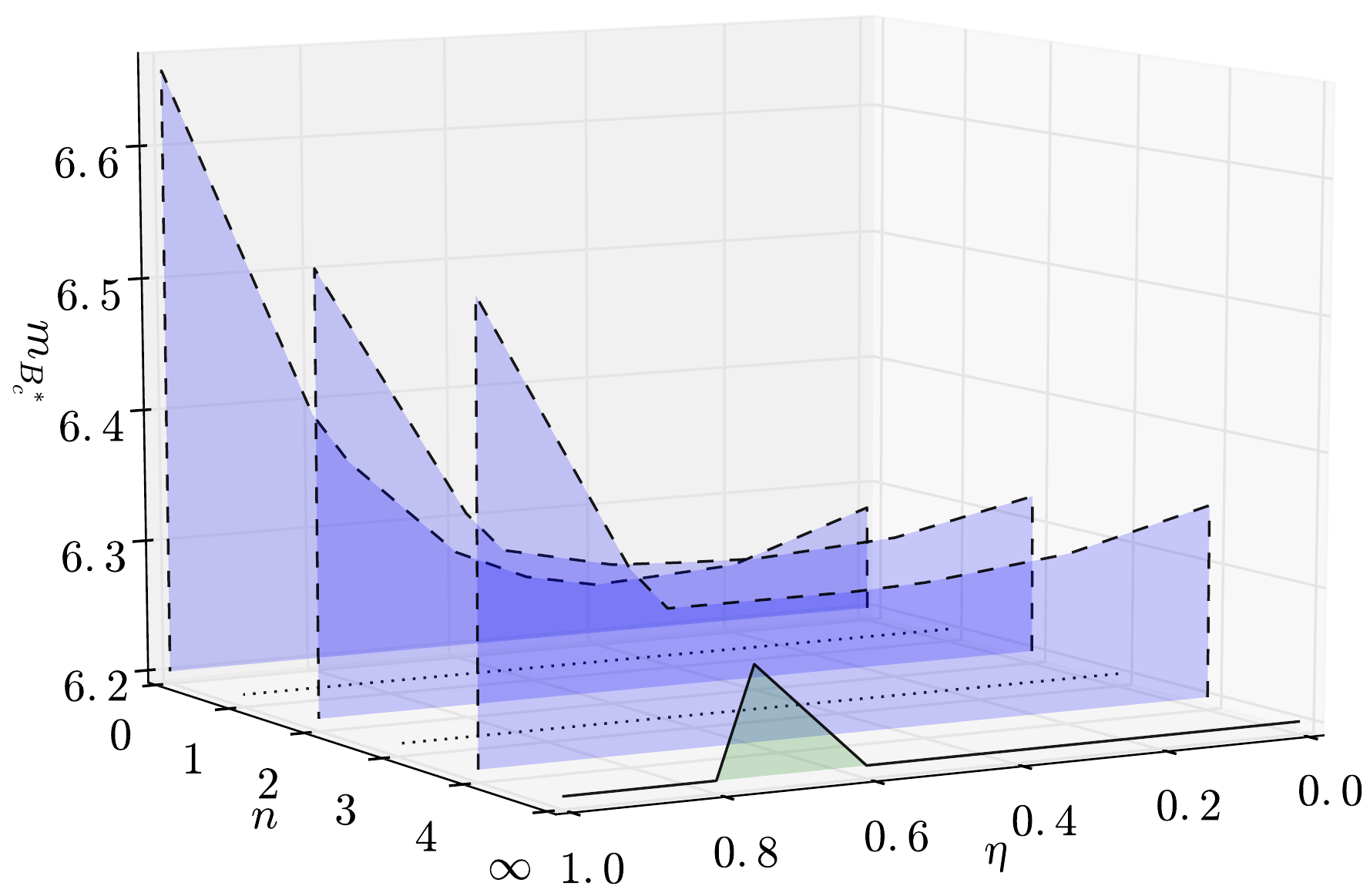}
\caption{\label{fig:etauneven}
Same as Fig.~\ref{fig:etaeven}: Left upper panel: $K^*$; right upper panel: $D^*$;
left center panel: $B^*$; right center panel: $D_s^*$; 
left lower panel: $B_s^*$; right lower panel: $B_c^*$.}
\end{figure*}

In this appendix we collect data and plots about the details of the
dependence of the meson masses on the momentum-partitioning parameter $\eta$
as a function of the order $n$ in our scheme. The corresponding plots are shown
in the various panels of Figs.~\ref{fig:etaeven} and \ref{fig:etauneven} for the
equal-  and unequal-mass cases, respectively. The alternating pattern of convergence 
of the odd and even $n$ numbers described earlier in \cite{Gomez-Rocha:2015qga} is 
difficult to observe herein, since there is, again, a distinct lack of solutions for 
odd $n$ on our main $\eta$ grid points. Still, convergence with $n$ is observed as well as
a pronounced $\eta$ asymmetry for the heavy-light case, which is the source
of the large error bars plotted in Fig.~\ref{fig:expcomp}. Despite this artificial behavior
a detailed study such as ours remains true to the systematic character of both the approach 
and the truncation scheme presented here and, in particular, validates the qualitative
as well as quantitative statements made above.

\section{vector kernel details}\label{sec:vekerneldetails}

Following Refs.~\cite{Bender:2002as,Bhagwat:2004hn,Gomez-Rocha:2015qga} we collect the details
of the BSA, the correction term $\Lambda$, and the QGV in this appendix. 
The recursion relations for the QGV $\Gamma_\mu$, Eq.~(\ref{eq:qgvrecursion}) and the
BSE correction term $\Lambda_{M\mu}$, Eq.~(\ref{eq:recbiglambda}) are detailed, in 
particular with respect to Dirac structures.

From the 12 covariant structures of the full QGV, Eq.~(\ref{eq:mnmodel}) reduces this set to
three nonzero ones:
\begin{equation}\label{eq:qgvstructure}
\Gamma_\mu(p)=\alpha_1(p^2)\gamma_\mu + \alpha_2(p^2)\gamma\cdot p\; p_\mu - i \alpha_3(p^2) p_\mu\;.
\end{equation}
With the initial condition that the QGV be bare
\begin{equation}\label{eq:qgvinitial}
\Gamma_\mu(p)^0=\gamma_\mu 
\end{equation}
one can construct the QGV via its recursion relation at any order by expressing the functions
$(\alpha_1,\alpha_2,\alpha_3)$ given in Eq.~(\ref{eq:qgvstructure}) in terms of the quark propagator 
dressing functions $A$ and $B$. By inserting the result back into the quark DSE one obtains algebraic 
equations for $A(s)$ and $B(s)$ via Dirac-trace projections onto the two covariant quark propagator structures.

In order to compute $\Lambda_{M\mu}(P)$ a suitable decomposition in terms of Dirac covariants 
has to be found, depending on the quantum numbers appropriate for the meson $M$ under
consideration, in our case vector. In our setup the vector BSA has 2 non-vanishing components
from the eight general structures, namely:
\begin{equation}\label{eq:psbsa}
\Gamma_{1^-}^\xi(P)=f_1(P^2)\;\gamma\cdot \varepsilon^\xi(P) -  
f_2(P^2)\,\sigma_{\mu\nu}\;\varepsilon^\xi_\mu(P) \hat{P}_\nu
\end{equation}
with the unit vector $\hat{P}:=P/\sqrt{P^2}$ and 
\begin{equation}
\sigma_{\mu\nu}\;a_\mu b_\nu:=\frac{i}{2}(\gamma\cdot a \;\gamma\cdot b- \gamma\cdot b\;\gamma\cdot a)\;.
\end{equation}
$\varepsilon^\xi_\mu(P)$, $\xi=1,2,3$ are polarization vectors with respect to $P$. The corresponding
Dirac projections are
\begin{eqnarray}
\mathcal{F}_{1}^\xi&:=&\frac{1}{12}\gamma\cdot \varepsilon^\xi(P)\;, \\
\mathcal{F}_{2}^\xi&:=&-\frac{1}{12}\sigma_{\mu\nu}\;\varepsilon^\xi_\mu(P) \hat{P}_\nu  \;,
\end{eqnarray}
such that 
\begin{equation}
\Tr \sum_\xi\left[\mathcal{F}_{j}^\xi \Gamma_{1^-}^\xi(P) = f_j(P^2)\right],\quad j=1,2\;.
\end{equation}

We construct $\Lambda_{1^{-}\mu}$ following \cite{Bender:2002as} for direct comparability as
\begin{eqnarray}\nonumber
\Lambda_\mu^\xi(P)&=&\beta_1(P^2)\,\varepsilon^\xi_\mu(P)\\\nonumber 
&+& \beta_2(P^2)\,i\varepsilon^\xi_\mu(P)\;\gamma\cdot \hat{P} \\ \nonumber
&+& \beta_3(P^2) \,i\hat{P}_\mu\;\gamma\cdot\varepsilon^\xi(P) \\ \nonumber  
&-&\beta_4(P^2)\,\sigma_{\tau\nu}\;\varepsilon^\xi_\tau(P) \hat{P}_\nu\;\gamma_\mu\; \\  \nonumber
&-&\beta_5(P^2)\,i\sigma_{\tau\nu}\;\varepsilon^\xi_\tau(P) \hat{P}_\nu\;\hat{P}_\mu\;\\  
&+&\beta_6(P^2)\,\gamma\cdot\varepsilon^\xi(P)\; \gamma_\mu\;,\label{eq:lambdastructure}
\end{eqnarray}
where the subscript denoting the vector case has been omitted.
The corresponding Dirac projections are
\begin{eqnarray}
\mathcal{P}^\xi_{\mu,1}&:=&\frac{1}{4} \varepsilon^\xi_\mu(P)\;, \\
\mathcal{P}^\xi_{\mu,2}&:=&-\frac{i}{4} \varepsilon^\xi_\mu(P)\;\gamma\cdot \hat{P}\;, \\
\mathcal{P}^\xi_{\mu,3}&:=&-\frac{i}{4} \hat{P}_\mu\;\gamma\cdot\varepsilon^\xi(P) \;,\\
\mathcal{P}^\xi_{\mu,4}&:=&-\frac{1}{4} \sigma_{\tau\nu}\;\varepsilon^\xi_\tau(P) \hat{P}_\nu\;\gamma_\mu \;,\\ 
\mathcal{P}^\xi_{\mu,5}&:=&\frac{i}{4} \sigma_{\tau\nu}\;\varepsilon^\xi_\tau(P) \hat{P}_\nu\;\hat{P}_\mu \;,\\ 
\mathcal{P}^\xi_{\mu,6}&:=&\frac{1}{4} \gamma\cdot\varepsilon^\xi(P)\; \gamma_\mu  \;,
\end{eqnarray}
such that
\begin{equation} \label{eq:betamatrix} 
\beta_j = (\mathcal{M})_{jk} \mathrm{Tr}\sum_\xi\left[\mathcal{P}^\xi_{\mu,k} \Lambda^\xi_{\mu} \right] \;,
\end{equation} 
where the matrix $\mathcal{M}$ (the vector case $1^-$ is assumed implicitly from now on) is given by
\begin{equation}
\mathcal{M}=\frac{1}{2}
\left(
\begin{array}{cccccc}
 3 & 0 & 0 & 0 & 1 & -1 \\
 0 & 3 & -1 & -1 & 0 & 0 \\
 0 & -1 & 3 & 1 & 0 & 0 \\
 0 & -1 & 1 & 1 & 0 & 0 \\
 1 & 0 & 0 & 0 & 3 & -1 \\
 -1 & 0 & 0 & 0 & -1 & 1 \\
\end{array}
\right)\;.
\end{equation}

The scalar functions $\vec{\beta}:=\{\beta_j\}$, $j=1,\ldots,6$ are obtained at a particular order $n$
in the truncation via the recursion relation (\ref{eq:recbiglambda}), resulting in 
\begin{equation} 
\vec{\beta}^i = \mathcal{M}\left(\mathbf{G}_-\,\vec{\alpha}_-^{\,i-1}  +\mathbf{G}_+\,\vec{\alpha}_+^{\,i-1} + 
\mathbf{L}\, \vec{\beta}^{\,i-1}\right), 
\end{equation} 
which can be evaluated when the matrices $\mathbf{G}_\pm$ and $\mathbf{L}$ are known.
$\vec{\alpha}_+$ and $\vec{\alpha}_-$ denote the coefficients of the QGV decomposition
corresponding to the ${}_+$ and  ${}_-$ arguments appearing in their defining quark 
propagators as given above.
With the definitions
\begin{eqnarray}
 B_- &:=&  B(p_-^2) \;,\\
 B_+ &:=&  B(p_+^2)\;, \\
 A_- &:=&  (\eta -1)\,\sqrt{P^2} A(p_-^2)\;, \\
 A_+ &:=&  \eta \, \sqrt{P^2}   A(p_+^2) \;,
\end{eqnarray}
and
\begin{eqnarray}
\Delta_- &:=& A_-^2 + B_-^2 \;, \\
\Delta_+ &:=& A_+^2 + B_+^2 \;, 
\end{eqnarray}
as well as
\begin{eqnarray}
\Theta_- &:=& A_- A_+-B_- B_+  \;, \\
\Theta_+ &:=& A_- A_++B_- B_+  \;, \\
\Xi_- &:=& A_- B_+-A_+ B_-  \;, \\
\Xi_+ &:=& A_- B_+ +  A_+ B_-\;,
\end{eqnarray}
and
\begin{eqnarray}
\Phi_- &:=& B_+ A_-^2-2 A_- A_+ B_--B_+ B_-^2  \;, \\
\Phi_+ &:=& A_+ A_-^2+2 A_- B_- B_+-A_+ B_-^2  \;, \\
\Psi_- &:=& B_- A_+^2-2 A_- A_+ B_+-B_- B_+^2  \;, \\
\Psi_+ &:=& A_- A_+^2+2 A_+ B_- B_+-A_- B_+^2\;,
\end{eqnarray}
one obtains
\begin{widetext}
\begin{equation}
\mathcal{M}\mathbf{L}=
\frac{2\mathcal{C}}{\Delta_- \, \Delta_+ }
\left(
\begin{array}{cccccc}
 2 \Theta _- & -2 \Xi _+ & 0 & -2 \Xi _+ & 0 & 2 \Theta _- \\
 -\Xi _+ & -\Theta _- & 0 & -2 \Theta _- & 0 & -2 \Xi _+ \\
 0 & 0 & \Theta _+ & -2 B_- B_+ & -\Xi _- & 2 B_- A_+ \\
 0 & 0 & 0 & \Theta _- & 0 & \Xi _+ \\
 0 & 0 & 0 & 0 & 0 & 0 \\
 0 & 0 & 0 & 0 & 0 & 0 \\
\end{array}
\right)\;.
\end{equation}
In the equal-mass case, as presented in \cite{Bender:2002as} one obtains
\begin{equation}
\mathcal{M}\mathbf{L}=
\frac{1}{4\Delta^2}
\left(
\begin{array}{cccccc}
 2 \Delta  & 0 & 0 & 0 & 0 & 2 \Delta  \\
 0 & -\Delta  & 0 & -2 \Delta  & 0 & 0 \\
 0 & 0 & A^2 Q^2-B^2 & 2 B^2 & -2 A B \sqrt{Q^2} & -2 A B \sqrt{Q^2} \\
 0 & 0 & 0 & \Delta  & 0 & 0 \\
 0 & 0 & 0 & 0 & 0 & 0 \\
 0 & 0 & 0 & 0 & 0 & 0 \\
\end{array}
\right)
\end{equation}
with the replacements $\mathcal{C}\rightarrow-\frac{1}{8}$ and $P^2\rightarrow 4 Q^2$, which
is identical to the result given in \cite{Bender:2002as} except for a factor of $2$ in element
$(3,3)$ of this matrix.

The two matrices $\mathbf{G}_-$ and $\mathbf{G}_+$ are associated with the corresponding quark propagators with the 
${}_+$ and  ${}_-$ arguments as defined above:


\begin{equation}
\mathcal{M}\mathbf{G}_-=\frac{-\mathcal{C}}{\Delta_-^2 \Delta_+}
\left(
\begin{array}{ccc}
 -4 \Delta _- B_+ f_1-4 \Delta _- A_+ f_2 & 0 & 0 \\
 -4 \Delta _- A_+ f_1+4 \Delta _- B_+ f_2 & 0 & 0 \\
 -4 B_- \Xi _- f_1-4 B_- \Theta _+ f_2 & -2 q_-^2 \Phi _+ f_1+2 q_-^2 \Phi _- f_2 & -2 \sqrt{q_-^2}
   \Phi _- f_1-2 \sqrt{q_-^2} \Phi _+ f_2 \\
 2 \Delta _- A_+ f_1-2 \Delta _- B_+ f_2 & 0 & 0 \\
 0 & 0 & 0 \\
 0 & 0 & 0 \\
\end{array}
\right)\;,
\end{equation}
which is to be understood as a $6\times 3$ matrix,
and its corresponding analog
\begin{equation}
\mathcal{M}\mathbf{G}_+=\frac{-\mathcal{C}}{\Delta_- \Delta_+^2}
\left(
\begin{array}{ccc}
 -4 B_- \Delta _+ f_1+4 A_- \Delta _+ f_2 & 0 & 0 \\
 0 & 0 & 0 \\
 -4 A_+ \Theta _+ f_1+4 \Xi _- A_+ f_2 & -2 q_+^2\Psi _+ f_1 -2 q_+^2\Psi _- f_2  & 2 \sqrt{q_+^2}\Psi _-
    f_1-2 \sqrt{q_+^2} \Psi _+ f_2 \\
 -2 A_- \Delta _+ f_1-2 B_- \Delta _+ f_2 & 0 & 0 \\
 0 & 0 & 0 \\
 0 & 0 & 0 \\
\end{array}
\right)\;.
\end{equation}
\end{widetext}

In the equal-mass case, again with the replacements $\mathcal{C}\rightarrow-\frac{1}{8}$ 
and $P^2\rightarrow 4 Q^2$ this becomes
\begin{equation}
\mathcal{M}(\mathbf{G}_-+\mathbf{G}_+)=\frac{1}{2 \Delta^2 }
\left(
\begin{array}{ccc}
 2 \left(B f_1+A \sqrt{Q^2} f_2\right) & 0 & 0 \\
 A \sqrt{Q^2} f_1-B f_2 & 0 & 0 \\
 -A \sqrt{Q^2} f_1+B f_2 & 0 & 0 \\
 -A \sqrt{Q^2} f_1+B f_2 & 0 & 0 \\
 0 & 0 & 0 \\
 0 & 0 & 0 \\
\end{array}
\right)\;,
\end{equation}
which is identical to the result given in \cite{Bender:2002as} except for an overall factor of $1/\Delta$.

\section{Proof of Kernel Construction}\label{sec:ProofAposteriori}

In this appendix we present a short proof that our kernel construction in fact satisfies the 
Axial-Vector Ward-Takahashi Identity (AVWTI), as required by the general setup of the truncation scheme.
The AVWTI can be written in its integral form as \cite{Eichmann:2009zx}
\begin{eqnarray}
    && \int_q^\Lambda \left\{ S(q_+) \gamma_5 + \gamma_5 S(q_-)\right\}_{GH} K^{GH}_{EF}(k,q,P)
    \nonumber
    \\&& = \{ \Sigma(k_+) \gamma_5 + \gamma_5 \Sigma(k_-)\}_{EF}
    \;,
\end{eqnarray}
where $K$ is the quark-antiquark scattering kernel used in the meson BSE, $q_\pm$ are the (anti)quark momenta,
and $EFGH$ denote color, Dirac, and flavor indices.

In the following, we show that the kernel in Eq.~(\ref{eq:mngenbsekernel}) satisfies this equation. Alternatively, one can
in principle also reverse the argument to arrive at the kernel construction starting out from the AVWTI.

The gluon-loop dressed QGV correction term to the BSE kernel, as defined in Eqs.\ \eqref{eq:lambda} and \eqref{eq:recbiglambda}, leads to a corresponding correction to the AVWTI of the form
\begin{equation}\label{eq:lambda_bar}
    \bar\Lambda^M_\nu(P) = \sum_{i=0}^\infty \bar\Lambda_{\nu,i}^M(P)\,,
\end{equation}
with
\begin{eqnarray}\nonumber
    && \frac{1}{\mathcal{C}}\bar\Lambda_{\nu,n}^M(P) =
    \\\nonumber
    && -\gamma_\rho \left\{S(q_+)\gamma_5+\gamma_5S(q_-)\right\} \Gamma_{\nu,n-1}^\mathcal{C}(q_-)S(q_-) \gamma_\rho
    \\\nonumber
    &&- \gamma_\rho S(q_+) \Gamma_{\nu,n-1}^\mathcal{C}(q_+) \left\{S(q_+)\gamma_5+\gamma_5S(q_-)\right\} \gamma_\rho
    \\
    &&- \gamma_\rho S(q_+) \bar\Lambda_{\nu,n-1}^M(P) S(q_-) \gamma_\rho
\;.
\label{eq:recbiglambda_bar}
\end{eqnarray}

Using the recursion relation for the QGV \eqref{eq:qgvrecursion}, this becomes
\begin{eqnarray}\nonumber
    && \frac{1}{\mathcal{C}}\bar\Lambda_{\nu,n}^M(P) =
    \\\nonumber
    && -\gamma_\rho S(q_+)\gamma_5 \Gamma_{\nu,n-1}^\mathcal{C}(q_-)S(q_-) \gamma_\rho
    - \gamma_5 \Gamma_{\nu,n}^\mathcal{C}(q_-)
    \\\nonumber
    && - \Gamma_{\nu,n}^\mathcal{C}(q_+) \gamma_5
    - \gamma_\rho S(q_+) \Gamma_{\nu,n-1}^\mathcal{C}(q_+)\gamma_5S(q_-) \gamma_\rho
    \\
    &&- \gamma_\rho S(q_+) \bar\Lambda_{\nu,n-1}^M(P) S(q_-) \gamma_\rho
\;.
\label{eq:recbiglambda2}
\end{eqnarray}

Upon insertion of the lower-order correction term $\bar\Lambda_{\nu,n-1}^M(P)$ in Eq.\ \eqref{eq:recbiglambda2}, it can be seen that the first and fourth term in Eq.\ \eqref{eq:recbiglambda2} are canceled by the second and third term of the $\bar\Lambda_{\nu,n-1}^M(P)$ contribution.

Therefore, the recursion \eqref{eq:recbiglambda2} collapses to
\begin{eqnarray}\nonumber
    \frac{1}{\mathcal{C}}\bar\Lambda_{\nu,n}^M(P) =
    && -\gamma_\rho S(q_+)\gamma_5 \Gamma_{\nu,0}^\mathcal{C}(q_-)S(q_-) \gamma_\rho
    - \gamma_5 \Gamma_{\nu,n}^\mathcal{C}(q_-)
    \\\nonumber
    && - \Gamma_{\nu,n}^\mathcal{C}(q_+) \gamma_5
    - \gamma_\rho S(q_+) \Gamma_{\nu,0}^\mathcal{C}(q_+)\gamma_5S(q_-) \gamma_\rho
    \\
    &&- \gamma_\rho S(q_+) \bar\Lambda_{\nu,0}^M(P) S(q_-) \gamma_\rho
\;.
\label{eq:recbiglambda3}
\end{eqnarray}

From $\Lambda_{\nu,0}^M(P) = 0$, it follows that $\bar\Lambda_{\nu,0}^M(P) = 0$.

Because of $\Gamma_{\nu,0}^\mathcal{C}(q_\pm)=\gamma_\nu$ and the anticommutation properties 
of the Clifford-algebra, the first and fourth term cancel each other and Eq.\ \eqref{eq:recbiglambda3} reduces to
\begin{equation}
    \frac{1}{\mathcal{C}}\bar\Lambda_{\nu,n}^M(P) =
    - \gamma_5 \Gamma_{\nu,n}^\mathcal{C}(q_-)
    - \Gamma_{\nu,n}^\mathcal{C}(q_+) \gamma_5
\;.
\label{eq:reclambda_bar}
\end{equation}

Hence,
\begin{equation}
    \frac{1}{\mathcal{C}}\bar\Lambda_{\nu}^M(P) =
    - \gamma_5 \Gamma_{\nu}^\mathcal{C}(q_-)
    - \Gamma_{\nu}^\mathcal{C}(q_+) \gamma_5
\;.
\label{eq:lambda_bar_expl}
\end{equation}

The AVWTI becomes
\begin{eqnarray}
    && \int_q^\Lambda \left\{ S(q_+) \gamma_5 + \gamma_5 S(q_-)\right\}_{GH} K^{GH}_{EF}(k,q,P)
    \nonumber
    \\&& = -\frac12 \Bigl[
        \gamma_\nu \left\{S(k_+)\gamma_5+\gamma_5S(k_-)\right\} \Gamma_{\nu}^\mathcal{C}(k_-)
        \Bigr. \nonumber \\ \nonumber &&
        + \Gamma_{\nu}^\mathcal{C}(k_+) \left\{S(k_+)\gamma_5+\gamma_5S(k_-)\right\} \gamma_\nu
        \\ \Bigl. &&
        + \gamma_\nu S(k_+) \bar\Lambda_{\nu}^M(P)
        + \bar\Lambda_{\nu}^M(P) S(k_-) \gamma_\nu
        \Bigr]_{EF}
    \\
    && = \{\gamma_\nu S(k_+) \Gamma_{\nu}^\mathcal{C}(k_+) \gamma_5
        + \gamma_5  \gamma_\nu S(k_-) \Gamma_{\nu}^\mathcal{C}(k_-)\}_{EF}
    \\
    && = \{\Sigma(k_+) \gamma_5 + \gamma_5 \Sigma(k_-)\}_{EF}
    \;,
\end{eqnarray}
which is the desired result, where
\begin{equation}
    \gamma_\mu S(p) \Gamma_\mu(p) = \Gamma_\mu(p) S(p) \gamma_\mu
\end{equation}
has been used.

\section{Algebraic gap equations}\label{sec:algebraicgapequation}

In our recursive setup, the coupled
equations for the various dressing functions contain polynomials of increasing order
with increasing order in the recursion. The fully summed solution is obtained via 
a geometric sum and thus produces equations involving polynomials of a finite order
as well. 

Here we present the algebraic equations resulting for $A$ and $B$ at the orders used
in our study, namely $n=0,1,2,3,4,\infty$, including explicitly the current-quark mass $m$,
the coupling $\mathcal{G}$ as well as the strength parameter $\mathcal{C}$.

For $n=0$ one has RL truncation and the Dirac-projected gap equations for $A$ and $B$ read
\begin{eqnarray}
A&=& 1+\frac{2 A \mathcal{G}}{\Delta}\;,\\
B&=& m+\frac{4 B \mathcal{G}}{\Delta}\;.
\end{eqnarray}

For $n=1$ the Dirac-projected gap equations for $A$ and $B$ read
\begin{eqnarray}
A&=& 1+\frac{2 A \mathcal{G}}{\Delta }
      +\frac{8 A \mathcal{C} \mathcal{G}^2}{\Delta ^2}
      +\frac{4 A B^2 \mathcal{C} \mathcal{G}^2}{\Delta ^3}\;,\\
B&=& m+\frac{4 B \mathcal{G}}{\Delta }
      +\frac{12 B \mathcal{C} \mathcal{G}^2}{\Delta ^2}  
      -\frac{4 B^3 \mathcal{C} \mathcal{G}^2}{\Delta ^3}\;.
\end{eqnarray}

For $n=2$ the Dirac-projected gap equations for $A$ and $B$ read
\begin{widetext}
\begin{eqnarray}
A&=& 1+\frac{2 A \mathcal{G}}{\Delta }
      +\frac{8 A \mathcal{C} \mathcal{G}^2}{\Delta ^2}
      +\frac{4 A \mathcal{C} \mathcal{G}^2 \left(B^2+2 \mathcal{C} \mathcal{G}\right)}{\Delta ^3}
      -\frac{16 A B^4 \mathcal{C}^2 \mathcal{G}^3}{\Delta ^5}\;,\\
B&=& m+\frac{4 B \mathcal{G}}{\Delta }
      +\frac{12 B \mathcal{C} \mathcal{G}^2}{\Delta ^2}
      -\frac{4 B \mathcal{C} \mathcal{G}^2 \left(B^2-8 \mathcal{C} \mathcal{G}\right)}{\Delta ^3}
      -\frac{32 B^3 \mathcal{C}^2 \mathcal{G}^3}{\Delta ^4}
      +\frac{16 B^5 \mathcal{C}^2 \mathcal{G}^3}{\Delta ^5}\;.
\end{eqnarray}

For $n=3$ the Dirac-projected gap equations for $A$ and $B$ read
\begin{eqnarray}\nonumber
A&=& 1+\frac{2 A \mathcal{G}}{\Delta }
      +\frac{8 A \mathcal{C} \mathcal{G}^2}{\Delta ^2}
      +\frac{4 A \mathcal{C} \mathcal{G}^2 \left(B^2+2 \mathcal{C} \mathcal{G}\right)}{\Delta ^3}
      +\frac{32 A \mathcal{C}^3 \mathcal{G}^4}{\Delta ^4}
      -\frac{16 A B^2 \mathcal{C}^2 \mathcal{G}^3 \left(B^2-3 \mathcal{C} \mathcal{G}\right)}{\Delta ^5}\\
   && -\frac{32 A B^4 \mathcal{C}^3 \mathcal{G}^4}{\Delta^6}
      +\frac{64 A B^6 \mathcal{C}^3 \mathcal{G}^4}{\Delta ^7}\;,\\
B&=& m+\frac{4 B \mathcal{G}}{\Delta }\nonumber
      +\frac{12 B \mathcal{C} \mathcal{G}^2}{\Delta^2}
      -\frac{4 \left(B^3 \mathcal{C} \mathcal{G}^2-8 B \mathcal{C}^2 \mathcal{G}^3\right)}{\Delta ^3}
      -\frac{16 \left(2 B^3 \mathcal{C}^2 \mathcal{G}^3-7 B \mathcal{C}^3 \mathcal{G}^4\right)}{\Delta ^4}\\
   && +\frac{16 \left(B^5 \mathcal{C}^2 \mathcal{G}^3-11 B^3 \mathcal{C}^3 \mathcal{G}^4\right)}{\Delta ^5}
      +\frac{160 B^5 \mathcal{C}^3 \mathcal{G}^4}{\Delta ^6}
      -\frac{64 B^7 \mathcal{C}^3 \mathcal{G}^4}{\Delta ^7}\;.
\end{eqnarray}

For $n=4$ the Dirac-projected gap equations for $A$ and $B$ read
\begin{eqnarray}\nonumber
A&=& 1+\frac{2 A \mathcal{G}}{\Delta }
      +\frac{8 A \mathcal{C} \mathcal{G}^2}{\Delta ^2}
      +\frac{4 \left(A B^2 \mathcal{C} \mathcal{G}^2+2 A \mathcal{C}^2 \mathcal{G}^3\right)}{\Delta ^3}
      +\frac{32 A \mathcal{C}^3 \mathcal{G}^4}{\Delta ^4}
      +\frac{16 \left(-A B^4 \mathcal{C}^2 \mathcal{G}^3+3 A B^2 \mathcal{C}^3 
       \mathcal{G}^4+2 A \mathcal{C}^4 \mathcal{G}^5\right)}{\Delta^5}\\
   && +\frac{32 \left(2 A B^2 \mathcal{C}^4 \mathcal{G}^5-A B^4 \mathcal{C}^3 \mathcal{G}^4\right)}{\Delta ^6}
      -\frac{64 \left(6 A B^4 \mathcal{C}^4 \mathcal{G}^5-A B^6 \mathcal{C}^3 \mathcal{G}^4\right)}{\Delta ^7}
      -\frac{256 A B^8 \mathcal{C}^4 \mathcal{G}^5}{\Delta ^9}+\frac{256 A B^6 \mathcal{C}^4 \mathcal{G}^5}{\Delta^8}\;,\\
B&=& m+\frac{4 B \mathcal{G}}{\Delta } \nonumber
      +\frac{12 B \mathcal{C} \mathcal{G}^2}{\Delta ^2}
      -\frac{4 \left(B^3 \mathcal{C} \mathcal{G}^2-8 B \mathcal{C}^2 \mathcal{G}^3\right)}{\Delta ^3}
      -\frac{16 \left(2 B^3 \mathcal{C}^2 \mathcal{G}^3-7 B \mathcal{C}^3 \mathcal{G}^4\right)}{\Delta ^4}\\ \nonumber
   && +\frac{16 \left(B^5 \mathcal{C}^2 \mathcal{G}^3-11 B^3 \mathcal{C}^3 \mathcal{G}^4+24 B \mathcal{C}^4 \mathcal{G}^5\right)}{\Delta ^5}
      +\frac{160 \left(B^5 \mathcal{C}^3 \mathcal{G}^4-6 B^3 \mathcal{C}^4 \mathcal{G}^5\right)}{\Delta ^6}
      -\frac{64 \left(B^7 \mathcal{C}^3 \mathcal{G}^4-18 B^5 \mathcal{C}^4 \mathcal{G}^5\right)}{\Delta ^7} \\
   && -\frac{768 B^7 \mathcal{C}^4 \mathcal{G}^5}{\Delta ^8}
      +\frac{256 B^9 \mathcal{C}^4 \mathcal{G}^5}{\Delta ^9}\;.
\end{eqnarray}

Finally, for the fully summed vertex, the resulting gap equations are
\begin{eqnarray}
A&=&1+\frac{3 A \mathcal{G}}{A^2 s+B^2-2 \mathcal{C} \mathcal{G}}-\frac{A \mathcal{G} \left(A^2 s+B^2-4 \mathcal{C}
   \mathcal{G}\right)}{A^4 s^2+2 A^2 s \left(B^2-\mathcal{C} \mathcal{G}\right)+B^4+2 B^2 \mathcal{C} \mathcal{G}-8
   \mathcal{C}^2 \mathcal{G}^2}\;,\\
B&=&m+B \mathcal{G} \left(\frac{3}{A^2 s+B^2-2 \mathcal{C} \mathcal{G}}+\frac{A^2 s+B^2+4 \mathcal{C} \mathcal{G}}{A^4
   s^2+2 A^2 s \left(B^2-\mathcal{C} \mathcal{G}\right)+B^4+2 B^2 \mathcal{C} \mathcal{G}-8 \mathcal{C}^2
   \mathcal{G}^2}\right)  \;.
\end{eqnarray}
\end{widetext}

\bibliography{/Users/ank/Documents/library/had_nucl_graz}

\begin{thebibliography}{156}%
\makeatletter
\providecommand \@ifxundefined [1]{%
 \@ifx{#1\undefined}
}%
\providecommand \@ifnum [1]{%
 \ifnum #1\expandafter \@firstoftwo
 \else \expandafter \@secondoftwo
 \fi
}%
\providecommand \@ifx [1]{%
 \ifx #1\expandafter \@firstoftwo
 \else \expandafter \@secondoftwo
 \fi
}%
\providecommand \natexlab [1]{#1}%
\providecommand \enquote  [1]{``#1''}%
\providecommand \bibnamefont  [1]{#1}%
\providecommand \bibfnamefont [1]{#1}%
\providecommand \citenamefont [1]{#1}%
\providecommand \href@noop [0]{\@secondoftwo}%
\providecommand \href [0]{\begingroup \@sanitize@url \@href}%
\providecommand \@href[1]{\@@startlink{#1}\@@href}%
\providecommand \@@href[1]{\endgroup#1\@@endlink}%
\providecommand \@sanitize@url [0]{\catcode `\\12\catcode `\$12\catcode
  `\&12\catcode `\#12\catcode `\^12\catcode `\_12\catcode `\%12\relax}%
\providecommand \@@startlink[1]{}%
\providecommand \@@endlink[0]{}%
\providecommand \url  [0]{\begingroup\@sanitize@url \@url }%
\providecommand \@url [1]{\endgroup\@href {#1}{\urlprefix }}%
\providecommand \urlprefix  [0]{URL }%
\providecommand \Eprint [0]{\href }%
\providecommand \doibase [0]{http://dx.doi.org/}%
\providecommand \selectlanguage [0]{\@gobble}%
\providecommand \bibinfo  [0]{\@secondoftwo}%
\providecommand \bibfield  [0]{\@secondoftwo}%
\providecommand \translation [1]{[#1]}%
\providecommand \BibitemOpen [0]{}%
\providecommand \bibitemStop [0]{}%
\providecommand \bibitemNoStop [0]{.\EOS\space}%
\providecommand \EOS [0]{\spacefactor3000\relax}%
\providecommand \BibitemShut  [1]{\csname bibitem#1\endcsname}%
\let\auto@bib@innerbib\@empty
\bibitem [{\citenamefont {Roberts}\ \emph {et~al.}(2007)\citenamefont
  {Roberts}, \citenamefont {Bhagwat}, \citenamefont {Holl},\ and\ \citenamefont
  {Wright}}]{Roberts:2007jh}%
  \BibitemOpen
  \bibfield  {author} {\bibinfo {author} {\bibfnamefont {C.~D.}\ \bibnamefont
  {Roberts}}, \bibinfo {author} {\bibfnamefont {M.~S.}\ \bibnamefont
  {Bhagwat}}, \bibinfo {author} {\bibfnamefont {A.}~\bibnamefont {Holl}}, \
  and\ \bibinfo {author} {\bibfnamefont {S.~V.}\ \bibnamefont {Wright}},\
  }\href {\doibase 10.1140/epjst/e2007-00003-5} {\bibfield  {journal} {\bibinfo
   {journal} {Eur. Phys. J. Special Topics}\ }\textbf {\bibinfo {volume}
  {140}},\ \bibinfo {pages} {53} (\bibinfo {year} {2007})}\BibitemShut
  {NoStop}%
\bibitem [{\citenamefont {Fischer}(2006)}]{Fischer:2006ub}%
  \BibitemOpen
  \bibfield  {author} {\bibinfo {author} {\bibfnamefont {C.~S.}\ \bibnamefont
  {Fischer}},\ }\href {\doibase doi:10.1088/0954-3899/32/8/R02} {\bibfield
  {journal} {\bibinfo  {journal} {J. Phys. G}\ }\textbf {\bibinfo {volume}
  {32}},\ \bibinfo {pages} {R253} (\bibinfo {year} {2006})}\BibitemShut
  {NoStop}%
\bibitem [{\citenamefont {Alkofer}\ and\ \citenamefont {von
  Smekal}(2001)}]{Alkofer:2000wg}%
  \BibitemOpen
  \bibfield  {author} {\bibinfo {author} {\bibfnamefont {R.}~\bibnamefont
  {Alkofer}}\ and\ \bibinfo {author} {\bibfnamefont {L.}~\bibnamefont {von
  Smekal}},\ }\href {\doibase doi:10.1016/S0370-1573(01)00010-2} {\bibfield
  {journal} {\bibinfo  {journal} {Phys. Rept.}\ }\textbf {\bibinfo {volume}
  {353}},\ \bibinfo {pages} {281} (\bibinfo {year} {2001})}\BibitemShut
  {NoStop}%
\bibitem [{\citenamefont {Sanchis-Alepuz}\ and\ \citenamefont
  {Williams}(2015{\natexlab{a}})}]{Sanchis-Alepuz:2015tha}%
  \BibitemOpen
  \bibfield  {author} {\bibinfo {author} {\bibfnamefont {H.}~\bibnamefont
  {Sanchis-Alepuz}}\ and\ \bibinfo {author} {\bibfnamefont {R.}~\bibnamefont
  {Williams}},\ }\href {\doibase 10.1088/1742-6596/631/1/012064} {\bibfield
  {journal} {\bibinfo  {journal} {J. Phys. Conf. Ser.}\ }\textbf {\bibinfo
  {volume} {631}},\ \bibinfo {pages} {012064} (\bibinfo {year}
  {2015}{\natexlab{a}})}\BibitemShut {NoStop}%
\bibitem [{\citenamefont {Dudek}\ \emph {et~al.}(2008)\citenamefont {Dudek},
  \citenamefont {Edwards}, \citenamefont {Mathur},\ and\ \citenamefont
  {Richards}}]{Dudek:2007wv}%
  \BibitemOpen
  \bibfield  {author} {\bibinfo {author} {\bibfnamefont {J.~J.}\ \bibnamefont
  {Dudek}}, \bibinfo {author} {\bibfnamefont {R.~G.}\ \bibnamefont {Edwards}},
  \bibinfo {author} {\bibfnamefont {N.}~\bibnamefont {Mathur}}, \ and\ \bibinfo
  {author} {\bibfnamefont {D.~G.}\ \bibnamefont {Richards}},\ }\href {\doibase
  10.1103/PhysRevD.77.034501} {\bibfield  {journal} {\bibinfo  {journal} {Phys.
  Rev. D}\ }\textbf {\bibinfo {volume} {77}},\ \bibinfo {pages} {034501}
  (\bibinfo {year} {2008})}\BibitemShut {NoStop}%
\bibitem [{\citenamefont {Lang}\ \emph {et~al.}(2011)\citenamefont {Lang},
  \citenamefont {Mohler}, \citenamefont {Prelovsek},\ and\ \citenamefont
  {Vidmar}}]{Lang:2011mn}%
  \BibitemOpen
  \bibfield  {author} {\bibinfo {author} {\bibfnamefont {C.~B.}\ \bibnamefont
  {Lang}}, \bibinfo {author} {\bibfnamefont {D.}~\bibnamefont {Mohler}},
  \bibinfo {author} {\bibfnamefont {S.}~\bibnamefont {Prelovsek}}, \ and\
  \bibinfo {author} {\bibfnamefont {M.}~\bibnamefont {Vidmar}},\ }\href
  {\doibase 10.1103/PhysRevD.89.059903, 10.1103/PhysRevD.84.054503} {\bibfield
  {journal} {\bibinfo  {journal} {Phys. Rev.}\ }\textbf {\bibinfo {volume}
  {D84}},\ \bibinfo {pages} {054503} (\bibinfo {year} {2011})},\ \bibinfo
  {note} {[Erratum: Phys. Rev.D89,no.5,059903(2014)]}\BibitemShut {NoStop}%
\bibitem [{\citenamefont {Liu}\ \emph {et~al.}(2012)\citenamefont {Liu},
  \citenamefont {Moir}, \citenamefont {Peardon}, \citenamefont {Ryan},
  \citenamefont {Thomas}, \citenamefont {Vilaseca}, \citenamefont {Dudek},
  \citenamefont {Edwards}, \citenamefont {Joo},\ and\ \citenamefont
  {Richards}}]{Liu:2012ze}%
  \BibitemOpen
  \bibfield  {author} {\bibinfo {author} {\bibfnamefont {L.}~\bibnamefont
  {Liu}}, \bibinfo {author} {\bibfnamefont {G.}~\bibnamefont {Moir}}, \bibinfo
  {author} {\bibfnamefont {M.}~\bibnamefont {Peardon}}, \bibinfo {author}
  {\bibfnamefont {S.~M.}\ \bibnamefont {Ryan}}, \bibinfo {author}
  {\bibfnamefont {C.~E.}\ \bibnamefont {Thomas}}, \bibinfo {author}
  {\bibfnamefont {P.}~\bibnamefont {Vilaseca}}, \bibinfo {author}
  {\bibfnamefont {J.~J.}\ \bibnamefont {Dudek}}, \bibinfo {author}
  {\bibfnamefont {R.~G.}\ \bibnamefont {Edwards}}, \bibinfo {author}
  {\bibfnamefont {B.}~\bibnamefont {Joo}}, \ and\ \bibinfo {author}
  {\bibfnamefont {D.~G.}\ \bibnamefont {Richards}} (\bibinfo {collaboration}
  {Hadron Spectrum Collaboration}),\ }\href {\doibase 10.1007/JHEP07(2012)126}
  {\bibfield  {journal} {\bibinfo  {journal} {J. High Energy Phys.}\ }\textbf
  {\bibinfo {volume} {07}},\ \bibinfo {pages} {126} (\bibinfo {year}
  {2012})}\BibitemShut {NoStop}%
\bibitem [{\citenamefont {Thomas}(2013)}]{Thomas:2014dpa}%
  \BibitemOpen
  \bibfield  {author} {\bibinfo {author} {\bibfnamefont {C.}~\bibnamefont
  {Thomas}},\ }\href@noop {} {\bibfield  {journal} {\bibinfo  {journal} {Proc.
  Sci.}\ }\textbf {\bibinfo {volume} {LATTICE2013}},\ \bibinfo {pages} {003}
  (\bibinfo {year} {2013})}\BibitemShut {NoStop}%
\bibitem [{\citenamefont {Flynn}\ \emph {et~al.}(2015)\citenamefont {Flynn},
  \citenamefont {Izubuchi}, \citenamefont {Kawanai}, \citenamefont {Lehner},
  \citenamefont {Soni} \emph {et~al.}}]{Flynn:2015mha}%
  \BibitemOpen
  \bibfield  {author} {\bibinfo {author} {\bibfnamefont {J.}~\bibnamefont
  {Flynn}}, \bibinfo {author} {\bibfnamefont {T.}~\bibnamefont {Izubuchi}},
  \bibinfo {author} {\bibfnamefont {T.}~\bibnamefont {Kawanai}}, \bibinfo
  {author} {\bibfnamefont {C.}~\bibnamefont {Lehner}}, \bibinfo {author}
  {\bibfnamefont {A.}~\bibnamefont {Soni}},  \emph {et~al.},\ }\href@noop {} {\
  }\Eprint {http://arxiv.org/abs/1501.05373} {1501.05373 [hep-lat]}
  \BibitemShut {NoStop}%
\bibitem [{\citenamefont {Lang}\ \emph {et~al.}(2015)\citenamefont {Lang},
  \citenamefont {Mohler}, \citenamefont {Prelovsek},\ and\ \citenamefont
  {Woloshyn}}]{Lang:2015hza}%
  \BibitemOpen
  \bibfield  {author} {\bibinfo {author} {\bibfnamefont {C.~B.}\ \bibnamefont
  {Lang}}, \bibinfo {author} {\bibfnamefont {D.}~\bibnamefont {Mohler}},
  \bibinfo {author} {\bibfnamefont {S.}~\bibnamefont {Prelovsek}}, \ and\
  \bibinfo {author} {\bibfnamefont {R.~M.}\ \bibnamefont {Woloshyn}},\ }\href
  {\doibase 10.1016/j.physletb.2015.08.038} {\bibfield  {journal} {\bibinfo
  {journal} {Phys. Lett. B}\ }\textbf {\bibinfo {volume} {750}},\ \bibinfo
  {pages} {17} (\bibinfo {year} {2015})}\BibitemShut {NoStop}%
\bibitem [{\citenamefont {Maris}\ and\ \citenamefont
  {Tandy}(1999)}]{Maris:1999nt}%
  \BibitemOpen
  \bibfield  {author} {\bibinfo {author} {\bibfnamefont {P.}~\bibnamefont
  {Maris}}\ and\ \bibinfo {author} {\bibfnamefont {P.~C.}\ \bibnamefont
  {Tandy}},\ }\href {\doibase 10.1103/PhysRevC.60.055214} {\bibfield  {journal}
  {\bibinfo  {journal} {Phys. Rev. C}\ }\textbf {\bibinfo {volume} {60}},\
  \bibinfo {pages} {055214} (\bibinfo {year} {1999})}\BibitemShut {NoStop}%
\bibitem [{\citenamefont {Holl}\ \emph {et~al.}(2003)\citenamefont {Holl},
  \citenamefont {Krassnigg},\ and\ \citenamefont {Roberts}}]{Holl:2003dq}%
  \BibitemOpen
  \bibfield  {author} {\bibinfo {author} {\bibfnamefont {A.}~\bibnamefont
  {Holl}}, \bibinfo {author} {\bibfnamefont {A.}~\bibnamefont {Krassnigg}}, \
  and\ \bibinfo {author} {\bibfnamefont {C.~D.}\ \bibnamefont {Roberts}},\
  }\href@noop {} {\ }\Eprint {http://arxiv.org/abs/nucl-th/0311033}
  {nucl-th/0311033} \BibitemShut {NoStop}%
\bibitem [{\citenamefont {Holl}\ \emph {et~al.}(2004)\citenamefont {Holl},
  \citenamefont {Krassnigg},\ and\ \citenamefont {Roberts}}]{Holl:2004fr}%
  \BibitemOpen
  \bibfield  {author} {\bibinfo {author} {\bibfnamefont {A.}~\bibnamefont
  {Holl}}, \bibinfo {author} {\bibfnamefont {A.}~\bibnamefont {Krassnigg}}, \
  and\ \bibinfo {author} {\bibfnamefont {C.~D.}\ \bibnamefont {Roberts}},\
  }\href {\doibase 10.1103/PhysRevC.70.042203} {\bibfield  {journal} {\bibinfo
  {journal} {Phys. Rev. C}\ }\textbf {\bibinfo {volume} {70}},\ \bibinfo
  {pages} {042203(R)} (\bibinfo {year} {2004})}\BibitemShut {NoStop}%
\bibitem [{\citenamefont {Krassnigg}\ and\ \citenamefont
  {Maris}(2005)}]{Krassnigg:2004if}%
  \BibitemOpen
  \bibfield  {author} {\bibinfo {author} {\bibfnamefont {A.}~\bibnamefont
  {Krassnigg}}\ and\ \bibinfo {author} {\bibfnamefont {P.}~\bibnamefont
  {Maris}},\ }\href {\doibase doi:10.1088/1742-6596/9/1/029} {\bibfield
  {journal} {\bibinfo  {journal} {J. Phys. Conf. Ser.}\ }\textbf {\bibinfo
  {volume} {9}},\ \bibinfo {pages} {153} (\bibinfo {year} {2005})}\BibitemShut
  {NoStop}%
\bibitem [{\citenamefont {Holl}\ \emph
  {et~al.}(2005{\natexlab{a}})\citenamefont {Holl}, \citenamefont {Krassnigg},
  \citenamefont {Maris}, \citenamefont {Roberts},\ and\ \citenamefont
  {Wright}}]{Holl:2005vu}%
  \BibitemOpen
  \bibfield  {author} {\bibinfo {author} {\bibfnamefont {A.}~\bibnamefont
  {Holl}}, \bibinfo {author} {\bibfnamefont {A.}~\bibnamefont {Krassnigg}},
  \bibinfo {author} {\bibfnamefont {P.}~\bibnamefont {Maris}}, \bibinfo
  {author} {\bibfnamefont {C.~D.}\ \bibnamefont {Roberts}}, \ and\ \bibinfo
  {author} {\bibfnamefont {S.~V.}\ \bibnamefont {Wright}},\ }\href {\doibase
  10.1103/PhysRevC.71.065204} {\bibfield  {journal} {\bibinfo  {journal} {Phys.
  Rev. C}\ }\textbf {\bibinfo {volume} {71}},\ \bibinfo {pages} {065204}
  (\bibinfo {year} {2005}{\natexlab{a}})}\BibitemShut {NoStop}%
\bibitem [{\citenamefont {Alkofer}\ \emph {et~al.}(2006)\citenamefont
  {Alkofer}, \citenamefont {Kloker}, \citenamefont {Krassnigg},\ and\
  \citenamefont {Wagenbrunn}}]{Alkofer:2005ug}%
  \BibitemOpen
  \bibfield  {author} {\bibinfo {author} {\bibfnamefont {R.}~\bibnamefont
  {Alkofer}}, \bibinfo {author} {\bibfnamefont {M.}~\bibnamefont {Kloker}},
  \bibinfo {author} {\bibfnamefont {A.}~\bibnamefont {Krassnigg}}, \ and\
  \bibinfo {author} {\bibfnamefont {R.~F.}\ \bibnamefont {Wagenbrunn}},\ }\href
  {\doibase 10.1103/PhysRevLett.96.022001} {\bibfield  {journal} {\bibinfo
  {journal} {Phys. Rev. Lett.}\ }\textbf {\bibinfo {volume} {96}},\ \bibinfo
  {pages} {022001} (\bibinfo {year} {2006})}\BibitemShut {NoStop}%
\bibitem [{\citenamefont {Eichmann}\ \emph
  {et~al.}(2008{\natexlab{a}})\citenamefont {Eichmann}, \citenamefont
  {Alkofer}, \citenamefont {Krassnigg},\ and\ \citenamefont
  {Nicmorus}}]{Eichmann:2008kk}%
  \BibitemOpen
  \bibfield  {author} {\bibinfo {author} {\bibfnamefont {G.}~\bibnamefont
  {Eichmann}}, \bibinfo {author} {\bibfnamefont {R.}~\bibnamefont {Alkofer}},
  \bibinfo {author} {\bibfnamefont {A.}~\bibnamefont {Krassnigg}}, \ and\
  \bibinfo {author} {\bibfnamefont {D.}~\bibnamefont {Nicmorus}},\ }\bibfield
  {booktitle} {\emph {\bibinfo {booktitle} {{Proceedings, 8th Conference on
  Quark Confinement and the Hadron Spectrum (Confinement8)}}},\ }\href
  {http://pos.sissa.it//archive/conferences/077/077/Confinement8_077.pdf}
  {\bibfield  {journal} {\bibinfo  {journal} {PoS}\ }\textbf {\bibinfo {volume}
  {CONFINEMENT8}},\ \bibinfo {pages} {077} (\bibinfo {year}
  {2008}{\natexlab{a}})}\BibitemShut {NoStop}%
\bibitem [{\citenamefont {Eichmann}\ \emph {et~al.}(2009)\citenamefont
  {Eichmann}, \citenamefont {Cloet}, \citenamefont {Alkofer}, \citenamefont
  {Krassnigg},\ and\ \citenamefont {Roberts}}]{Eichmann:2008ef}%
  \BibitemOpen
  \bibfield  {author} {\bibinfo {author} {\bibfnamefont {G.}~\bibnamefont
  {Eichmann}}, \bibinfo {author} {\bibfnamefont {I.~C.}\ \bibnamefont {Cloet}},
  \bibinfo {author} {\bibfnamefont {R.}~\bibnamefont {Alkofer}}, \bibinfo
  {author} {\bibfnamefont {A.}~\bibnamefont {Krassnigg}}, \ and\ \bibinfo
  {author} {\bibfnamefont {C.~D.}\ \bibnamefont {Roberts}},\ }\href {\doibase
  10.1103/PhysRevC.79.012202} {\bibfield  {journal} {\bibinfo  {journal} {Phys.
  Rev. C}\ }\textbf {\bibinfo {volume} {79}},\ \bibinfo {pages} {012202(R)}
  (\bibinfo {year} {2009})}\BibitemShut {NoStop}%
\bibitem [{\citenamefont {Eichmann}\ \emph
  {et~al.}(2008{\natexlab{b}})\citenamefont {Eichmann}, \citenamefont
  {Alkofer}, \citenamefont {Cloet}, \citenamefont {Krassnigg},\ and\
  \citenamefont {Roberts}}]{Eichmann:2008ae}%
  \BibitemOpen
  \bibfield  {author} {\bibinfo {author} {\bibfnamefont {G.}~\bibnamefont
  {Eichmann}}, \bibinfo {author} {\bibfnamefont {R.}~\bibnamefont {Alkofer}},
  \bibinfo {author} {\bibfnamefont {I.~C.}\ \bibnamefont {Cloet}}, \bibinfo
  {author} {\bibfnamefont {A.}~\bibnamefont {Krassnigg}}, \ and\ \bibinfo
  {author} {\bibfnamefont {C.~D.}\ \bibnamefont {Roberts}},\ }\href {\doibase
  10.1103/PhysRevC.77.042202} {\bibfield  {journal} {\bibinfo  {journal} {Phys.
  Rev. C}\ }\textbf {\bibinfo {volume} {77}},\ \bibinfo {pages} {042202(R)}
  (\bibinfo {year} {2008}{\natexlab{b}})}\BibitemShut {NoStop}%
\bibitem [{\citenamefont {Krassnigg}(2009)}]{Krassnigg:2009zh}%
  \BibitemOpen
  \bibfield  {author} {\bibinfo {author} {\bibfnamefont {A.}~\bibnamefont
  {Krassnigg}},\ }\href {\doibase 10.1103/PhysRevD.80.114010} {\bibfield
  {journal} {\bibinfo  {journal} {Phys. Rev. D}\ }\textbf {\bibinfo {volume}
  {80}},\ \bibinfo {pages} {114010} (\bibinfo {year} {2009})}\BibitemShut
  {NoStop}%
\bibitem [{\citenamefont {Eichmann}\ \emph {et~al.}(2010)\citenamefont
  {Eichmann}, \citenamefont {Alkofer}, \citenamefont {Krassnigg},\ and\
  \citenamefont {Nicmorus}}]{Eichmann:2009qa}%
  \BibitemOpen
  \bibfield  {author} {\bibinfo {author} {\bibfnamefont {G.}~\bibnamefont
  {Eichmann}}, \bibinfo {author} {\bibfnamefont {R.}~\bibnamefont {Alkofer}},
  \bibinfo {author} {\bibfnamefont {A.}~\bibnamefont {Krassnigg}}, \ and\
  \bibinfo {author} {\bibfnamefont {D.}~\bibnamefont {Nicmorus}},\ }\href
  {\doibase 10.1103/PhysRevLett.104.201601} {\bibfield  {journal} {\bibinfo
  {journal} {Phys. Rev. Lett.}\ }\textbf {\bibinfo {volume} {104}},\ \bibinfo
  {pages} {201601} (\bibinfo {year} {2010})}\BibitemShut {NoStop}%
\bibitem [{\citenamefont {Alkofer}\ \emph {et~al.}(2010)\citenamefont
  {Alkofer}, \citenamefont {Eichmann}, \citenamefont {Krassnigg},\ and\
  \citenamefont {Nicmorus}}]{Alkofer:2009jk}%
  \BibitemOpen
  \bibfield  {author} {\bibinfo {author} {\bibfnamefont {R.}~\bibnamefont
  {Alkofer}}, \bibinfo {author} {\bibfnamefont {G.}~\bibnamefont {Eichmann}},
  \bibinfo {author} {\bibfnamefont {A.}~\bibnamefont {Krassnigg}}, \ and\
  \bibinfo {author} {\bibfnamefont {D.}~\bibnamefont {Nicmorus}},\ }\href@noop
  {} {\bibfield  {journal} {\bibinfo  {journal} {Chinese Physics C}\ }\textbf
  {\bibinfo {volume} {34}},\ \bibinfo {pages} {1175} (\bibinfo {year}
  {2010})}\BibitemShut {NoStop}%
\bibitem [{\citenamefont {Krassnigg}\ and\ \citenamefont
  {Blank}(2011)}]{Krassnigg:2010mh}%
  \BibitemOpen
  \bibfield  {author} {\bibinfo {author} {\bibfnamefont {A.}~\bibnamefont
  {Krassnigg}}\ and\ \bibinfo {author} {\bibfnamefont {M.}~\bibnamefont
  {Blank}},\ }\href {\doibase 10.1103/PhysRevD.83.096006} {\bibfield  {journal}
  {\bibinfo  {journal} {Phys. Rev. D}\ }\textbf {\bibinfo {volume} {83}},\
  \bibinfo {pages} {096006} (\bibinfo {year} {2011})}\BibitemShut {NoStop}%
\bibitem [{\citenamefont {Dorkin}\ \emph {et~al.}(2011)\citenamefont {Dorkin},
  \citenamefont {Hilger}, \citenamefont {Kaptari},\ and\ \citenamefont
  {K{\"a}mpfer}}]{Dorkin:2010ut}%
  \BibitemOpen
  \bibfield  {author} {\bibinfo {author} {\bibfnamefont {S.~M.}\ \bibnamefont
  {Dorkin}}, \bibinfo {author} {\bibfnamefont {T.}~\bibnamefont {Hilger}},
  \bibinfo {author} {\bibfnamefont {L.~P.}\ \bibnamefont {Kaptari}}, \ and\
  \bibinfo {author} {\bibfnamefont {B.}~\bibnamefont {K{\"a}mpfer}},\ }\href
  {\doibase 10.1007/s00601-010-0108-6} {\bibfield  {journal} {\bibinfo
  {journal} {Few Body Syst.}\ }\textbf {\bibinfo {volume} {49}},\ \bibinfo
  {pages} {247} (\bibinfo {year} {2011})}\BibitemShut {NoStop}%
\bibitem [{\citenamefont {Blank}\ and\ \citenamefont
  {Krassnigg}(2010)}]{Blank:2010bz}%
  \BibitemOpen
  \bibfield  {author} {\bibinfo {author} {\bibfnamefont {M.}~\bibnamefont
  {Blank}}\ and\ \bibinfo {author} {\bibfnamefont {A.}~\bibnamefont
  {Krassnigg}},\ }\href {\doibase 10.1103/PhysRevD.82.034006} {\bibfield
  {journal} {\bibinfo  {journal} {Phys. Rev. D}\ }\textbf {\bibinfo {volume}
  {82}},\ \bibinfo {pages} {034006} (\bibinfo {year} {2010})}\BibitemShut
  {NoStop}%
\bibitem [{\citenamefont {Blank}\ \emph {et~al.}(2011)\citenamefont {Blank},
  \citenamefont {Krassnigg},\ and\ \citenamefont {Maas}}]{Blank:2010pa}%
  \BibitemOpen
  \bibfield  {author} {\bibinfo {author} {\bibfnamefont {M.}~\bibnamefont
  {Blank}}, \bibinfo {author} {\bibfnamefont {A.}~\bibnamefont {Krassnigg}}, \
  and\ \bibinfo {author} {\bibfnamefont {A.}~\bibnamefont {Maas}},\ }\href
  {\doibase 10.1103/PhysRevD.83.034020} {\bibfield  {journal} {\bibinfo
  {journal} {Phys. Rev. D}\ }\textbf {\bibinfo {volume} {83}},\ \bibinfo
  {pages} {034020} (\bibinfo {year} {2011})}\BibitemShut {NoStop}%
\bibitem [{\citenamefont {Mader}\ \emph {et~al.}(2011)\citenamefont {Mader},
  \citenamefont {Eichmann}, \citenamefont {Blank},\ and\ \citenamefont
  {Krassnigg}}]{Mader:2011zf}%
  \BibitemOpen
  \bibfield  {author} {\bibinfo {author} {\bibfnamefont {V.}~\bibnamefont
  {Mader}}, \bibinfo {author} {\bibfnamefont {G.}~\bibnamefont {Eichmann}},
  \bibinfo {author} {\bibfnamefont {M.}~\bibnamefont {Blank}}, \ and\ \bibinfo
  {author} {\bibfnamefont {A.}~\bibnamefont {Krassnigg}},\ }\href@noop {}
  {\bibfield  {journal} {\bibinfo  {journal} {Phys. Rev. D}\ }\textbf {\bibinfo
  {volume} {84}},\ \bibinfo {pages} {034012} (\bibinfo {year}
  {2011})}\BibitemShut {NoStop}%
\bibitem [{\citenamefont {Blank}\ and\ \citenamefont
  {Krassnigg}(2011{\natexlab{a}})}]{Blank:2011ha}%
  \BibitemOpen
  \bibfield  {author} {\bibinfo {author} {\bibfnamefont {M.}~\bibnamefont
  {Blank}}\ and\ \bibinfo {author} {\bibfnamefont {A.}~\bibnamefont
  {Krassnigg}},\ }\href {\doibase 10.1103/PhysRevD.84.096014} {\bibfield
  {journal} {\bibinfo  {journal} {Phys. Rev. D}\ }\textbf {\bibinfo {volume}
  {84}},\ \bibinfo {pages} {096014} (\bibinfo {year}
  {2011}{\natexlab{a}})}\BibitemShut {NoStop}%
\bibitem [{\citenamefont {Hilger}(2012)}]{UweHilger:2012uua}%
  \BibitemOpen
  \bibfield  {author} {\bibinfo {author} {\bibfnamefont {T.~U.}\ \bibnamefont
  {Hilger}},\ }\emph {\bibinfo {title} {{Medium Modifications of Mesons}}},\
  \href@noop {} {Ph.D. thesis},\ \bibinfo  {school} {TU Dresden} (\bibinfo
  {year} {2012})\BibitemShut {NoStop}%
\bibitem [{\citenamefont {Popovici}\ \emph {et~al.}(2015)\citenamefont
  {Popovici}, \citenamefont {Hilger}, \citenamefont {Gomez-Rocha},\ and\
  \citenamefont {Krassnigg}}]{Popovici:2014pha}%
  \BibitemOpen
  \bibfield  {author} {\bibinfo {author} {\bibfnamefont {C.}~\bibnamefont
  {Popovici}}, \bibinfo {author} {\bibfnamefont {T.}~\bibnamefont {Hilger}},
  \bibinfo {author} {\bibfnamefont {M.}~\bibnamefont {Gomez-Rocha}}, \ and\
  \bibinfo {author} {\bibfnamefont {A.}~\bibnamefont {Krassnigg}},\ }\href
  {\doibase 10.1007/s00601-014-0934-z} {\bibfield  {journal} {\bibinfo
  {journal} {Few Body Syst.}\ }\textbf {\bibinfo {volume} {56}},\ \bibinfo
  {pages} {481} (\bibinfo {year} {2015})}\BibitemShut {NoStop}%
\bibitem [{\citenamefont {Hilger}\ \emph
  {et~al.}(2015{\natexlab{a}})\citenamefont {Hilger}, \citenamefont {Popovici},
  \citenamefont {Gomez-Rocha},\ and\ \citenamefont
  {Krassnigg}}]{Hilger:2014nma}%
  \BibitemOpen
  \bibfield  {author} {\bibinfo {author} {\bibfnamefont {T.}~\bibnamefont
  {Hilger}}, \bibinfo {author} {\bibfnamefont {C.}~\bibnamefont {Popovici}},
  \bibinfo {author} {\bibfnamefont {M.}~\bibnamefont {Gomez-Rocha}}, \ and\
  \bibinfo {author} {\bibfnamefont {A.}~\bibnamefont {Krassnigg}},\ }\href
  {\doibase 10.1103/PhysRevD.91.034013} {\bibfield  {journal} {\bibinfo
  {journal} {Phys. Rev. D}\ }\textbf {\bibinfo {volume} {91}},\ \bibinfo
  {pages} {034013} (\bibinfo {year} {2015}{\natexlab{a}})}\BibitemShut
  {NoStop}%
\bibitem [{\citenamefont {Fischer}\ \emph {et~al.}(2014)\citenamefont
  {Fischer}, \citenamefont {Kubrak},\ and\ \citenamefont
  {Williams}}]{Fischer:2014xha}%
  \BibitemOpen
  \bibfield  {author} {\bibinfo {author} {\bibfnamefont {C.~S.}\ \bibnamefont
  {Fischer}}, \bibinfo {author} {\bibfnamefont {S.}~\bibnamefont {Kubrak}}, \
  and\ \bibinfo {author} {\bibfnamefont {R.}~\bibnamefont {Williams}},\ }\href
  {\doibase 10.1140/epja/i2014-14126-6} {\bibfield  {journal} {\bibinfo
  {journal} {Eur. Phys. J. A}\ }\textbf {\bibinfo {volume} {50}},\ \bibinfo
  {pages} {126} (\bibinfo {year} {2014})}\BibitemShut {NoStop}%
\bibitem [{\citenamefont {Fischer}\ \emph {et~al.}(2015)\citenamefont
  {Fischer}, \citenamefont {Kubrak},\ and\ \citenamefont
  {Williams}}]{Fischer:2014cfa}%
  \BibitemOpen
  \bibfield  {author} {\bibinfo {author} {\bibfnamefont {C.~S.}\ \bibnamefont
  {Fischer}}, \bibinfo {author} {\bibfnamefont {S.}~\bibnamefont {Kubrak}}, \
  and\ \bibinfo {author} {\bibfnamefont {R.}~\bibnamefont {Williams}},\ }\href
  {\doibase 10.1140/epja/i2015-15010-7} {\bibfield  {journal} {\bibinfo
  {journal} {Eur.Phys.J. A}\ }\textbf {\bibinfo {volume} {51}},\ \bibinfo
  {pages} {10} (\bibinfo {year} {2015})}\BibitemShut {NoStop}%
\bibitem [{\citenamefont {Hilger}\ \emph
  {et~al.}(2015{\natexlab{b}})\citenamefont {Hilger}, \citenamefont
  {Gomez-Rocha},\ and\ \citenamefont {Krassnigg}}]{Hilger:2015hka}%
  \BibitemOpen
  \bibfield  {author} {\bibinfo {author} {\bibfnamefont {T.}~\bibnamefont
  {Hilger}}, \bibinfo {author} {\bibfnamefont {M.}~\bibnamefont {Gomez-Rocha}},
  \ and\ \bibinfo {author} {\bibfnamefont {A.}~\bibnamefont {Krassnigg}},\
  }\href {\doibase 10.1103/PhysRevD.91.114004} {\bibfield  {journal} {\bibinfo
  {journal} {Phys. Rev. D}\ }\textbf {\bibinfo {volume} {91}},\ \bibinfo
  {pages} {114004} (\bibinfo {year} {2015}{\natexlab{b}})}\BibitemShut
  {NoStop}%
\bibitem [{\citenamefont {Hilger}\ \emph
  {et~al.}(2015{\natexlab{c}})\citenamefont {Hilger}, \citenamefont
  {Gomez-Rocha},\ and\ \citenamefont {Krassnigg}}]{Hilger:2015ora}%
  \BibitemOpen
  \bibfield  {author} {\bibinfo {author} {\bibfnamefont {T.}~\bibnamefont
  {Hilger}}, \bibinfo {author} {\bibfnamefont {M.}~\bibnamefont {Gomez-Rocha}},
  \ and\ \bibinfo {author} {\bibfnamefont {A.}~\bibnamefont {Krassnigg}},\
  }\href@noop {} {\ }\Eprint {http://arxiv.org/abs/1508.07183} {1508.07183
  [hep-ph]} \BibitemShut {NoStop}%
\bibitem [{\citenamefont {Hilger}(2015)}]{Hilger:2015zva}%
  \BibitemOpen
  \bibfield  {author} {\bibinfo {author} {\bibfnamefont {T.}~\bibnamefont
  {Hilger}},\ }\href@noop {} {\ }\Eprint {http://arxiv.org/abs/1510.08288}
  {1510.08288 [hep-ph]} \BibitemShut {NoStop}%
\bibitem [{\citenamefont {Raya}\ \emph {et~al.}(2015)\citenamefont {Raya},
  \citenamefont {Chang}, \citenamefont {Bashir}, \citenamefont
  {Cobos-Martinez}, \citenamefont {Gutiérrez-Guerrero}, \citenamefont
  {Roberts},\ and\ \citenamefont {Tandy}}]{Raya:2015gva}%
  \BibitemOpen
  \bibfield  {author} {\bibinfo {author} {\bibfnamefont {K.}~\bibnamefont
  {Raya}}, \bibinfo {author} {\bibfnamefont {L.}~\bibnamefont {Chang}},
  \bibinfo {author} {\bibfnamefont {A.}~\bibnamefont {Bashir}}, \bibinfo
  {author} {\bibfnamefont {J.~J.}\ \bibnamefont {Cobos-Martinez}}, \bibinfo
  {author} {\bibfnamefont {L.~X.}\ \bibnamefont {Gutiérrez-Guerrero}},
  \bibinfo {author} {\bibfnamefont {C.~D.}\ \bibnamefont {Roberts}}, \ and\
  \bibinfo {author} {\bibfnamefont {P.~C.}\ \bibnamefont {Tandy}},\ }\href@noop
  {} {\ }\Eprint {http://arxiv.org/abs/1510.02799} {1510.02799 [nucl-th]}
  \BibitemShut {NoStop}%
\bibitem [{\citenamefont {Bender}\ \emph {et~al.}(1996)\citenamefont {Bender},
  \citenamefont {Roberts},\ and\ \citenamefont {Von~Smekal}}]{Bender:1996bb}%
  \BibitemOpen
  \bibfield  {author} {\bibinfo {author} {\bibfnamefont {A.}~\bibnamefont
  {Bender}}, \bibinfo {author} {\bibfnamefont {C.~D.}\ \bibnamefont {Roberts}},
  \ and\ \bibinfo {author} {\bibfnamefont {L.}~\bibnamefont {Von~Smekal}},\
  }\href {\doibase 10.1016/0370-2693(96)00372-3} {\bibfield  {journal}
  {\bibinfo  {journal} {Phys. Lett. B}\ }\textbf {\bibinfo {volume} {380}},\
  \bibinfo {pages} {7} (\bibinfo {year} {1996})}\BibitemShut {NoStop}%
\bibitem [{\citenamefont {Bhagwat}\ \emph
  {et~al.}(2007{\natexlab{a}})\citenamefont {Bhagwat}, \citenamefont {Hoell},
  \citenamefont {Krassnigg}, \citenamefont {Roberts},\ and\ \citenamefont
  {Wright}}]{Bhagwat:2007rj}%
  \BibitemOpen
  \bibfield  {author} {\bibinfo {author} {\bibfnamefont {M.~S.}\ \bibnamefont
  {Bhagwat}}, \bibinfo {author} {\bibfnamefont {A.}~\bibnamefont {Hoell}},
  \bibinfo {author} {\bibfnamefont {A.}~\bibnamefont {Krassnigg}}, \bibinfo
  {author} {\bibfnamefont {C.~D.}\ \bibnamefont {Roberts}}, \ and\ \bibinfo
  {author} {\bibfnamefont {S.~V.}\ \bibnamefont {Wright}},\ }\href {\doibase
  10.1007/s00601-007-0174-6} {\bibfield  {journal} {\bibinfo  {journal}
  {Few-Body Syst.}\ }\textbf {\bibinfo {volume} {40}},\ \bibinfo {pages} {209}
  (\bibinfo {year} {2007}{\natexlab{a}})}\BibitemShut {NoStop}%
\bibitem [{\citenamefont {Krassnigg}(2008)}]{Krassnigg:2008gd}%
  \BibitemOpen
  \bibfield  {author} {\bibinfo {author} {\bibfnamefont {A.}~\bibnamefont
  {Krassnigg}},\ }\bibfield  {booktitle} {\emph {\bibinfo {booktitle}
  {{Proceedings, 8th Conference on Quark Confinement and the Hadron Spectrum
  (Confinement8)}}},\ }\href@noop {} {\bibfield  {journal} {\bibinfo  {journal}
  {{PoS}}\ }\textbf {\bibinfo {volume} {{CONFINEMENT8}}},\ \bibinfo {pages}
  {075} (\bibinfo {year} {2008})}\BibitemShut {NoStop}%
\bibitem [{\citenamefont {Blank}\ and\ \citenamefont
  {Krassnigg}(2011{\natexlab{b}})}]{Blank:2010bp}%
  \BibitemOpen
  \bibfield  {author} {\bibinfo {author} {\bibfnamefont {M.}~\bibnamefont
  {Blank}}\ and\ \bibinfo {author} {\bibfnamefont {A.}~\bibnamefont
  {Krassnigg}},\ }\href {\doibase 10.1016/j.cpc.2011.03.003} {\bibfield
  {journal} {\bibinfo  {journal} {Comput. Phys. Commun.}\ }\textbf {\bibinfo
  {volume} {182}},\ \bibinfo {pages} {1391} (\bibinfo {year}
  {2011}{\natexlab{b}})}\BibitemShut {NoStop}%
\bibitem [{\citenamefont {Blank}\ and\ \citenamefont
  {Krassnigg}(2011{\natexlab{c}})}]{Blank:2010sn}%
  \BibitemOpen
  \bibfield  {author} {\bibinfo {author} {\bibfnamefont {M.}~\bibnamefont
  {Blank}}\ and\ \bibinfo {author} {\bibfnamefont {A.}~\bibnamefont
  {Krassnigg}},\ }\href {\doibase 10.1063/1.3575026} {\bibfield  {journal}
  {\bibinfo  {journal} {AIP Conf. Proc.}\ }\textbf {\bibinfo {volume} {1343}},\
  \bibinfo {pages} {349} (\bibinfo {year} {2011}{\natexlab{c}})}\BibitemShut
  {NoStop}%
\bibitem [{\citenamefont {Blank}(2011)}]{Blank:2011qk}%
  \BibitemOpen
  \bibfield  {author} {\bibinfo {author} {\bibfnamefont {M.}~\bibnamefont
  {Blank}},\ }\emph {\bibinfo {title} {{Properties of quarks and mesons in the
  Dyson-Schwinger/Bethe-Salpeter approach}}},\ \href@noop {} {Ph.D. thesis},\
  \bibinfo  {school} {University of Graz} (\bibinfo {year} {2011}),\ \Eprint
  {http://arxiv.org/abs/1106.4843} {1106.4843 [hep-ph]} \BibitemShut {NoStop}%
\bibitem [{\citenamefont {Watson}\ and\ \citenamefont
  {Cassing}(2004)}]{Watson:2004jq}%
  \BibitemOpen
  \bibfield  {author} {\bibinfo {author} {\bibfnamefont {P.}~\bibnamefont
  {Watson}}\ and\ \bibinfo {author} {\bibfnamefont {W.}~\bibnamefont
  {Cassing}},\ }\href {\doibase 10.1007/s00601-004-0063-1} {\bibfield
  {journal} {\bibinfo  {journal} {Few-Body Syst.}\ }\textbf {\bibinfo {volume}
  {35}},\ \bibinfo {pages} {99} (\bibinfo {year} {2004})}\BibitemShut {NoStop}%
\bibitem [{\citenamefont {Watson}\ \emph {et~al.}(2004)\citenamefont {Watson},
  \citenamefont {Cassing},\ and\ \citenamefont {Tandy}}]{Watson:2004kd}%
  \BibitemOpen
  \bibfield  {author} {\bibinfo {author} {\bibfnamefont {P.}~\bibnamefont
  {Watson}}, \bibinfo {author} {\bibfnamefont {W.}~\bibnamefont {Cassing}}, \
  and\ \bibinfo {author} {\bibfnamefont {P.~C.}\ \bibnamefont {Tandy}},\ }\href
  {\doibase 10.1007/s00601-004-0067-x} {\bibfield  {journal} {\bibinfo
  {journal} {Few-Body Syst.}\ }\textbf {\bibinfo {volume} {35}},\ \bibinfo
  {pages} {129} (\bibinfo {year} {2004})}\BibitemShut {NoStop}%
\bibitem [{\citenamefont {Fischer}\ \emph {et~al.}(2005)\citenamefont
  {Fischer}, \citenamefont {Watson},\ and\ \citenamefont
  {Cassing}}]{Fischer:2005en}%
  \BibitemOpen
  \bibfield  {author} {\bibinfo {author} {\bibfnamefont {C.~S.}\ \bibnamefont
  {Fischer}}, \bibinfo {author} {\bibfnamefont {P.}~\bibnamefont {Watson}}, \
  and\ \bibinfo {author} {\bibfnamefont {W.}~\bibnamefont {Cassing}},\ }\href
  {\doibase 10.1103/PhysRevD.72.094025} {\bibfield  {journal} {\bibinfo
  {journal} {Phys. Rev. D}\ }\textbf {\bibinfo {volume} {72}},\ \bibinfo
  {pages} {094025} (\bibinfo {year} {2005})}\BibitemShut {NoStop}%
\bibitem [{\citenamefont {Fischer}\ and\ \citenamefont
  {Williams}(2008)}]{Fischer:2008wy}%
  \BibitemOpen
  \bibfield  {author} {\bibinfo {author} {\bibfnamefont {C.~S.}\ \bibnamefont
  {Fischer}}\ and\ \bibinfo {author} {\bibfnamefont {R.}~\bibnamefont
  {Williams}},\ }\href {\doibase 10.1103/PhysRevD.78.074006} {\bibfield
  {journal} {\bibinfo  {journal} {Phys. Rev. D}\ }\textbf {\bibinfo {volume}
  {78}},\ \bibinfo {pages} {074006} (\bibinfo {year} {2008})}\BibitemShut
  {NoStop}%
\bibitem [{\citenamefont {Fischer}\ and\ \citenamefont
  {Williams}(2009)}]{Fischer:2009jm}%
  \BibitemOpen
  \bibfield  {author} {\bibinfo {author} {\bibfnamefont {C.~S.}\ \bibnamefont
  {Fischer}}\ and\ \bibinfo {author} {\bibfnamefont {R.}~\bibnamefont
  {Williams}},\ }\href {\doibase 10.1103/PhysRevLett.103.122001} {\bibfield
  {journal} {\bibinfo  {journal} {Phys. Rev. Lett.}\ }\textbf {\bibinfo
  {volume} {103}},\ \bibinfo {pages} {122001} (\bibinfo {year}
  {2009})}\BibitemShut {NoStop}%
\bibitem [{\citenamefont {Williams}(2010)}]{Williams:2009wx}%
  \BibitemOpen
  \bibfield  {author} {\bibinfo {author} {\bibfnamefont {R.}~\bibnamefont
  {Williams}},\ }\href {\doibase 10.1051/epjconf/20100303005} {\bibfield
  {journal} {\bibinfo  {journal} {EPJ Web Conf.}\ }\textbf {\bibinfo {volume}
  {3}},\ \bibinfo {pages} {03005} (\bibinfo {year} {2010})}\BibitemShut
  {NoStop}%
\bibitem [{\citenamefont {Williams}(2015)}]{Williams:2014iea}%
  \BibitemOpen
  \bibfield  {author} {\bibinfo {author} {\bibfnamefont {R.}~\bibnamefont
  {Williams}},\ }\href {\doibase 10.1140/epja/i2015-15057-4} {\bibfield
  {journal} {\bibinfo  {journal} {Eur.Phys.J.}\ }\textbf {\bibinfo {volume}
  {A51}},\ \bibinfo {pages} {57} (\bibinfo {year} {2015})}\BibitemShut
  {NoStop}%
\bibitem [{\citenamefont {Sanchis-Alepuz}\ \emph {et~al.}(2014)\citenamefont
  {Sanchis-Alepuz}, \citenamefont {Fischer},\ and\ \citenamefont
  {Kubrak}}]{Sanchis-Alepuz:2014wea}%
  \BibitemOpen
  \bibfield  {author} {\bibinfo {author} {\bibfnamefont {H.}~\bibnamefont
  {Sanchis-Alepuz}}, \bibinfo {author} {\bibfnamefont {C.~S.}\ \bibnamefont
  {Fischer}}, \ and\ \bibinfo {author} {\bibfnamefont {S.}~\bibnamefont
  {Kubrak}},\ }\href {\doibase 10.1016/j.physletb.2014.04.031} {\bibfield
  {journal} {\bibinfo  {journal} {Phys. Lett. B}\ }\textbf {\bibinfo {volume}
  {733}},\ \bibinfo {pages} {151} (\bibinfo {year} {2014})}\BibitemShut
  {NoStop}%
\bibitem [{\citenamefont {Chang}\ and\ \citenamefont
  {Roberts}(2009)}]{Chang:2009zb}%
  \BibitemOpen
  \bibfield  {author} {\bibinfo {author} {\bibfnamefont {L.}~\bibnamefont
  {Chang}}\ and\ \bibinfo {author} {\bibfnamefont {C.~D.}\ \bibnamefont
  {Roberts}},\ }\href {\doibase 10.1103/PhysRevLett.103.081601} {\bibfield
  {journal} {\bibinfo  {journal} {Phys. Rev. Lett.}\ }\textbf {\bibinfo
  {volume} {103}},\ \bibinfo {pages} {081601} (\bibinfo {year}
  {2009})}\BibitemShut {NoStop}%
\bibitem [{\citenamefont {Heupel}\ \emph {et~al.}(2014)\citenamefont {Heupel},
  \citenamefont {Goecke},\ and\ \citenamefont {Fischer}}]{Heupel:2014ina}%
  \BibitemOpen
  \bibfield  {author} {\bibinfo {author} {\bibfnamefont {W.}~\bibnamefont
  {Heupel}}, \bibinfo {author} {\bibfnamefont {T.}~\bibnamefont {Goecke}}, \
  and\ \bibinfo {author} {\bibfnamefont {C.~S.}\ \bibnamefont {Fischer}},\
  }\href {\doibase 10.1140/epja/i2014-14085-x} {\bibfield  {journal} {\bibinfo
  {journal} {Eur. Phys. J. A}\ }\textbf {\bibinfo {volume} {50}},\ \bibinfo
  {pages} {85} (\bibinfo {year} {2014})}\BibitemShut {NoStop}%
\bibitem [{\citenamefont {Sanchis-Alepuz}\ and\ \citenamefont
  {Williams}(2015{\natexlab{b}})}]{Sanchis-Alepuz:2015qra}%
  \BibitemOpen
  \bibfield  {author} {\bibinfo {author} {\bibfnamefont {H.}~\bibnamefont
  {Sanchis-Alepuz}}\ and\ \bibinfo {author} {\bibfnamefont {R.}~\bibnamefont
  {Williams}},\ }\href {\doibase 10.1016/j.physletb.2015.08.067} {\bibfield
  {journal} {\bibinfo  {journal} {Phys. Lett. B}\ }\textbf {\bibinfo {volume}
  {749}},\ \bibinfo {pages} {592} (\bibinfo {year}
  {2015}{\natexlab{b}})}\BibitemShut {NoStop}%
\bibitem [{\citenamefont {Williams}\ \emph {et~al.}(2015)\citenamefont
  {Williams}, \citenamefont {Fischer},\ and\ \citenamefont
  {Heupel}}]{Williams:2015cvx}%
  \BibitemOpen
  \bibfield  {author} {\bibinfo {author} {\bibfnamefont {R.}~\bibnamefont
  {Williams}}, \bibinfo {author} {\bibfnamefont {C.~S.}\ \bibnamefont
  {Fischer}}, \ and\ \bibinfo {author} {\bibfnamefont {W.}~\bibnamefont
  {Heupel}},\ }\href@noop {} {\ }\Eprint {http://arxiv.org/abs/1512.00455}
  {1512.00455 [hep-ph]} \BibitemShut {NoStop}%
\bibitem [{\citenamefont {Fu}\ and\ \citenamefont {Wang}(2016)}]{Fu:2015tdu}%
  \BibitemOpen
  \bibfield  {author} {\bibinfo {author} {\bibfnamefont {H.-F.}\ \bibnamefont
  {Fu}}\ and\ \bibinfo {author} {\bibfnamefont {Q.}~\bibnamefont {Wang}},\
  }\href {\doibase 10.1103/PhysRevD.93.014013} {\bibfield  {journal} {\bibinfo
  {journal} {Phys. Rev. D}\ }\textbf {\bibinfo {volume} {93}},\ \bibinfo
  {pages} {014013} (\bibinfo {year} {2016})}\BibitemShut {NoStop}%
\bibitem [{\citenamefont {Binosi}\ \emph {et~al.}(2016)\citenamefont {Binosi},
  \citenamefont {Chang}, \citenamefont {Papavassiliou}, \citenamefont {Qin},\
  and\ \citenamefont {Roberts}}]{Binosi:2016rxz}%
  \BibitemOpen
  \bibfield  {author} {\bibinfo {author} {\bibfnamefont {D.}~\bibnamefont
  {Binosi}}, \bibinfo {author} {\bibfnamefont {L.}~\bibnamefont {Chang}},
  \bibinfo {author} {\bibfnamefont {J.}~\bibnamefont {Papavassiliou}}, \bibinfo
  {author} {\bibfnamefont {S.-X.}\ \bibnamefont {Qin}}, \ and\ \bibinfo
  {author} {\bibfnamefont {C.~D.}\ \bibnamefont {Roberts}},\ }\href@noop {} {\
  }\Eprint {http://arxiv.org/abs/1601.05441} {1601.05441 [nucl-th]}
  \BibitemShut {NoStop}%
\bibitem [{\citenamefont {Qin}(2016)}]{Qin:2016fbu}%
  \BibitemOpen
  \bibfield  {author} {\bibinfo {author} {\bibfnamefont {S.-x.}\ \bibnamefont
  {Qin}},\ }\href
  {http://inspirehep.net/record/1415121/files/arXiv:1601.03134.pdf} {\ }\Eprint
  {http://arxiv.org/abs/1601.03134} {1601.03134 [nucl-th]} \BibitemShut
  {NoStop}%
\bibitem [{\citenamefont {Horvatic}\ \emph {et~al.}(2007)\citenamefont
  {Horvatic}, \citenamefont {Klabucar},\ and\ \citenamefont
  {Radzhabov}}]{Horvatic:2007qs}%
  \BibitemOpen
  \bibfield  {author} {\bibinfo {author} {\bibfnamefont {D.}~\bibnamefont
  {Horvatic}}, \bibinfo {author} {\bibfnamefont {D.}~\bibnamefont {Klabucar}},
  \ and\ \bibinfo {author} {\bibfnamefont {A.~E.}\ \bibnamefont {Radzhabov}},\
  }\href {\doibase 10.1103/PhysRevD.76.096009} {\bibfield  {journal} {\bibinfo
  {journal} {Phys. Rev. D}\ }\textbf {\bibinfo {volume} {76}},\ \bibinfo
  {pages} {096009} (\bibinfo {year} {2007})}\BibitemShut {NoStop}%
\bibitem [{\citenamefont {Horvatic}\ \emph {et~al.}(2008)\citenamefont
  {Horvatic}, \citenamefont {Blaschke}, \citenamefont {Klabucar},\ and\
  \citenamefont {Radzhabov}}]{Horvatic:2007wu}%
  \BibitemOpen
  \bibfield  {author} {\bibinfo {author} {\bibfnamefont {D.}~\bibnamefont
  {Horvatic}}, \bibinfo {author} {\bibfnamefont {D.}~\bibnamefont {Blaschke}},
  \bibinfo {author} {\bibfnamefont {D.}~\bibnamefont {Klabucar}}, \ and\
  \bibinfo {author} {\bibfnamefont {A.~E.}\ \bibnamefont {Radzhabov}},\ }\href
  {\doibase 10.1134/S1063779608070095} {\bibfield  {journal} {\bibinfo
  {journal} {Phys. Part. Nucl.}\ }\textbf {\bibinfo {volume} {39}},\ \bibinfo
  {pages} {1033} (\bibinfo {year} {2008})}\BibitemShut {NoStop}%
\bibitem [{\citenamefont {Horvatic}\ \emph {et~al.}(2011)\citenamefont
  {Horvatic}, \citenamefont {Blaschke}, \citenamefont {Klabucar},\ and\
  \citenamefont {Kaczmarek}}]{Horvatic:2010md}%
  \BibitemOpen
  \bibfield  {author} {\bibinfo {author} {\bibfnamefont {D.}~\bibnamefont
  {Horvatic}}, \bibinfo {author} {\bibfnamefont {D.}~\bibnamefont {Blaschke}},
  \bibinfo {author} {\bibfnamefont {D.}~\bibnamefont {Klabucar}}, \ and\
  \bibinfo {author} {\bibfnamefont {O.}~\bibnamefont {Kaczmarek}},\ }\href
  {\doibase 10.1103/PhysRevD.84.016005} {\bibfield  {journal} {\bibinfo
  {journal} {Phys. Rev. D}\ }\textbf {\bibinfo {volume} {84}},\ \bibinfo
  {pages} {016005} (\bibinfo {year} {2011})}\BibitemShut {NoStop}%
\bibitem [{\citenamefont {Roberts}\ \emph {et~al.}(2011)\citenamefont
  {Roberts}, \citenamefont {Bashir}, \citenamefont {Gutierrez-Guerrero},
  \citenamefont {Roberts},\ and\ \citenamefont {Wilson}}]{Roberts:2011wy}%
  \BibitemOpen
  \bibfield  {author} {\bibinfo {author} {\bibfnamefont {H.~L.~L.}\
  \bibnamefont {Roberts}}, \bibinfo {author} {\bibfnamefont {A.}~\bibnamefont
  {Bashir}}, \bibinfo {author} {\bibfnamefont {L.~X.}\ \bibnamefont
  {Gutierrez-Guerrero}}, \bibinfo {author} {\bibfnamefont {C.~D.}\ \bibnamefont
  {Roberts}}, \ and\ \bibinfo {author} {\bibfnamefont {D.~J.}\ \bibnamefont
  {Wilson}},\ }\href {\doibase 10.1103/PhysRevC.83.065206} {\bibfield
  {journal} {\bibinfo  {journal} {Phys. Rev.}\ }\textbf {\bibinfo {volume}
  {C83}},\ \bibinfo {pages} {065206} (\bibinfo {year} {2011})}\BibitemShut
  {NoStop}%
\bibitem [{\citenamefont {Gutierrez-Guerrero}\ \emph
  {et~al.}(2010)\citenamefont {Gutierrez-Guerrero}, \citenamefont {Bashir},
  \citenamefont {Cloet},\ and\ \citenamefont
  {Roberts}}]{GutierrezGuerrero:2010md}%
  \BibitemOpen
  \bibfield  {author} {\bibinfo {author} {\bibfnamefont {L.~X.}\ \bibnamefont
  {Gutierrez-Guerrero}}, \bibinfo {author} {\bibfnamefont {A.}~\bibnamefont
  {Bashir}}, \bibinfo {author} {\bibfnamefont {I.~C.}\ \bibnamefont {Cloet}}, \
  and\ \bibinfo {author} {\bibfnamefont {C.~D.}\ \bibnamefont {Roberts}},\
  }\href {\doibase 10.1103/PhysRevC.81.065202} {\bibfield  {journal} {\bibinfo
  {journal} {Phys. Rev. C}\ }\textbf {\bibinfo {volume} {81}},\ \bibinfo
  {pages} {065202} (\bibinfo {year} {2010})}\BibitemShut {NoStop}%
\bibitem [{\citenamefont {Bedolla}\ \emph {et~al.}(2015)\citenamefont
  {Bedolla}, \citenamefont {Cobos-Martínez},\ and\ \citenamefont
  {Bashir}}]{Bedolla:2015mpa}%
  \BibitemOpen
  \bibfield  {author} {\bibinfo {author} {\bibfnamefont {M.~A.}\ \bibnamefont
  {Bedolla}}, \bibinfo {author} {\bibfnamefont {J.}~\bibnamefont
  {Cobos-Martínez}}, \ and\ \bibinfo {author} {\bibfnamefont {A.}~\bibnamefont
  {Bashir}},\ }\href {\doibase 10.1103/PhysRevD.92.054031} {\bibfield
  {journal} {\bibinfo  {journal} {Phys. Rev. D}\ }\textbf {\bibinfo {volume}
  {92}},\ \bibinfo {pages} {054031} (\bibinfo {year} {2015})}\BibitemShut
  {NoStop}%
\bibitem [{\citenamefont {Segovia}(2016)}]{Segovia:2016iaf}%
  \BibitemOpen
  \bibfield  {author} {\bibinfo {author} {\bibfnamefont {J.}~\bibnamefont
  {Segovia}},\ }\href
  {http://inspirehep.net/record/1420548/files/arXiv:1602.02768.pdf} {\ }\Eprint
  {http://arxiv.org/abs/1602.02768} {1602.02768 [nucl-th]} \BibitemShut
  {NoStop}%
\bibitem [{\citenamefont {Munczek}\ and\ \citenamefont
  {Nemirovsky}(1983)}]{Munczek:1983dx}%
  \BibitemOpen
  \bibfield  {author} {\bibinfo {author} {\bibfnamefont {H.~J.}\ \bibnamefont
  {Munczek}}\ and\ \bibinfo {author} {\bibfnamefont {A.~M.}\ \bibnamefont
  {Nemirovsky}},\ }\href {\doibase 10.1103/PhysRevD.28.181} {\bibfield
  {journal} {\bibinfo  {journal} {Phys. Rev. D}\ }\textbf {\bibinfo {volume}
  {28}},\ \bibinfo {pages} {181} (\bibinfo {year} {1983})}\BibitemShut
  {NoStop}%
\bibitem [{\citenamefont {Bender}\ \emph {et~al.}(2002)\citenamefont {Bender},
  \citenamefont {Detmold}, \citenamefont {Roberts},\ and\ \citenamefont
  {Thomas}}]{Bender:2002as}%
  \BibitemOpen
  \bibfield  {author} {\bibinfo {author} {\bibfnamefont {A.}~\bibnamefont
  {Bender}}, \bibinfo {author} {\bibfnamefont {W.}~\bibnamefont {Detmold}},
  \bibinfo {author} {\bibfnamefont {C.~D.}\ \bibnamefont {Roberts}}, \ and\
  \bibinfo {author} {\bibfnamefont {A.~W.}\ \bibnamefont {Thomas}},\ }\href
  {\doibase 10.1103/PhysRevC.65.065203} {\bibfield  {journal} {\bibinfo
  {journal} {Phys. Rev. C}\ }\textbf {\bibinfo {volume} {65}},\ \bibinfo
  {pages} {065203} (\bibinfo {year} {2002})}\BibitemShut {NoStop}%
\bibitem [{\citenamefont {Krassnigg}\ and\ \citenamefont
  {Roberts}(2004)}]{Krassnigg:2003dr}%
  \BibitemOpen
  \bibfield  {author} {\bibinfo {author} {\bibfnamefont {A.}~\bibnamefont
  {Krassnigg}}\ and\ \bibinfo {author} {\bibfnamefont {C.~D.}\ \bibnamefont
  {Roberts}},\ }\href {\doibase 10.1016/S0375-9474(04)00291-X} {\bibfield
  {journal} {\bibinfo  {journal} {Nucl. Phys. A}\ }\textbf {\bibinfo {volume}
  {737}},\ \bibinfo {pages} {7} (\bibinfo {year} {2004})}\BibitemShut {NoStop}%
\bibitem [{\citenamefont {Bhagwat}\ \emph {et~al.}(2004)\citenamefont
  {Bhagwat}, \citenamefont {Holl}, \citenamefont {Krassnigg}, \citenamefont
  {Roberts},\ and\ \citenamefont {Tandy}}]{Bhagwat:2004hn}%
  \BibitemOpen
  \bibfield  {author} {\bibinfo {author} {\bibfnamefont {M.~S.}\ \bibnamefont
  {Bhagwat}}, \bibinfo {author} {\bibfnamefont {A.}~\bibnamefont {Holl}},
  \bibinfo {author} {\bibfnamefont {A.}~\bibnamefont {Krassnigg}}, \bibinfo
  {author} {\bibfnamefont {C.~D.}\ \bibnamefont {Roberts}}, \ and\ \bibinfo
  {author} {\bibfnamefont {P.~C.}\ \bibnamefont {Tandy}},\ }\href {\doibase
  10.1103/PhysRevC.70.035205} {\bibfield  {journal} {\bibinfo  {journal} {Phys.
  Rev. C}\ }\textbf {\bibinfo {volume} {70}},\ \bibinfo {pages} {035205}
  (\bibinfo {year} {2004})}\BibitemShut {NoStop}%
\bibitem [{\citenamefont {Holl}\ \emph
  {et~al.}(2005{\natexlab{b}})\citenamefont {Holl}, \citenamefont {Krassnigg},\
  and\ \citenamefont {Roberts}}]{Holl:2004qn}%
  \BibitemOpen
  \bibfield  {author} {\bibinfo {author} {\bibfnamefont {A.}~\bibnamefont
  {Holl}}, \bibinfo {author} {\bibfnamefont {A.}~\bibnamefont {Krassnigg}}, \
  and\ \bibinfo {author} {\bibfnamefont {C.~D.}\ \bibnamefont {Roberts}},\
  }\href {\doibase 10.1016/j.nuclphysbps.2004.12.009} {\bibfield  {journal}
  {\bibinfo  {journal} {Nucl. Phys. Proc. Suppl.}\ }\textbf {\bibinfo {volume}
  {141}},\ \bibinfo {pages} {47} (\bibinfo {year}
  {2005}{\natexlab{b}})}\BibitemShut {NoStop}%
\bibitem [{\citenamefont {Matevosyan}\ \emph
  {et~al.}(2007{\natexlab{a}})\citenamefont {Matevosyan}, \citenamefont
  {Thomas},\ and\ \citenamefont {Tandy}}]{Matevosyan:2006bk}%
  \BibitemOpen
  \bibfield  {author} {\bibinfo {author} {\bibfnamefont {H.~H.}\ \bibnamefont
  {Matevosyan}}, \bibinfo {author} {\bibfnamefont {A.~W.}\ \bibnamefont
  {Thomas}}, \ and\ \bibinfo {author} {\bibfnamefont {P.~C.}\ \bibnamefont
  {Tandy}},\ }\href {\doibase 10.1103/PhysRevC.75.045201} {\bibfield  {journal}
  {\bibinfo  {journal} {Phys. Rev. C}\ }\textbf {\bibinfo {volume} {75}},\
  \bibinfo {pages} {045201} (\bibinfo {year} {2007}{\natexlab{a}})}\BibitemShut
  {NoStop}%
\bibitem [{\citenamefont {Matevosyan}\ \emph
  {et~al.}(2007{\natexlab{b}})\citenamefont {Matevosyan}, \citenamefont
  {Thomas},\ and\ \citenamefont {Tandy}}]{Matevosyan:2007cx}%
  \BibitemOpen
  \bibfield  {author} {\bibinfo {author} {\bibfnamefont {H.~H.}\ \bibnamefont
  {Matevosyan}}, \bibinfo {author} {\bibfnamefont {A.~W.}\ \bibnamefont
  {Thomas}}, \ and\ \bibinfo {author} {\bibfnamefont {P.~C.}\ \bibnamefont
  {Tandy}},\ }\href {\doibase 10.1088/0954-3899/34/10/005} {\bibfield
  {journal} {\bibinfo  {journal} {J. Phys. G}\ }\textbf {\bibinfo {volume}
  {34}},\ \bibinfo {pages} {2153} (\bibinfo {year}
  {2007}{\natexlab{b}})}\BibitemShut {NoStop}%
\bibitem [{\citenamefont {Jinno}\ \emph {et~al.}(2015)\citenamefont {Jinno},
  \citenamefont {Kitahara},\ and\ \citenamefont {Mishima}}]{Jinno:2015sea}%
  \BibitemOpen
  \bibfield  {author} {\bibinfo {author} {\bibfnamefont {R.}~\bibnamefont
  {Jinno}}, \bibinfo {author} {\bibfnamefont {T.}~\bibnamefont {Kitahara}}, \
  and\ \bibinfo {author} {\bibfnamefont {G.}~\bibnamefont {Mishima}},\ }\href
  {\doibase 10.1103/PhysRevD.91.076011} {\bibfield  {journal} {\bibinfo
  {journal} {Phys. Rev. D}\ }\textbf {\bibinfo {volume} {91}},\ \bibinfo
  {pages} {076011} (\bibinfo {year} {2015})}\BibitemShut {NoStop}%
\bibitem [{\citenamefont {Gomez-Rocha}\ \emph
  {et~al.}(2015{\natexlab{a}})\citenamefont {Gomez-Rocha}, \citenamefont
  {Hilger},\ and\ \citenamefont {Krassnigg}}]{Gomez-Rocha:2014vsa}%
  \BibitemOpen
  \bibfield  {author} {\bibinfo {author} {\bibfnamefont {M.}~\bibnamefont
  {Gomez-Rocha}}, \bibinfo {author} {\bibfnamefont {T.}~\bibnamefont {Hilger}},
  \ and\ \bibinfo {author} {\bibfnamefont {A.}~\bibnamefont {Krassnigg}},\
  }\href {\doibase 10.1007/s00601-014-0938-8} {\bibfield  {journal} {\bibinfo
  {journal} {Few Body Syst.}\ }\textbf {\bibinfo {volume} {56}},\ \bibinfo
  {pages} {475} (\bibinfo {year} {2015}{\natexlab{a}})}\BibitemShut {NoStop}%
\bibitem [{\citenamefont {Gomez-Rocha}\ \emph
  {et~al.}(2015{\natexlab{b}})\citenamefont {Gomez-Rocha}, \citenamefont
  {Hilger},\ and\ \citenamefont {Krassnigg}}]{Gomez-Rocha:2015qga}%
  \BibitemOpen
  \bibfield  {author} {\bibinfo {author} {\bibfnamefont {M.}~\bibnamefont
  {Gomez-Rocha}}, \bibinfo {author} {\bibfnamefont {T.}~\bibnamefont {Hilger}},
  \ and\ \bibinfo {author} {\bibfnamefont {A.}~\bibnamefont {Krassnigg}},\
  }\href {\doibase 10.1103/PhysRevD.92.054030} {\bibfield  {journal} {\bibinfo
  {journal} {Phys. Rev. D}\ }\textbf {\bibinfo {volume} {92}},\ \bibinfo
  {pages} {054030} (\bibinfo {year} {2015}{\natexlab{b}})}\BibitemShut
  {NoStop}%
\bibitem [{\citenamefont {Nguyen}\ \emph {et~al.}(2009)\citenamefont {Nguyen},
  \citenamefont {Souchlas},\ and\ \citenamefont {Tandy}}]{Nguyen:2009if}%
  \BibitemOpen
  \bibfield  {author} {\bibinfo {author} {\bibfnamefont {T.}~\bibnamefont
  {Nguyen}}, \bibinfo {author} {\bibfnamefont {N.~A.}\ \bibnamefont
  {Souchlas}}, \ and\ \bibinfo {author} {\bibfnamefont {P.~C.}\ \bibnamefont
  {Tandy}},\ }\href {\doibase 10.1063/1.3131570} {\bibfield  {journal}
  {\bibinfo  {journal} {AIP Conf. Proc.}\ }\textbf {\bibinfo {volume} {1116}},\
  \bibinfo {pages} {327} (\bibinfo {year} {2009})}\BibitemShut {NoStop}%
\bibitem [{\citenamefont {Souchlas}\ and\ \citenamefont
  {Stratakis}(2010)}]{Souchlas:2010zz}%
  \BibitemOpen
  \bibfield  {author} {\bibinfo {author} {\bibfnamefont {N.}~\bibnamefont
  {Souchlas}}\ and\ \bibinfo {author} {\bibfnamefont {D.}~\bibnamefont
  {Stratakis}},\ }\href {\doibase 10.1103/PhysRevD.81.114019} {\bibfield
  {journal} {\bibinfo  {journal} {Phys. Rev. D}\ }\textbf {\bibinfo {volume}
  {81}},\ \bibinfo {pages} {114019} (\bibinfo {year} {2010})}\BibitemShut
  {NoStop}%
\bibitem [{\citenamefont {Nguyen}\ \emph {et~al.}(2011)\citenamefont {Nguyen},
  \citenamefont {Souchlas},\ and\ \citenamefont {Tandy}}]{Nguyen:2010yh}%
  \BibitemOpen
  \bibfield  {author} {\bibinfo {author} {\bibfnamefont {T.}~\bibnamefont
  {Nguyen}}, \bibinfo {author} {\bibfnamefont {N.~A.}\ \bibnamefont
  {Souchlas}}, \ and\ \bibinfo {author} {\bibfnamefont {P.~C.}\ \bibnamefont
  {Tandy}},\ }\href {\doibase 10.1063/1.3622693} {\bibfield  {journal}
  {\bibinfo  {journal} {AIP Conf.Proc.}\ }\textbf {\bibinfo {volume} {1361}},\
  \bibinfo {pages} {142} (\bibinfo {year} {2011})}\BibitemShut {NoStop}%
\bibitem [{\citenamefont {Ivanov}\ \emph
  {et~al.}(1998{\natexlab{a}})\citenamefont {Ivanov}, \citenamefont
  {Kalinovsky}, \citenamefont {Maris},\ and\ \citenamefont
  {Roberts}}]{Ivanov:1997iu}%
  \BibitemOpen
  \bibfield  {author} {\bibinfo {author} {\bibfnamefont {M.~A.}\ \bibnamefont
  {Ivanov}}, \bibinfo {author} {\bibfnamefont {Y.~L.}\ \bibnamefont
  {Kalinovsky}}, \bibinfo {author} {\bibfnamefont {P.}~\bibnamefont {Maris}}, \
  and\ \bibinfo {author} {\bibfnamefont {C.~D.}\ \bibnamefont {Roberts}},\
  }\href {\doibase 10.1103/PhysRevC.57.1991} {\bibfield  {journal} {\bibinfo
  {journal} {Phys. Rev. C}\ }\textbf {\bibinfo {volume} {57}},\ \bibinfo
  {pages} {1991} (\bibinfo {year} {1998}{\natexlab{a}})}\BibitemShut {NoStop}%
\bibitem [{\citenamefont {Ivanov}\ \emph
  {et~al.}(1998{\natexlab{b}})\citenamefont {Ivanov}, \citenamefont
  {Kalinovsky}, \citenamefont {Maris},\ and\ \citenamefont
  {Roberts}}]{Ivanov:1997yg}%
  \BibitemOpen
  \bibfield  {author} {\bibinfo {author} {\bibfnamefont {M.~A.}\ \bibnamefont
  {Ivanov}}, \bibinfo {author} {\bibfnamefont {Y.~L.}\ \bibnamefont
  {Kalinovsky}}, \bibinfo {author} {\bibfnamefont {P.}~\bibnamefont {Maris}}, \
  and\ \bibinfo {author} {\bibfnamefont {C.~D.}\ \bibnamefont {Roberts}},\
  }\href {\doibase 10.1016/S0370-2693(97)01323-3} {\bibfield  {journal}
  {\bibinfo  {journal} {Phys. Lett. B}\ }\textbf {\bibinfo {volume} {416}},\
  \bibinfo {pages} {29} (\bibinfo {year} {1998}{\natexlab{b}})}\BibitemShut
  {NoStop}%
\bibitem [{\citenamefont {Ivanov}\ \emph {et~al.}(1999)\citenamefont {Ivanov},
  \citenamefont {Kalinovsky},\ and\ \citenamefont {Roberts}}]{Ivanov:1998ms}%
  \BibitemOpen
  \bibfield  {author} {\bibinfo {author} {\bibfnamefont {M.~A.}\ \bibnamefont
  {Ivanov}}, \bibinfo {author} {\bibfnamefont {Y.~L.}\ \bibnamefont
  {Kalinovsky}}, \ and\ \bibinfo {author} {\bibfnamefont {C.~D.}\ \bibnamefont
  {Roberts}},\ }\href {\doibase 10.1103/PhysRevD.60.034018} {\bibfield
  {journal} {\bibinfo  {journal} {Phys. Rev. D}\ }\textbf {\bibinfo {volume}
  {60}},\ \bibinfo {pages} {034018} (\bibinfo {year} {1999})}\BibitemShut
  {NoStop}%
\bibitem [{\citenamefont {Blaschke}\ \emph {et~al.}(2000)\citenamefont
  {Blaschke}, \citenamefont {Burau}, \citenamefont {Ivanov}, \citenamefont
  {Kalinovsky},\ and\ \citenamefont {Tandy}}]{Blaschke:2000zm}%
  \BibitemOpen
  \bibfield  {author} {\bibinfo {author} {\bibfnamefont {D.~B.}\ \bibnamefont
  {Blaschke}}, \bibinfo {author} {\bibfnamefont {G.~R.~G.}\ \bibnamefont
  {Burau}}, \bibinfo {author} {\bibfnamefont {M.~A.}\ \bibnamefont {Ivanov}},
  \bibinfo {author} {\bibfnamefont {Y.~L.}\ \bibnamefont {Kalinovsky}}, \ and\
  \bibinfo {author} {\bibfnamefont {P.~C.}\ \bibnamefont {Tandy}},\ }\href@noop
  {} {\ }\Eprint {http://arxiv.org/abs/hep-ph/0002047} {hep-ph/0002047}
  \BibitemShut {NoStop}%
\bibitem [{\citenamefont {Bhagwat}\ \emph
  {et~al.}(2007{\natexlab{b}})\citenamefont {Bhagwat}, \citenamefont
  {Krassnigg}, \citenamefont {Maris},\ and\ \citenamefont
  {Roberts}}]{Bhagwat:2006xi}%
  \BibitemOpen
  \bibfield  {author} {\bibinfo {author} {\bibfnamefont {M.~S.}\ \bibnamefont
  {Bhagwat}}, \bibinfo {author} {\bibfnamefont {A.}~\bibnamefont {Krassnigg}},
  \bibinfo {author} {\bibfnamefont {P.}~\bibnamefont {Maris}}, \ and\ \bibinfo
  {author} {\bibfnamefont {C.~D.}\ \bibnamefont {Roberts}},\ }\href {\doibase
  10.1140/epja/i2006-10271-9} {\bibfield  {journal} {\bibinfo  {journal} {Eur.
  Phys. J.}\ }\textbf {\bibinfo {volume} {A31}},\ \bibinfo {pages} {630}
  (\bibinfo {year} {2007}{\natexlab{b}})}\BibitemShut {NoStop}%
\bibitem [{\citenamefont {Ivanov}\ \emph {et~al.}(2007)\citenamefont {Ivanov},
  \citenamefont {Korner}, \citenamefont {Kovalenko},\ and\ \citenamefont
  {Roberts}}]{Ivanov:2007cw}%
  \BibitemOpen
  \bibfield  {author} {\bibinfo {author} {\bibfnamefont {M.~A.}\ \bibnamefont
  {Ivanov}}, \bibinfo {author} {\bibfnamefont {J.~G.}\ \bibnamefont {Korner}},
  \bibinfo {author} {\bibfnamefont {S.~G.}\ \bibnamefont {Kovalenko}}, \ and\
  \bibinfo {author} {\bibfnamefont {C.~D.}\ \bibnamefont {Roberts}},\ }\href
  {\doibase 10.1103/PhysRevD.76.034018} {\bibfield  {journal} {\bibinfo
  {journal} {Phys. Rev. D}\ }\textbf {\bibinfo {volume} {76}},\ \bibinfo
  {pages} {034018} (\bibinfo {year} {2007})}\BibitemShut {NoStop}%
\bibitem [{\citenamefont {El-Bennich}\ \emph {et~al.}(2010)\citenamefont
  {El-Bennich}, \citenamefont {Ivanov},\ and\ \citenamefont
  {Roberts}}]{ElBennich:2009vx}%
  \BibitemOpen
  \bibfield  {author} {\bibinfo {author} {\bibfnamefont {B.}~\bibnamefont
  {El-Bennich}}, \bibinfo {author} {\bibfnamefont {M.~A.}\ \bibnamefont
  {Ivanov}}, \ and\ \bibinfo {author} {\bibfnamefont {C.~D.}\ \bibnamefont
  {Roberts}},\ }\href {\doibase 10.1016/j.nuclphysbps.2010.02.026} {\bibfield
  {journal} {\bibinfo  {journal} {Nucl.Phys.Proc.Suppl.}\ }\textbf {\bibinfo
  {volume} {199}},\ \bibinfo {pages} {184} (\bibinfo {year}
  {2010})}\BibitemShut {NoStop}%
\bibitem [{\citenamefont {El-Bennich}\ \emph {et~al.}(2011)\citenamefont
  {El-Bennich}, \citenamefont {Ivanov},\ and\ \citenamefont
  {Roberts}}]{ElBennich:2010ha}%
  \BibitemOpen
  \bibfield  {author} {\bibinfo {author} {\bibfnamefont {B.}~\bibnamefont
  {El-Bennich}}, \bibinfo {author} {\bibfnamefont {M.~A.}\ \bibnamefont
  {Ivanov}}, \ and\ \bibinfo {author} {\bibfnamefont {C.~D.}\ \bibnamefont
  {Roberts}},\ }\href {\doibase 10.1103/PhysRevC.83.025205} {\bibfield
  {journal} {\bibinfo  {journal} {Phys. Rev. C}\ }\textbf {\bibinfo {volume}
  {83}},\ \bibinfo {pages} {025205} (\bibinfo {year} {2011})}\BibitemShut
  {NoStop}%
\bibitem [{\citenamefont {Neubert}(1994)}]{Neubert:1993mb}%
  \BibitemOpen
  \bibfield  {author} {\bibinfo {author} {\bibfnamefont {M.}~\bibnamefont
  {Neubert}},\ }\href {\doibase 10.1016/0370-1573(94)90091-4} {\bibfield
  {journal} {\bibinfo  {journal} {Phys.Rept.}\ }\textbf {\bibinfo {volume}
  {245}},\ \bibinfo {pages} {259} (\bibinfo {year} {1994})}\BibitemShut
  {NoStop}%
\bibitem [{\citenamefont {Keister}\ and\ \citenamefont
  {Polyzou}(1991)}]{Keister:1991sb}%
  \BibitemOpen
  \bibfield  {author} {\bibinfo {author} {\bibfnamefont {B.~D.}\ \bibnamefont
  {Keister}}\ and\ \bibinfo {author} {\bibfnamefont {W.~N.}\ \bibnamefont
  {Polyzou}},\ }\href@noop {} {\bibfield  {journal} {\bibinfo  {journal} {Adv.
  Nucl. Phys.}\ }\textbf {\bibinfo {volume} {20}},\ \bibinfo {pages} {225}
  (\bibinfo {year} {1991})}\BibitemShut {NoStop}%
\bibitem [{\citenamefont {Krassnigg}\ \emph {et~al.}(2003)\citenamefont
  {Krassnigg}, \citenamefont {Schweiger},\ and\ \citenamefont
  {Klink}}]{Krassnigg:2003gh}%
  \BibitemOpen
  \bibfield  {author} {\bibinfo {author} {\bibfnamefont {A.}~\bibnamefont
  {Krassnigg}}, \bibinfo {author} {\bibfnamefont {W.}~\bibnamefont
  {Schweiger}}, \ and\ \bibinfo {author} {\bibfnamefont {W.~H.}\ \bibnamefont
  {Klink}},\ }\href {\doibase 10.1103/PhysRevC.67.064003} {\bibfield  {journal}
  {\bibinfo  {journal} {Phys. Rev. C}\ }\textbf {\bibinfo {volume} {67}},\
  \bibinfo {pages} {064003} (\bibinfo {year} {2003})}\BibitemShut {NoStop}%
\bibitem [{\citenamefont {Krassnigg}(2005)}]{Krassnigg:2004sp}%
  \BibitemOpen
  \bibfield  {author} {\bibinfo {author} {\bibfnamefont {A.}~\bibnamefont
  {Krassnigg}},\ }\href {\doibase 10.1103/PhysRevC.72.028201} {\bibfield
  {journal} {\bibinfo  {journal} {Phys. Rev. C}\ }\textbf {\bibinfo {volume}
  {72}},\ \bibinfo {pages} {028201} (\bibinfo {year} {2005})}\BibitemShut
  {NoStop}%
\bibitem [{\citenamefont {Polyzou}(2015)}]{Polyzou:2015rra}%
  \BibitemOpen
  \bibfield  {author} {\bibinfo {author} {\bibfnamefont {W.}~\bibnamefont
  {Polyzou}},\ }\href
  {http://inspirehep.net/record/1391667/files/arXiv:1509.00928.pdf} {\ }\Eprint
  {http://arxiv.org/abs/1509.00928} {1509.00928 [nucl-th]} \BibitemShut
  {NoStop}%
\bibitem [{\citenamefont {Blankenbecler}\ and\ \citenamefont
  {Sugar}(1966)}]{Blankenbecler:1965gx}%
  \BibitemOpen
  \bibfield  {author} {\bibinfo {author} {\bibfnamefont {R.}~\bibnamefont
  {Blankenbecler}}\ and\ \bibinfo {author} {\bibfnamefont {R.}~\bibnamefont
  {Sugar}},\ }\href {\doibase 10.1103/PhysRev.142.1051} {\bibfield  {journal}
  {\bibinfo  {journal} {Phys. Rev.}\ }\textbf {\bibinfo {volume} {142}},\
  \bibinfo {pages} {1051} (\bibinfo {year} {1966})}\BibitemShut {NoStop}%
\bibitem [{\citenamefont {Gross}(1969)}]{Gross:1969rv}%
  \BibitemOpen
  \bibfield  {author} {\bibinfo {author} {\bibfnamefont {F.}~\bibnamefont
  {Gross}},\ }\href {\doibase 10.1103/PhysRev.186.1448} {\bibfield  {journal}
  {\bibinfo  {journal} {Phys. Rev.}\ }\textbf {\bibinfo {volume} {186}},\
  \bibinfo {pages} {1448} (\bibinfo {year} {1969})}\BibitemShut {NoStop}%
\bibitem [{\citenamefont {Gomez-Rocha}\ and\ \citenamefont
  {Schweiger}(2012)}]{GomezRocha:2012zd}%
  \BibitemOpen
  \bibfield  {author} {\bibinfo {author} {\bibfnamefont {M.}~\bibnamefont
  {Gomez-Rocha}}\ and\ \bibinfo {author} {\bibfnamefont {W.}~\bibnamefont
  {Schweiger}},\ }\href {\doibase 10.1103/PhysRevD.86.053010} {\bibfield
  {journal} {\bibinfo  {journal} {Phys.Rev. D}\ }\textbf {\bibinfo {volume}
  {86}},\ \bibinfo {pages} {053010} (\bibinfo {year} {2012})}\BibitemShut
  {NoStop}%
\bibitem [{\citenamefont {Gomez-Rocha}(2014)}]{Gomez-Rocha:2014aoa}%
  \BibitemOpen
  \bibfield  {author} {\bibinfo {author} {\bibfnamefont {M.}~\bibnamefont
  {Gomez-Rocha}},\ }\href@noop {} {\bibfield  {journal} {\bibinfo  {journal}
  {Phys. Rev. D}\ }\textbf {\bibinfo {volume} {90}},\ \bibinfo {pages} {076003}
  (\bibinfo {year} {2014})}\BibitemShut {NoStop}%
\bibitem [{\citenamefont {Li}\ \emph {et~al.}(2015)\citenamefont {Li},
  \citenamefont {Maris}, \citenamefont {Zhao},\ and\ \citenamefont
  {Vary}}]{Li:2015zda}%
  \BibitemOpen
  \bibfield  {author} {\bibinfo {author} {\bibfnamefont {Y.}~\bibnamefont
  {Li}}, \bibinfo {author} {\bibfnamefont {P.}~\bibnamefont {Maris}}, \bibinfo
  {author} {\bibfnamefont {X.}~\bibnamefont {Zhao}}, \ and\ \bibinfo {author}
  {\bibfnamefont {J.~P.}\ \bibnamefont {Vary}},\ }\href@noop {} {\ }\Eprint
  {http://arxiv.org/abs/1509.07212} {1509.07212 [hep-ph]} \BibitemShut
  {NoStop}%
\bibitem [{\citenamefont {Leitão}\ \emph {et~al.}(2015)\citenamefont
  {Leitão}, \citenamefont {Stadler}, \citenamefont {Peña},\ and\
  \citenamefont {Biernat}}]{Leitao:2015cxa}%
  \BibitemOpen
  \bibfield  {author} {\bibinfo {author} {\bibfnamefont {S.}~\bibnamefont
  {Leitão}}, \bibinfo {author} {\bibfnamefont {A.}~\bibnamefont {Stadler}},
  \bibinfo {author} {\bibfnamefont {M.~T.}\ \bibnamefont {Peña}}, \ and\
  \bibinfo {author} {\bibfnamefont {E.~P.}\ \bibnamefont {Biernat}},\ }\href
  {http://inspirehep.net/record/1391808/files/arXiv:1509.01497.pdf} {\ }\Eprint
  {http://arxiv.org/abs/1509.01497} {1509.01497 [hep-ph]} \BibitemShut
  {NoStop}%
\bibitem [{\citenamefont {Thomas}\ \emph {et~al.}(2008)\citenamefont {Thomas},
  \citenamefont {Hilger},\ and\ \citenamefont {Kampfer}}]{Thomas:2007es}%
  \BibitemOpen
  \bibfield  {author} {\bibinfo {author} {\bibfnamefont {R.}~\bibnamefont
  {Thomas}}, \bibinfo {author} {\bibfnamefont {T.}~\bibnamefont {Hilger}}, \
  and\ \bibinfo {author} {\bibfnamefont {B.}~\bibnamefont {Kampfer}},\
  }\bibfield  {booktitle} {\emph {\bibinfo {booktitle} {{Quarks in hadrons and
  nuclei. Proceedings, International Workshop on Nuclear Physics, 29th Course,
  Erice, Italy, September 16-24, 2007}}},\ }\href {\doibase
  10.1016/j.ppnp.2007.12.028} {\bibfield  {journal} {\bibinfo  {journal} {Prog.
  Part. Nucl. Phys.}\ }\textbf {\bibinfo {volume} {61}},\ \bibinfo {pages}
  {297} (\bibinfo {year} {2008})}\BibitemShut {NoStop}%
\bibitem [{\citenamefont {Hilger}\ and\ \citenamefont
  {K{\"a}mpfer}(2009)}]{Hilger:2009kn}%
  \BibitemOpen
  \bibfield  {author} {\bibinfo {author} {\bibfnamefont {T.}~\bibnamefont
  {Hilger}}\ and\ \bibinfo {author} {\bibfnamefont {B.}~\bibnamefont
  {K{\"a}mpfer}},\ }\href@noop {} {\ }\Eprint {http://arxiv.org/abs/0904.3491}
  {0904.3491 [nucl-th]} \BibitemShut {NoStop}%
\bibitem [{\citenamefont {Hilger}\ \emph {et~al.}(2010)\citenamefont {Hilger},
  \citenamefont {Schulze},\ and\ \citenamefont {K{\"a}mpfer}}]{Hilger:2010zb}%
  \BibitemOpen
  \bibfield  {author} {\bibinfo {author} {\bibfnamefont {T.}~\bibnamefont
  {Hilger}}, \bibinfo {author} {\bibfnamefont {R.}~\bibnamefont {Schulze}}, \
  and\ \bibinfo {author} {\bibfnamefont {B.}~\bibnamefont {K{\"a}mpfer}},\
  }\href {\doibase 10.1088/0954-3899/37/9/094054} {\bibfield  {journal}
  {\bibinfo  {journal} {J.Phys. G}\ }\textbf {\bibinfo {volume} {37}},\
  \bibinfo {pages} {094054} (\bibinfo {year} {2010})}\BibitemShut {NoStop}%
\bibitem [{\citenamefont {Hilger}\ \emph
  {et~al.}(2012{\natexlab{a}})\citenamefont {Hilger}, \citenamefont {Thomas},
  \citenamefont {K{\"a}mpfer},\ and\ \citenamefont {Leupold}}]{Hilger:2010cn}%
  \BibitemOpen
  \bibfield  {author} {\bibinfo {author} {\bibfnamefont {T.}~\bibnamefont
  {Hilger}}, \bibinfo {author} {\bibfnamefont {R.}~\bibnamefont {Thomas}},
  \bibinfo {author} {\bibfnamefont {B.}~\bibnamefont {K{\"a}mpfer}}, \ and\
  \bibinfo {author} {\bibfnamefont {S.}~\bibnamefont {Leupold}},\ }\href
  {\doibase 10.1016/j.physletb.2012.02.007} {\bibfield  {journal} {\bibinfo
  {journal} {Phys.Lett. B}\ }\textbf {\bibinfo {volume} {709}},\ \bibinfo
  {pages} {200} (\bibinfo {year} {2012}{\natexlab{a}})}\BibitemShut {NoStop}%
\bibitem [{\citenamefont {Hilger}\ \emph {et~al.}(2011)\citenamefont {Hilger},
  \citenamefont {K{\"a}mpfer},\ and\ \citenamefont {Leupold}}]{Hilger:2011cq}%
  \BibitemOpen
  \bibfield  {author} {\bibinfo {author} {\bibfnamefont {T.}~\bibnamefont
  {Hilger}}, \bibinfo {author} {\bibfnamefont {B.}~\bibnamefont {K{\"a}mpfer}},
  \ and\ \bibinfo {author} {\bibfnamefont {S.}~\bibnamefont {Leupold}},\ }\href
  {\doibase 10.1103/PhysRevC.84.045202} {\bibfield  {journal} {\bibinfo
  {journal} {Phys. Rev. C}\ }\textbf {\bibinfo {volume} {84}},\ \bibinfo
  {pages} {045202} (\bibinfo {year} {2011})}\BibitemShut {NoStop}%
\bibitem [{\citenamefont {Hilger}\ \emph
  {et~al.}(2012{\natexlab{b}})\citenamefont {Hilger}, \citenamefont {Buchheim},
  \citenamefont {K{\"a}mpfer},\ and\ \citenamefont {Leupold}}]{Hilger:2012db}%
  \BibitemOpen
  \bibfield  {author} {\bibinfo {author} {\bibfnamefont {T.}~\bibnamefont
  {Hilger}}, \bibinfo {author} {\bibfnamefont {T.}~\bibnamefont {Buchheim}},
  \bibinfo {author} {\bibfnamefont {B.}~\bibnamefont {K{\"a}mpfer}}, \ and\
  \bibinfo {author} {\bibfnamefont {S.}~\bibnamefont {Leupold}},\ }\href
  {\doibase 10.1016/j.ppnp.2011.12.016} {\bibfield  {journal} {\bibinfo
  {journal} {Prog.Part.Nucl.Phys.}\ }\textbf {\bibinfo {volume} {67}},\
  \bibinfo {pages} {188} (\bibinfo {year} {2012}{\natexlab{b}})}\BibitemShut
  {NoStop}%
\bibitem [{\citenamefont {Buchheim}\ \emph
  {et~al.}(2015{\natexlab{a}})\citenamefont {Buchheim}, \citenamefont
  {Hilger},\ and\ \citenamefont {K{\"a}mpfer}}]{Buchheim:2014rpa}%
  \BibitemOpen
  \bibfield  {author} {\bibinfo {author} {\bibfnamefont {T.}~\bibnamefont
  {Buchheim}}, \bibinfo {author} {\bibfnamefont {T.}~\bibnamefont {Hilger}}, \
  and\ \bibinfo {author} {\bibfnamefont {B.}~\bibnamefont {K{\"a}mpfer}},\
  }\href {\doibase 10.1103/PhysRevC.91.015205} {\bibfield  {journal} {\bibinfo
  {journal} {Phys. Rev. C}\ }\textbf {\bibinfo {volume} {91}},\ \bibinfo
  {pages} {015205} (\bibinfo {year} {2015}{\natexlab{a}})}\BibitemShut
  {NoStop}%
\bibitem [{\citenamefont {Gubler}\ and\ \citenamefont
  {Ohtani}(2014)}]{Gubler:2014pta}%
  \BibitemOpen
  \bibfield  {author} {\bibinfo {author} {\bibfnamefont {P.}~\bibnamefont
  {Gubler}}\ and\ \bibinfo {author} {\bibfnamefont {K.}~\bibnamefont
  {Ohtani}},\ }\href {\doibase 10.1103/PhysRevD.90.094002} {\bibfield
  {journal} {\bibinfo  {journal} {Phys. Rev. D}\ }\textbf {\bibinfo {volume}
  {90}},\ \bibinfo {pages} {094002} (\bibinfo {year} {2014})}\BibitemShut
  {NoStop}%
\bibitem [{\citenamefont {Buchheim}\ \emph
  {et~al.}(2015{\natexlab{b}})\citenamefont {Buchheim}, \citenamefont
  {Kampfer},\ and\ \citenamefont {Hilger}}]{Buchheim:2015yyc}%
  \BibitemOpen
  \bibfield  {author} {\bibinfo {author} {\bibfnamefont {T.}~\bibnamefont
  {Buchheim}}, \bibinfo {author} {\bibfnamefont {B.}~\bibnamefont {Kampfer}}, \
  and\ \bibinfo {author} {\bibfnamefont {T.}~\bibnamefont {Hilger}},\
  }\href@noop {} {\ }\Eprint {http://arxiv.org/abs/1511.06234} {1511.06234
  [nucl-th]} \BibitemShut {NoStop}%
\bibitem [{\citenamefont {Buchheim}\ \emph
  {et~al.}(2015{\natexlab{c}})\citenamefont {Buchheim}, \citenamefont
  {Hilger},\ and\ \citenamefont {K\"ampfer}}]{Buchheim:2015xka}%
  \BibitemOpen
  \bibfield  {author} {\bibinfo {author} {\bibfnamefont {T.}~\bibnamefont
  {Buchheim}}, \bibinfo {author} {\bibfnamefont {T.}~\bibnamefont {Hilger}}, \
  and\ \bibinfo {author} {\bibfnamefont {B.}~\bibnamefont {K\"ampfer}},\
  }\href@noop {} {\ }\Eprint {http://arxiv.org/abs/1509.06144} {1509.06144
  [nucl-th]} \BibitemShut {NoStop}%
\bibitem [{\citenamefont {Gubler}\ and\ \citenamefont
  {Weise}(2015)}]{Gubler:2015yna}%
  \BibitemOpen
  \bibfield  {author} {\bibinfo {author} {\bibfnamefont {P.}~\bibnamefont
  {Gubler}}\ and\ \bibinfo {author} {\bibfnamefont {W.}~\bibnamefont {Weise}},\
  }\href {\doibase 10.1016/j.physletb.2015.10.068} {\bibfield  {journal}
  {\bibinfo  {journal} {Phys. Lett. B}\ }\textbf {\bibinfo {volume} {751}},\
  \bibinfo {pages} {396} (\bibinfo {year} {2015})}\BibitemShut {NoStop}%
\bibitem [{\citenamefont {Bowman}\ \emph {et~al.}(2003)\citenamefont {Bowman},
  \citenamefont {Heller}, \citenamefont {Leinweber},\ and\ \citenamefont
  {Williams}}]{Bowman:2002kn}%
  \BibitemOpen
  \bibfield  {author} {\bibinfo {author} {\bibfnamefont {P.~O.}\ \bibnamefont
  {Bowman}}, \bibinfo {author} {\bibfnamefont {U.~M.}\ \bibnamefont {Heller}},
  \bibinfo {author} {\bibfnamefont {D.~B.}\ \bibnamefont {Leinweber}}, \ and\
  \bibinfo {author} {\bibfnamefont {A.~G.}\ \bibnamefont {Williams}},\
  }\bibfield  {booktitle} {\emph {\bibinfo {booktitle} {{Lattice field theory.
  Proceedings: 20th International Symposium, Lattice 2002, Cambridge, USA, Jun
  24-29, 2002}}},\ }\href {\doibase 10.1016/S0920-5632(03)01533-0} {\bibfield
  {journal} {\bibinfo  {journal} {Nucl. Phys. Proc. Suppl.}\ }\textbf {\bibinfo
  {volume} {119}},\ \bibinfo {pages} {323} (\bibinfo {year} {2003})},\ \bibinfo
  {note} {[,323(2002)]}\BibitemShut {NoStop}%
\bibitem [{\citenamefont {Bowman}\ \emph {et~al.}(2002)\citenamefont {Bowman},
  \citenamefont {Heller},\ and\ \citenamefont {Williams}}]{Bowman:2002bm}%
  \BibitemOpen
  \bibfield  {author} {\bibinfo {author} {\bibfnamefont {P.~O.}\ \bibnamefont
  {Bowman}}, \bibinfo {author} {\bibfnamefont {U.~M.}\ \bibnamefont {Heller}},
  \ and\ \bibinfo {author} {\bibfnamefont {A.~G.}\ \bibnamefont {Williams}},\
  }\href {\doibase 10.1103/PhysRevD.66.014505} {\bibfield  {journal} {\bibinfo
  {journal} {Phys. Rev. D}\ }\textbf {\bibinfo {volume} {66}},\ \bibinfo
  {pages} {014505} (\bibinfo {year} {2002})}\BibitemShut {NoStop}%
\bibitem [{\citenamefont {Bhagwat}\ \emph {et~al.}(2003)\citenamefont
  {Bhagwat}, \citenamefont {Pichowsky}, \citenamefont {Roberts},\ and\
  \citenamefont {Tandy}}]{Bhagwat:2003vw}%
  \BibitemOpen
  \bibfield  {author} {\bibinfo {author} {\bibfnamefont {M.~S.}\ \bibnamefont
  {Bhagwat}}, \bibinfo {author} {\bibfnamefont {M.~A.}\ \bibnamefont
  {Pichowsky}}, \bibinfo {author} {\bibfnamefont {C.~D.}\ \bibnamefont
  {Roberts}}, \ and\ \bibinfo {author} {\bibfnamefont {P.~C.}\ \bibnamefont
  {Tandy}},\ }\href {\doibase 10.1103/PhysRevC.68.015203} {\bibfield  {journal}
  {\bibinfo  {journal} {Phys. Rev. C}\ }\textbf {\bibinfo {volume} {68}},\
  \bibinfo {pages} {015203} (\bibinfo {year} {2003})}\BibitemShut {NoStop}%
\bibitem [{\citenamefont {Olive}\ and\ \citenamefont {others (Particle
  Data~Group)}(2014)}]{Olive:2014rpp}%
  \BibitemOpen
  \bibfield  {author} {\bibinfo {author} {\bibfnamefont {K.~A.}\ \bibnamefont
  {Olive}}\ and\ \bibinfo {author} {\bibnamefont {others (Particle
  Data~Group)}},\ }\href@noop {} {\bibfield  {journal} {\bibinfo  {journal}
  {Chin. Phys. C}\ }\textbf {\bibinfo {volume} {38}},\ \bibinfo {pages}
  {090001} (\bibinfo {year} {2014})}\BibitemShut {NoStop}%
\bibitem [{\citenamefont {Eichten}\ and\ \citenamefont
  {Feinberg}(1981)}]{Eichten:1980mw}%
  \BibitemOpen
  \bibfield  {author} {\bibinfo {author} {\bibfnamefont {E.}~\bibnamefont
  {Eichten}}\ and\ \bibinfo {author} {\bibfnamefont {F.}~\bibnamefont
  {Feinberg}},\ }\href {\doibase 10.1103/PhysRevD.23.2724} {\bibfield
  {journal} {\bibinfo  {journal} {Phys. Rev. D}\ }\textbf {\bibinfo {volume}
  {23}},\ \bibinfo {pages} {2724} (\bibinfo {year} {1981})}\BibitemShut
  {NoStop}%
\bibitem [{\citenamefont {Stanley}\ and\ \citenamefont
  {Robson}(1980)}]{Stanley:1980zm}%
  \BibitemOpen
  \bibfield  {author} {\bibinfo {author} {\bibfnamefont {D.~P.}\ \bibnamefont
  {Stanley}}\ and\ \bibinfo {author} {\bibfnamefont {D.}~\bibnamefont
  {Robson}},\ }\href {\doibase 10.1103/PhysRevD.21.3180} {\bibfield  {journal}
  {\bibinfo  {journal} {Phys. Rev. D}\ }\textbf {\bibinfo {volume} {21}},\
  \bibinfo {pages} {3180} (\bibinfo {year} {1980})}\BibitemShut {NoStop}%
\bibitem [{\citenamefont {Buchmuller}\ and\ \citenamefont
  {Tye}(1981)}]{Buchmuller:1980su}%
  \BibitemOpen
  \bibfield  {author} {\bibinfo {author} {\bibfnamefont {W.}~\bibnamefont
  {Buchmuller}}\ and\ \bibinfo {author} {\bibfnamefont {S.~H.~H.}\ \bibnamefont
  {Tye}},\ }\href {\doibase 10.1103/PhysRevD.24.132} {\bibfield  {journal}
  {\bibinfo  {journal} {Phys. Rev. D}\ }\textbf {\bibinfo {volume} {24}},\
  \bibinfo {pages} {132} (\bibinfo {year} {1981})}\BibitemShut {NoStop}%
\bibitem [{\citenamefont {Godfrey}\ and\ \citenamefont
  {Isgur}(1985)}]{Godfrey:1985xj}%
  \BibitemOpen
  \bibfield  {author} {\bibinfo {author} {\bibfnamefont {S.}~\bibnamefont
  {Godfrey}}\ and\ \bibinfo {author} {\bibfnamefont {N.}~\bibnamefont
  {Isgur}},\ }\href {\doibase 10.1103/PhysRevD.32.189} {\bibfield  {journal}
  {\bibinfo  {journal} {Phys. Rev. D}\ }\textbf {\bibinfo {volume} {32}},\
  \bibinfo {pages} {189} (\bibinfo {year} {1985})}\BibitemShut {NoStop}%
\bibitem [{\citenamefont {Gershtein}\ \emph {et~al.}(1988)\citenamefont
  {Gershtein}, \citenamefont {Kiselev}, \citenamefont {Likhoded}, \citenamefont
  {Slabospitsky},\ and\ \citenamefont {Tkabladze}}]{Gershtein:1987jj}%
  \BibitemOpen
  \bibfield  {author} {\bibinfo {author} {\bibfnamefont {S.~S.}\ \bibnamefont
  {Gershtein}}, \bibinfo {author} {\bibfnamefont {V.~V.}\ \bibnamefont
  {Kiselev}}, \bibinfo {author} {\bibfnamefont {A.~K.}\ \bibnamefont
  {Likhoded}}, \bibinfo {author} {\bibfnamefont {S.~R.}\ \bibnamefont
  {Slabospitsky}}, \ and\ \bibinfo {author} {\bibfnamefont {A.~V.}\
  \bibnamefont {Tkabladze}},\ }\href@noop {} {\bibfield  {journal} {\bibinfo
  {journal} {Sov. J. Nucl. Phys.}\ }\textbf {\bibinfo {volume} {48}},\ \bibinfo
  {pages} {327} (\bibinfo {year} {1988})},\ \bibinfo {note} {[Yad.
  Fiz.48,515(1988)]}\BibitemShut {NoStop}%
\bibitem [{\citenamefont {Kaidalov}\ and\ \citenamefont
  {Nogteva}(1988)}]{Kaidalov:1987gk}%
  \BibitemOpen
  \bibfield  {author} {\bibinfo {author} {\bibfnamefont {A.~B.}\ \bibnamefont
  {Kaidalov}}\ and\ \bibinfo {author} {\bibfnamefont {A.~V.}\ \bibnamefont
  {Nogteva}},\ }\href@noop {} {\bibfield  {journal} {\bibinfo  {journal} {Sov.
  J. Nucl. Phys.}\ }\textbf {\bibinfo {volume} {47}},\ \bibinfo {pages} {321}
  (\bibinfo {year} {1988})},\ \bibinfo {note} {[Yad.
  Fiz.47,505(1988)]}\BibitemShut {NoStop}%
\bibitem [{\citenamefont {Kwong}\ and\ \citenamefont
  {Rosner}(1991)}]{Kwong:1990am}%
  \BibitemOpen
  \bibfield  {author} {\bibinfo {author} {\bibfnamefont {W.-k.}\ \bibnamefont
  {Kwong}}\ and\ \bibinfo {author} {\bibfnamefont {J.~L.}\ \bibnamefont
  {Rosner}},\ }\href {\doibase 10.1103/PhysRevD.44.212} {\bibfield  {journal}
  {\bibinfo  {journal} {Phys. Rev. D}\ }\textbf {\bibinfo {volume} {44}},\
  \bibinfo {pages} {212} (\bibinfo {year} {1991})}\BibitemShut {NoStop}%
\bibitem [{\citenamefont {Baker}\ \emph {et~al.}(1992)\citenamefont {Baker},
  \citenamefont {Ball},\ and\ \citenamefont {Zachariasen}}]{Baker:1991ty}%
  \BibitemOpen
  \bibfield  {author} {\bibinfo {author} {\bibfnamefont {M.}~\bibnamefont
  {Baker}}, \bibinfo {author} {\bibfnamefont {J.~S.}\ \bibnamefont {Ball}}, \
  and\ \bibinfo {author} {\bibfnamefont {F.}~\bibnamefont {Zachariasen}},\
  }\href {\doibase 10.1103/PhysRevD.45.910} {\bibfield  {journal} {\bibinfo
  {journal} {Phys. Rev. D}\ }\textbf {\bibinfo {volume} {45}},\ \bibinfo
  {pages} {910} (\bibinfo {year} {1992})}\BibitemShut {NoStop}%
\bibitem [{\citenamefont {Chen}\ and\ \citenamefont
  {Kuang}(1992)}]{Chen:1992fq}%
  \BibitemOpen
  \bibfield  {author} {\bibinfo {author} {\bibfnamefont {Y.-Q.}\ \bibnamefont
  {Chen}}\ and\ \bibinfo {author} {\bibfnamefont {Y.-P.}\ \bibnamefont
  {Kuang}},\ }\href {\doibase 10.1103/PhysRevD.47.350,
  10.1103/PhysRevD.46.1165} {\bibfield  {journal} {\bibinfo  {journal} {Phys.
  Rev. D}\ }\textbf {\bibinfo {volume} {46}},\ \bibinfo {pages} {1165}
  (\bibinfo {year} {1992})},\ \bibinfo {note} {[Erratum: Phys.
  Rev.D47,350(1993)]}\BibitemShut {NoStop}%
\bibitem [{\citenamefont {Itoh}\ \emph {et~al.}(1992)\citenamefont {Itoh},
  \citenamefont {Minamikawa}, \citenamefont {Miura},\ and\ \citenamefont
  {Watanabe}}]{Itoh:1992sd}%
  \BibitemOpen
  \bibfield  {author} {\bibinfo {author} {\bibfnamefont {C.}~\bibnamefont
  {Itoh}}, \bibinfo {author} {\bibfnamefont {T.}~\bibnamefont {Minamikawa}},
  \bibinfo {author} {\bibfnamefont {K.}~\bibnamefont {Miura}}, \ and\ \bibinfo
  {author} {\bibfnamefont {T.}~\bibnamefont {Watanabe}},\ }\href {\doibase
  10.1007/BF02731983} {\bibfield  {journal} {\bibinfo  {journal} {Nuovo Cim.
  A}\ }\textbf {\bibinfo {volume} {105}},\ \bibinfo {pages} {1539} (\bibinfo
  {year} {1992})}\BibitemShut {NoStop}%
\bibitem [{\citenamefont {Eichten}\ and\ \citenamefont
  {Quigg}(1994)}]{Eichten:1994gt}%
  \BibitemOpen
  \bibfield  {author} {\bibinfo {author} {\bibfnamefont {E.~J.}\ \bibnamefont
  {Eichten}}\ and\ \bibinfo {author} {\bibfnamefont {C.}~\bibnamefont
  {Quigg}},\ }\href {\doibase 10.1103/PhysRevD.49.5845} {\bibfield  {journal}
  {\bibinfo  {journal} {Phys. Rev. D}\ }\textbf {\bibinfo {volume} {49}},\
  \bibinfo {pages} {5845} (\bibinfo {year} {1994})}\BibitemShut {NoStop}%
\bibitem [{\citenamefont {Bagan}\ \emph {et~al.}(1994)\citenamefont {Bagan},
  \citenamefont {Dosch}, \citenamefont {Gosdzinsky}, \citenamefont {Narison},\
  and\ \citenamefont {Richard}}]{Bagan:1994dy}%
  \BibitemOpen
  \bibfield  {author} {\bibinfo {author} {\bibfnamefont {E.}~\bibnamefont
  {Bagan}}, \bibinfo {author} {\bibfnamefont {H.~G.}\ \bibnamefont {Dosch}},
  \bibinfo {author} {\bibfnamefont {P.}~\bibnamefont {Gosdzinsky}}, \bibinfo
  {author} {\bibfnamefont {S.}~\bibnamefont {Narison}}, \ and\ \bibinfo
  {author} {\bibfnamefont {J.~M.}\ \bibnamefont {Richard}},\ }\href {\doibase
  10.1007/BF01557235} {\bibfield  {journal} {\bibinfo  {journal} {Z. Phys.}\
  }\textbf {\bibinfo {volume} {C64}},\ \bibinfo {pages} {57} (\bibinfo {year}
  {1994})}\BibitemShut {NoStop}%
\bibitem [{\citenamefont {Zeng}\ \emph {et~al.}(1995)\citenamefont {Zeng},
  \citenamefont {Van~Orden},\ and\ \citenamefont {Roberts}}]{Zeng:1994vj}%
  \BibitemOpen
  \bibfield  {author} {\bibinfo {author} {\bibfnamefont {J.}~\bibnamefont
  {Zeng}}, \bibinfo {author} {\bibfnamefont {J.~W.}\ \bibnamefont {Van~Orden}},
  \ and\ \bibinfo {author} {\bibfnamefont {W.}~\bibnamefont {Roberts}},\ }\href
  {\doibase 10.1103/PhysRevD.52.5229} {\bibfield  {journal} {\bibinfo
  {journal} {Phys. Rev. D}\ }\textbf {\bibinfo {volume} {52}},\ \bibinfo
  {pages} {5229} (\bibinfo {year} {1995})}\BibitemShut {NoStop}%
\bibitem [{\citenamefont {Roncaglia}\ \emph {et~al.}(1995)\citenamefont
  {Roncaglia}, \citenamefont {Dzierba}, \citenamefont {Lichtenberg},\ and\
  \citenamefont {Predazzi}}]{Roncaglia:1994ex}%
  \BibitemOpen
  \bibfield  {author} {\bibinfo {author} {\bibfnamefont {R.}~\bibnamefont
  {Roncaglia}}, \bibinfo {author} {\bibfnamefont {A.}~\bibnamefont {Dzierba}},
  \bibinfo {author} {\bibfnamefont {D.~B.}\ \bibnamefont {Lichtenberg}}, \ and\
  \bibinfo {author} {\bibfnamefont {E.}~\bibnamefont {Predazzi}},\ }\href
  {\doibase 10.1103/PhysRevD.51.1248} {\bibfield  {journal} {\bibinfo
  {journal} {Phys. Rev. D}\ }\textbf {\bibinfo {volume} {51}},\ \bibinfo
  {pages} {1248} (\bibinfo {year} {1995})}\BibitemShut {NoStop}%
\bibitem [{\citenamefont {Kiselev}\ \emph {et~al.}(1995)\citenamefont
  {Kiselev}, \citenamefont {Likhoded},\ and\ \citenamefont
  {Tkabladze}}]{Kiselev:1994rc}%
  \BibitemOpen
  \bibfield  {author} {\bibinfo {author} {\bibfnamefont {V.~V.}\ \bibnamefont
  {Kiselev}}, \bibinfo {author} {\bibfnamefont {A.~K.}\ \bibnamefont
  {Likhoded}}, \ and\ \bibinfo {author} {\bibfnamefont {A.~V.}\ \bibnamefont
  {Tkabladze}},\ }\href {\doibase 10.1103/PhysRevD.51.3613} {\bibfield
  {journal} {\bibinfo  {journal} {Phys. Rev. D}\ }\textbf {\bibinfo {volume}
  {51}},\ \bibinfo {pages} {3613} (\bibinfo {year} {1995})}\BibitemShut
  {NoStop}%
\bibitem [{\citenamefont {Gupta}\ and\ \citenamefont
  {Johnson}(1996)}]{Gupta:1995ps}%
  \BibitemOpen
  \bibfield  {author} {\bibinfo {author} {\bibfnamefont {S.~N.}\ \bibnamefont
  {Gupta}}\ and\ \bibinfo {author} {\bibfnamefont {J.~M.}\ \bibnamefont
  {Johnson}},\ }\href {\doibase 10.1103/PhysRevD.53.312} {\bibfield  {journal}
  {\bibinfo  {journal} {Phys. Rev. D}\ }\textbf {\bibinfo {volume} {53}},\
  \bibinfo {pages} {312} (\bibinfo {year} {1996})}\BibitemShut {NoStop}%
\bibitem [{\citenamefont {Fulcher}(1999)}]{Fulcher:1998ka}%
  \BibitemOpen
  \bibfield  {author} {\bibinfo {author} {\bibfnamefont {L.~P.}\ \bibnamefont
  {Fulcher}},\ }\href {\doibase 10.1103/PhysRevD.60.074006} {\bibfield
  {journal} {\bibinfo  {journal} {Phys. Rev. D}\ }\textbf {\bibinfo {volume}
  {60}},\ \bibinfo {pages} {074006} (\bibinfo {year} {1999})}\BibitemShut
  {NoStop}%
\bibitem [{\citenamefont {Ebert}\ \emph {et~al.}(2003)\citenamefont {Ebert},
  \citenamefont {Faustov},\ and\ \citenamefont {Galkin}}]{Ebert:2002pp}%
  \BibitemOpen
  \bibfield  {author} {\bibinfo {author} {\bibfnamefont {D.}~\bibnamefont
  {Ebert}}, \bibinfo {author} {\bibfnamefont {R.~N.}\ \bibnamefont {Faustov}},
  \ and\ \bibinfo {author} {\bibfnamefont {V.~O.}\ \bibnamefont {Galkin}},\
  }\href {\doibase 10.1103/PhysRevD.67.014027} {\bibfield  {journal} {\bibinfo
  {journal} {Phys. Rev. D}\ }\textbf {\bibinfo {volume} {67}},\ \bibinfo
  {pages} {014027} (\bibinfo {year} {2003})}\BibitemShut {NoStop}%
\bibitem [{\citenamefont {Ikhdair}\ and\ \citenamefont
  {Sever}(2003)}]{Ikhdair:2003tt}%
  \BibitemOpen
  \bibfield  {author} {\bibinfo {author} {\bibfnamefont {S.~M.}\ \bibnamefont
  {Ikhdair}}\ and\ \bibinfo {author} {\bibfnamefont {R.}~\bibnamefont
  {Sever}},\ }\href {\doibase 10.1142/S0217751X03015088} {\bibfield  {journal}
  {\bibinfo  {journal} {Int. J. Mod. Phys. A}\ }\textbf {\bibinfo {volume}
  {18}},\ \bibinfo {pages} {4215} (\bibinfo {year} {2003})}\BibitemShut
  {NoStop}%
\bibitem [{\citenamefont {Ikhdair}\ and\ \citenamefont
  {Sever}(2004)}]{Ikhdair:2003ry}%
  \BibitemOpen
  \bibfield  {author} {\bibinfo {author} {\bibfnamefont {S.~M.}\ \bibnamefont
  {Ikhdair}}\ and\ \bibinfo {author} {\bibfnamefont {R.}~\bibnamefont
  {Sever}},\ }\href {\doibase 10.1142/S0217751X0401780X} {\bibfield  {journal}
  {\bibinfo  {journal} {Int. J. Mod. Phys. A}\ }\textbf {\bibinfo {volume}
  {19}},\ \bibinfo {pages} {1771} (\bibinfo {year} {2004})}\BibitemShut
  {NoStop}%
\bibitem [{\citenamefont {Godfrey}(2004)}]{Godfrey:2004ya}%
  \BibitemOpen
  \bibfield  {author} {\bibinfo {author} {\bibfnamefont {S.}~\bibnamefont
  {Godfrey}},\ }\href {\doibase 10.1103/PhysRevD.70.054017} {\bibfield
  {journal} {\bibinfo  {journal} {Phys. Rev. D}\ }\textbf {\bibinfo {volume}
  {70}},\ \bibinfo {pages} {054017} (\bibinfo {year} {2004})}\BibitemShut
  {NoStop}%
\bibitem [{\citenamefont {Ikhdair}\ and\ \citenamefont
  {Sever}(2005{\natexlab{a}})}]{Ikhdair:2004hg}%
  \BibitemOpen
  \bibfield  {author} {\bibinfo {author} {\bibfnamefont {S.~M.}\ \bibnamefont
  {Ikhdair}}\ and\ \bibinfo {author} {\bibfnamefont {R.}~\bibnamefont
  {Sever}},\ }\href {\doibase 10.1142/S0217751X05022275} {\bibfield  {journal}
  {\bibinfo  {journal} {Int. J. Mod. Phys. A}\ }\textbf {\bibinfo {volume}
  {20}},\ \bibinfo {pages} {4035} (\bibinfo {year}
  {2005}{\natexlab{a}})}\BibitemShut {NoStop}%
\bibitem [{\citenamefont {Ebert}\ \emph {et~al.}(2011)\citenamefont {Ebert},
  \citenamefont {Faustov},\ and\ \citenamefont {Galkin}}]{Ebert:2011jc}%
  \BibitemOpen
  \bibfield  {author} {\bibinfo {author} {\bibfnamefont {D.}~\bibnamefont
  {Ebert}}, \bibinfo {author} {\bibfnamefont {R.~N.}\ \bibnamefont {Faustov}},
  \ and\ \bibinfo {author} {\bibfnamefont {V.~O.}\ \bibnamefont {Galkin}},\
  }\href {\doibase 10.1140/epjc/s10052-011-1825-9} {\bibfield  {journal}
  {\bibinfo  {journal} {Eur. Phys. J. C}\ }\textbf {\bibinfo {volume} {71}},\
  \bibinfo {pages} {1825} (\bibinfo {year} {2011})}\BibitemShut {NoStop}%
\bibitem [{\citenamefont {Frederico}\ \emph {et~al.}(2002)\citenamefont
  {Frederico}, \citenamefont {Pauli},\ and\ \citenamefont
  {Zhou}}]{Frederico:2002vs}%
  \BibitemOpen
  \bibfield  {author} {\bibinfo {author} {\bibfnamefont {T.}~\bibnamefont
  {Frederico}}, \bibinfo {author} {\bibfnamefont {H.-C.}\ \bibnamefont
  {Pauli}}, \ and\ \bibinfo {author} {\bibfnamefont {S.-G.}\ \bibnamefont
  {Zhou}},\ }\href {\doibase 10.1103/PhysRevD.66.116011} {\bibfield  {journal}
  {\bibinfo  {journal} {Phys. Rev. D}\ }\textbf {\bibinfo {volume} {66}},\
  \bibinfo {pages} {116011} (\bibinfo {year} {2002})}\BibitemShut {NoStop}%
\bibitem [{\citenamefont {Choi}\ and\ \citenamefont {Ji}(2009)}]{Choi:2009ai}%
  \BibitemOpen
  \bibfield  {author} {\bibinfo {author} {\bibfnamefont {H.-M.}\ \bibnamefont
  {Choi}}\ and\ \bibinfo {author} {\bibfnamefont {C.-R.}\ \bibnamefont {Ji}},\
  }\href {\doibase 10.1103/PhysRevD.80.054016} {\bibfield  {journal} {\bibinfo
  {journal} {Phys. Rev. D}\ }\textbf {\bibinfo {volume} {80}},\ \bibinfo
  {pages} {054016} (\bibinfo {year} {2009})}\BibitemShut {NoStop}%
\bibitem [{\citenamefont {Choi}\ \emph {et~al.}(2015)\citenamefont {Choi},
  \citenamefont {Ji}, \citenamefont {Li},\ and\ \citenamefont
  {Ryu}}]{Choi:2015ywa}%
  \BibitemOpen
  \bibfield  {author} {\bibinfo {author} {\bibfnamefont {H.-M.}\ \bibnamefont
  {Choi}}, \bibinfo {author} {\bibfnamefont {C.-R.}\ \bibnamefont {Ji}},
  \bibinfo {author} {\bibfnamefont {Z.}~\bibnamefont {Li}}, \ and\ \bibinfo
  {author} {\bibfnamefont {H.-Y.}\ \bibnamefont {Ryu}},\ }\href {\doibase
  10.1103/PhysRevC.92.055203} {\bibfield  {journal} {\bibinfo  {journal} {Phys.
  Rev. C}\ }\textbf {\bibinfo {volume} {92}},\ \bibinfo {pages} {055203}
  (\bibinfo {year} {2015})}\BibitemShut {NoStop}%
\bibitem [{\citenamefont {Abd El-Hady}\ \emph {et~al.}(1999)\citenamefont {Abd
  El-Hady}, \citenamefont {Lodhi},\ and\ \citenamefont
  {Vary}}]{AbdElHady:1998kc}%
  \BibitemOpen
  \bibfield  {author} {\bibinfo {author} {\bibfnamefont {A.}~\bibnamefont {Abd
  El-Hady}}, \bibinfo {author} {\bibfnamefont {M.~A.~K.}\ \bibnamefont
  {Lodhi}}, \ and\ \bibinfo {author} {\bibfnamefont {J.~P.}\ \bibnamefont
  {Vary}},\ }\href {\doibase 10.1103/PhysRevD.59.094001} {\bibfield  {journal}
  {\bibinfo  {journal} {Phys. Rev. D}\ }\textbf {\bibinfo {volume} {59}},\
  \bibinfo {pages} {094001} (\bibinfo {year} {1999})}\BibitemShut {NoStop}%
\bibitem [{\citenamefont {Baldicchi}\ and\ \citenamefont
  {Prosperi}(2000)}]{Baldicchi:2000cf}%
  \BibitemOpen
  \bibfield  {author} {\bibinfo {author} {\bibfnamefont {M.}~\bibnamefont
  {Baldicchi}}\ and\ \bibinfo {author} {\bibfnamefont {G.~M.}\ \bibnamefont
  {Prosperi}},\ }\href {\doibase 10.1103/PhysRevD.62.114024} {\bibfield
  {journal} {\bibinfo  {journal} {Phys. Rev. D}\ }\textbf {\bibinfo {volume}
  {62}},\ \bibinfo {pages} {114024} (\bibinfo {year} {2000})}\BibitemShut
  {NoStop}%
\bibitem [{\citenamefont {Ikhdair}\ and\ \citenamefont
  {Sever}(2005{\natexlab{b}})}]{Ikhdair:2004tj}%
  \BibitemOpen
  \bibfield  {author} {\bibinfo {author} {\bibfnamefont {S.~M.}\ \bibnamefont
  {Ikhdair}}\ and\ \bibinfo {author} {\bibfnamefont {R.}~\bibnamefont
  {Sever}},\ }\href {\doibase 10.1142/S0217751X05021294} {\bibfield  {journal}
  {\bibinfo  {journal} {Int. J. Mod. Phys. A}\ }\textbf {\bibinfo {volume}
  {20}},\ \bibinfo {pages} {6509} (\bibinfo {year}
  {2005}{\natexlab{b}})}\BibitemShut {NoStop}%
\bibitem [{\citenamefont {Penin}\ \emph {et~al.}(2004)\citenamefont {Penin},
  \citenamefont {Pineda}, \citenamefont {Smirnov},\ and\ \citenamefont
  {Steinhauser}}]{Penin:2004xi}%
  \BibitemOpen
  \bibfield  {author} {\bibinfo {author} {\bibfnamefont {A.~A.}\ \bibnamefont
  {Penin}}, \bibinfo {author} {\bibfnamefont {A.}~\bibnamefont {Pineda}},
  \bibinfo {author} {\bibfnamefont {V.~A.}\ \bibnamefont {Smirnov}}, \ and\
  \bibinfo {author} {\bibfnamefont {M.}~\bibnamefont {Steinhauser}},\ }\href
  {\doibase 10.1016/j.physletb.2009.05.036, 10.1016/j.physletb.2004.04.066,
  10.1016/j.physletb.2009.12.035} {\bibfield  {journal} {\bibinfo  {journal}
  {Phys. Lett. B}\ }\textbf {\bibinfo {volume} {593}},\ \bibinfo {pages} {124}
  (\bibinfo {year} {2004})},\ \bibinfo {note} {[Erratum: Phys.
  Lett.683,358(2010)]}\BibitemShut {NoStop}%
\bibitem [{\citenamefont {Aliev}\ and\ \citenamefont
  {Yilmaz}(1992)}]{Aliev:1992vp}%
  \BibitemOpen
  \bibfield  {author} {\bibinfo {author} {\bibfnamefont {T.~M.}\ \bibnamefont
  {Aliev}}\ and\ \bibinfo {author} {\bibfnamefont {O.}~\bibnamefont {Yilmaz}},\
  }\href {\doibase 10.1007/BF02799097} {\bibfield  {journal} {\bibinfo
  {journal} {Nuovo Cim. A}\ }\textbf {\bibinfo {volume} {105}},\ \bibinfo
  {pages} {827} (\bibinfo {year} {1992})}\BibitemShut {NoStop}%
\bibitem [{\citenamefont {Gershtein}\ \emph {et~al.}(1995)\citenamefont
  {Gershtein}, \citenamefont {Kiselev}, \citenamefont {Likhoded},\ and\
  \citenamefont {Tkabladze}}]{Gershtein:1994jw}%
  \BibitemOpen
  \bibfield  {author} {\bibinfo {author} {\bibfnamefont {S.~S.}\ \bibnamefont
  {Gershtein}}, \bibinfo {author} {\bibfnamefont {V.~V.}\ \bibnamefont
  {Kiselev}}, \bibinfo {author} {\bibfnamefont {A.~K.}\ \bibnamefont
  {Likhoded}}, \ and\ \bibinfo {author} {\bibfnamefont {A.~V.}\ \bibnamefont
  {Tkabladze}},\ }\href {\doibase 10.1070/PU1995v038n01ABEH000063} {\bibfield
  {journal} {\bibinfo  {journal} {Phys. Usp.}\ }\textbf {\bibinfo {volume}
  {38}},\ \bibinfo {pages} {1} (\bibinfo {year} {1995})},\ \bibinfo {note}
  {[Usp. Fiz. Nauk165,3(1995)]}\BibitemShut {NoStop}%
\bibitem [{\citenamefont {Wang}(2013)}]{Wang:2012kw}%
  \BibitemOpen
  \bibfield  {author} {\bibinfo {author} {\bibfnamefont {Z.-G.}\ \bibnamefont
  {Wang}},\ }\href {\doibase 10.1140/epja/i2013-13131-7} {\bibfield  {journal}
  {\bibinfo  {journal} {Eur. Phys. J. A}\ }\textbf {\bibinfo {volume} {49}},\
  \bibinfo {pages} {131} (\bibinfo {year} {2013})}\BibitemShut {NoStop}%
\bibitem [{\citenamefont {Davies}\ \emph {et~al.}(1996)\citenamefont {Davies},
  \citenamefont {Hornbostel}, \citenamefont {Lepage}, \citenamefont {Lidsey},
  \citenamefont {Shigemitsu},\ and\ \citenamefont {Sloan}}]{Davies:1996gi}%
  \BibitemOpen
  \bibfield  {author} {\bibinfo {author} {\bibfnamefont {C.~T.~H.}\
  \bibnamefont {Davies}}, \bibinfo {author} {\bibfnamefont {K.}~\bibnamefont
  {Hornbostel}}, \bibinfo {author} {\bibfnamefont {G.~P.}\ \bibnamefont
  {Lepage}}, \bibinfo {author} {\bibfnamefont {A.~J.}\ \bibnamefont {Lidsey}},
  \bibinfo {author} {\bibfnamefont {J.}~\bibnamefont {Shigemitsu}}, \ and\
  \bibinfo {author} {\bibfnamefont {J.~H.}\ \bibnamefont {Sloan}},\ }\href
  {\doibase 10.1016/0370-2693(96)00650-8} {\bibfield  {journal} {\bibinfo
  {journal} {Phys. Lett. B}\ }\textbf {\bibinfo {volume} {382}},\ \bibinfo
  {pages} {131} (\bibinfo {year} {1996})}\BibitemShut {NoStop}%
\bibitem [{\citenamefont {Chiu}\ and\ \citenamefont
  {Hsieh}(2007)}]{Chiu:2007bc}%
  \BibitemOpen
  \bibfield  {author} {\bibinfo {author} {\bibfnamefont {T.-W.}\ \bibnamefont
  {Chiu}}\ and\ \bibinfo {author} {\bibfnamefont {T.-H.}\ \bibnamefont {Hsieh}}
  (\bibinfo {collaboration} {TWQCD}),\ }\bibfield  {booktitle} {\emph {\bibinfo
  {booktitle} {{Proceedings, 24th International Symposium on Lattice Field
  Theory (Lattice 2006)}}},\ }\href@noop {} {\bibfield  {journal} {\bibinfo
  {journal} {PoS}\ }\textbf {\bibinfo {volume} {LAT2006}},\ \bibinfo {pages}
  {180} (\bibinfo {year} {2007})}\BibitemShut {NoStop}%
\bibitem [{\citenamefont {Gregory}\ \emph {et~al.}(2010)\citenamefont
  {Gregory}, \citenamefont {Davies}, \citenamefont {Follana}, \citenamefont
  {Gamiz}, \citenamefont {Kendall}, \citenamefont {Lepage}, \citenamefont {Na},
  \citenamefont {Shigemitsu},\ and\ \citenamefont {Wong}}]{Gregory:2009hq}%
  \BibitemOpen
  \bibfield  {author} {\bibinfo {author} {\bibfnamefont {E.~B.}\ \bibnamefont
  {Gregory}}, \bibinfo {author} {\bibfnamefont {C.~T.~H.}\ \bibnamefont
  {Davies}}, \bibinfo {author} {\bibfnamefont {E.}~\bibnamefont {Follana}},
  \bibinfo {author} {\bibfnamefont {E.}~\bibnamefont {Gamiz}}, \bibinfo
  {author} {\bibfnamefont {I.~D.}\ \bibnamefont {Kendall}}, \bibinfo {author}
  {\bibfnamefont {G.~P.}\ \bibnamefont {Lepage}}, \bibinfo {author}
  {\bibfnamefont {H.}~\bibnamefont {Na}}, \bibinfo {author} {\bibfnamefont
  {J.}~\bibnamefont {Shigemitsu}}, \ and\ \bibinfo {author} {\bibfnamefont
  {K.~Y.}\ \bibnamefont {Wong}},\ }\href {\doibase
  10.1103/PhysRevLett.104.022001} {\bibfield  {journal} {\bibinfo  {journal}
  {Phys. Rev. Lett.}\ }\textbf {\bibinfo {volume} {104}},\ \bibinfo {pages}
  {022001} (\bibinfo {year} {2010})}\BibitemShut {NoStop}%
\bibitem [{\citenamefont {Dowdall}\ \emph {et~al.}(2012)\citenamefont
  {Dowdall}, \citenamefont {Davies}, \citenamefont {Hammant},\ and\
  \citenamefont {Horgan}}]{Dowdall:2012ab}%
  \BibitemOpen
  \bibfield  {author} {\bibinfo {author} {\bibfnamefont {R.~J.}\ \bibnamefont
  {Dowdall}}, \bibinfo {author} {\bibfnamefont {C.~T.~H.}\ \bibnamefont
  {Davies}}, \bibinfo {author} {\bibfnamefont {T.~C.}\ \bibnamefont {Hammant}},
  \ and\ \bibinfo {author} {\bibfnamefont {R.~R.}\ \bibnamefont {Horgan}},\
  }\href {\doibase 10.1103/PhysRevD.86.094510} {\bibfield  {journal} {\bibinfo
  {journal} {Phys. Rev. D}\ }\textbf {\bibinfo {volume} {86}},\ \bibinfo
  {pages} {094510} (\bibinfo {year} {2012})}\BibitemShut {NoStop}%
\bibitem [{\citenamefont {Burch}(2015)}]{Burch:2015pka}%
  \BibitemOpen
  \bibfield  {author} {\bibinfo {author} {\bibfnamefont {T.}~\bibnamefont
  {Burch}},\ }\href@noop {} {\ }\Eprint {http://arxiv.org/abs/1502.00675}
  {1502.00675 [hep-lat]} \BibitemShut {NoStop}%
\bibitem [{\citenamefont {Maris}\ and\ \citenamefont
  {Tandy}(2000)}]{Maris:1999bh}%
  \BibitemOpen
  \bibfield  {author} {\bibinfo {author} {\bibfnamefont {P.}~\bibnamefont
  {Maris}}\ and\ \bibinfo {author} {\bibfnamefont {P.~C.}\ \bibnamefont
  {Tandy}},\ }\href {\doibase 10.1103/PhysRevC.61.045202} {\bibfield  {journal}
  {\bibinfo  {journal} {Phys. Rev. C}\ }\textbf {\bibinfo {volume} {61}},\
  \bibinfo {pages} {045202} (\bibinfo {year} {2000})}\BibitemShut {NoStop}%
\bibitem [{\citenamefont {Maris}\ and\ \citenamefont
  {Tandy}(2002)}]{Maris:2002mz}%
  \BibitemOpen
  \bibfield  {author} {\bibinfo {author} {\bibfnamefont {P.}~\bibnamefont
  {Maris}}\ and\ \bibinfo {author} {\bibfnamefont {P.~C.}\ \bibnamefont
  {Tandy}},\ }\href {\doibase 10.1103/PhysRevC.65.045211} {\bibfield  {journal}
  {\bibinfo  {journal} {Phys. Rev. C}\ }\textbf {\bibinfo {volume} {65}},\
  \bibinfo {pages} {045211} (\bibinfo {year} {2002})}\BibitemShut {NoStop}%
\bibitem [{\citenamefont {Maris}\ and\ \citenamefont
  {Tandy}(2006)}]{Maris:2005tt}%
  \BibitemOpen
  \bibfield  {author} {\bibinfo {author} {\bibfnamefont {P.}~\bibnamefont
  {Maris}}\ and\ \bibinfo {author} {\bibfnamefont {P.~C.}\ \bibnamefont
  {Tandy}},\ }\href {\doibase 10.1016/j.nuclphysbps.2006.08.012} {\bibfield
  {journal} {\bibinfo  {journal} {Nucl. Phys. B, Proc. Suppl.}\ }\textbf
  {\bibinfo {volume} {161}},\ \bibinfo {pages} {136} (\bibinfo {year}
  {2006})}\BibitemShut {NoStop}%
\bibitem [{\citenamefont {Eichmann}(2014)}]{Eichmann:2014qva}%
  \BibitemOpen
  \bibfield  {author} {\bibinfo {author} {\bibfnamefont {G.}~\bibnamefont
  {Eichmann}},\ }\href {\doibase 10.5506/APhysPolBSupp.7.597} {\bibfield
  {journal} {\bibinfo  {journal} {Acta Phys.Polon.Supp.}\ }\textbf {\bibinfo
  {volume} {7}},\ \bibinfo {pages} {597} (\bibinfo {year} {2014})}\BibitemShut
  {NoStop}%
\bibitem [{\citenamefont {Ball}\ and\ \citenamefont
  {Chiu}(1980)}]{Ball:1980ay}%
  \BibitemOpen
  \bibfield  {author} {\bibinfo {author} {\bibfnamefont {J.~S.}\ \bibnamefont
  {Ball}}\ and\ \bibinfo {author} {\bibfnamefont {T.-W.}\ \bibnamefont
  {Chiu}},\ }\href {\doibase 10.1103/PhysRevD.22.2542} {\bibfield  {journal}
  {\bibinfo  {journal} {Phys. Rev. D}\ }\textbf {\bibinfo {volume} {22}},\
  \bibinfo {pages} {2542} (\bibinfo {year} {1980})}\BibitemShut {NoStop}%
\bibitem [{\citenamefont {Eichmann}(2009)}]{Eichmann:2009zx}%
  \BibitemOpen
  \bibfield  {author} {\bibinfo {author} {\bibfnamefont {G.}~\bibnamefont
  {Eichmann}},\ }\emph {\bibinfo {title} {{Hadron properties from QCD
  bound-state equations}}},\ \href@noop {} {Ph.D. thesis},\ \bibinfo  {school}
  {University of Graz} (\bibinfo {year} {2009}),\ \Eprint
  {http://arxiv.org/abs/0909.0703} {0909.0703 [hep-ph]} \BibitemShut {NoStop}%
\end{thebibliography}%

\end{document}